\documentstyle[aabib]{l-aa}       
\input psfig
\newcommand{\bm}{\begin{math}}
\newcommand{\enm}{\end{math}}
\newcommand{\bdm}{\begin{displaymath}}
\newcommand{\edm}{\end{displaymath}}
\newcommand{\be}{\begin{equation}}
\newcommand{\ee}{\end{equation}}
\newcommand{\bea}{\begin{eqnarray}}
\newcommand{\eea}{\end{eqnarray}}
\newcommand{\ba}{\begin{array}}
\newcommand{\ea}{\end{array}}
\newcommand{\btbb}{\begin{tabbing}}
\newcommand{\etbb}{\end{tabbing}}
\newcommand{\btab}{\begin{tabular}}
\newcommand{\etab}{\end{tabular}}
\newcommand{\bfi}{\begin{figure}}
\newcommand{\efi}{\end{figure}}
\newcommand{\bfid}{\begin{figure*}}
\newcommand{\efid}{\end{figure*}}
\newcommand{\btabl}{\begin{table}}
\newcommand{\etabl}{\end{table}}
\newcommand{\btabld}{\begin{table*}}
\newcommand{\etabld}{\end{table*}}
\newcommand{\bc}{\begin{center}}
\newcommand{\ec}{\end{center}}
\newcommand{\bab}{\begin{abstract}}
\newcommand{\eab}{\end{abstract}}
\newcommand{\bi}{\begin{itemize}}
\newcommand{\ei}{\end{itemize}}
\newcommand{\pa}{\partial}
\newcommand{\bbib}{\begin{thebibliography}}
\newcommand{\ebib}{\end{thebibliography}}
\newcommand{\ed}{\end{document}}
\def\gapprox{\;\rlap{\lower 2.5pt            
 \hbox{$\sim$}}\raise 1.5pt\hbox{$>$}\;}       
\def\lapprox{\;\rlap{\lower 2.5pt            
 \hbox{$\sim$}}\raise 1.5pt\hbox{$<$}\;} 
\begin{document}
%
%
%
%
%
%
\thesaurus{08(02.08.1;02.09.1;02.19.1;09.11.1;03.13.4)}
\title{Radiative cooling instability in 1D colliding flows}
\author{Rolf Walder\inst{1} and Doris Folini \inst{1,2}}      
\institute{Institut f\"ur Astronomie,
           ETH-Zentrum, CH-8092 Z\"urich, Switzerland; \\
           \hspace{0.3cm} E-mail: walder@astro.phys.ethz.ch,
          \and Seminar f\"ur Angewandte Mathematik, 
           ETH-Zentrum, CH-8092 Z\"urich, Switzerland; \\
           \hspace{0.3cm} E-mail: folini@astro.phys.ethz.ch}
\offprints{R. Walder}
\date{Received ... ; accepted ...}
\maketitle
%
%
%
%
%
%
\begin{abstract} 
Radiative shock waves show a strong cooling instability at
temperatures above approximately $2 \cdot 10^5$K. We numerically
investigate this instability by simulating different astronomical
objects in which colliding flows play an outstanding role: Wind
bubbles, supernova remnants, and colliding winds. Computing the flow
of each object over a large part of its evolutionary time and
resolving all physically relevant scales, we find several
phenomenologically different types of this instability. If two smooth
flows collide, the instability follows a periodic limit cycle with two
modes being important. The connection between the radiative loss
function and the mode and type of the overstability is discussed. The
collision of non-smooth flows can temporarily result in an aperiodic
evolution of the system. After a characteristic relaxation time the
instability then becomes periodic again. Such disturbances as well as
violent types of the instability can excite oscillations of the thin
layer of cold compressed gas downstream of the shock, which in turn
can influence the stability of the radiative shock.
\end{abstract} 
%
%
%
%
%
\section{Introduction}
\label{sec:intro}
\subsection{Radiative shocks}
\label{subsec:intro-astro}
Shock waves become radiative if the dynamical time of the flow exceeds
the cooling time of the shocked matter. Radiative shocks are present
in a variety of astrophysical objects: e.g. in supernova remnants, in
colliding winds of binary star systems, in accretion processes, in all
wind driven structures, in young stellar objects, in galactic and
extragalactic jets, in star burst galaxies, in active galactic
nuclei. Usually, such shocks are sufficiently strong to substantially
ionize and heat the matter. As the heated matter cools, it is
compressed into thin dense layers. The radiative cooling itself and
the presence of a dense layer of cooled gas greatly influence the
emitted spectrum. Radiative cooling of shock heated matter leads to
emission mostly in the frequency range between X-ray and UV. The cold
layer emits mostly in the radio, IR, optical, and UV. Therefore, a
better understanding of radiative shock waves, therefore, is crucial
for the interpretation of the observed spectra.

Of similar significance is the influence of radiative shock waves on
the dynamical shaping of their environment. The cold layer is much
denser than the surrounding matter and often prey to
instabilities of various types. In addition, the neighboring shock
heated zones are often subject to thermal instabilities. Such effects
can be crucial for the distribution and the state of the matter in the
vicinity of radiative shocks.

In this paper, we study the thermal instability of radiative shock
waves in the context of colliding flows.  Radiative cooling is
included as $\dot{\epsilon} = N^{2} \Lambda(T)$, where $\Lambda(T)$ is
piecewise linear on a log-log scale. The necessary spatial resolution
to resolve all the relevant length and time scales is provided by an
adaptive mesh refinement algorithm. In contrast to previous
investigations, we follow the long term evolution of the entire
interaction zone. In the case of a decelerating interaction zone this
leads to an automatic scanning of wide ranges of the steady state
shock velocity. This enables us to make a systematic study of the
cooling instability. It also allows us to investigate the mutual
influence of the cooling layer and the cold dense layer. In agreement
with other authors, we find that for smooth flows the system is
overstable: the instability is oscillatory and two different modes are
important. The phenomenology of the overstability, however, may be
very different within one mode which leads us to the introduction of
phenomenological types. In the case of density disturbances upstream
of the interaction zone, a temporarily aperiodic evolution can result
for which we give constraints. Finally, we emphasize the importance of
the cold dense layer for the thermal instability of the cooling layer.
\subsection{The thermal cooling instability}
\label{subsec:basics-therm_inst}
Catastrophic cooling and the overstability of radiative shock waves
due to thermal instability of radiative cooling have been known for
several decades. First numerical investigations of catastrophic
cooling in shock heated plasmas were carried out by Falle~(1975,1981).
\nocite{radshock:falle-1} \nocite{radshock:falle-2} 
First numerical investigations of the overstability of radiative shock
waves were made by Langer et al.~(1981,1982)
\nocite{radshock:langer-chanmugam-shaviv-stab1}
\nocite{radshock:langer-chanmugam-shaviv-stab2}
for the case of accretion flows onto compact objects.

\cite*{radshock:cheva-ima-stab1} (CI in the following) performed a 
linear stability analysis for a planar flow against a wall as a model
for accretion flows. They applied a cooling law of the form
$\dot{\epsilon}\!=\!N^2\!\cdot\!T^{\beta}$, and found the cooling layer
between the accretion shock and the wall to be thermally unstable for
$\beta<\beta_{c}$. An infinite number of modes with successively
higher eigenfrequencies were found, among them a fundamental and a
first overtone mode, for which $\beta_{c}\lapprox$~0.4 and 0.8
respectively. Numerical simulations carried out by
\cite*{radshock:imamura-wolff-durisen} were in agreement with these
results. \cite*{radshock:wolff-gardner-wood-stabshock1} and
\cite*{radshock:imamura-wolff-stab1} extended these investigations to 
two-temperature models with heat conduction and different cooling
processes such as Compton cooling and relativistic Bremsstrahlung. They
still found the same modes of the overstable oscillation of the
radiative accretion shock. 

In a beautiful analytical paper, \cite*{radshock:bertsch-stab2}
extended the analysis to interstellar shocks. He showed that under
certain conditions there exists a piecewise self-similar solution in
spherical symmetry. The flow pattern is more complex than for
accretion flows: a leading shock, a cooling layer behind the shock, a
hot or cold interior, and a thin layer of already cooled, compressed
gas between the interior and the cooling layer. Performing a linear
stability analysis on this solution that allowed for disturbances in
all three spatial directions, he found two types of instabilities, a
thermal and a dynamical one. For the thermal one he found the same
unstable modes in the radial direction as
\cite*{radshock:cheva-ima-stab1}. Surprisingly, he found that no new
modes are introduced by nonradial disturbances. But modes stable to
radial perturbations can be unstable to nonradial perturbations.

On the numerical side, significant progress has been made by including
detailed radiation processes into the analysis. 
\cite*{radshock:innes-giddings-falle-2} (IGF in the following) and 
\cite*{radshock:gaetz-edgar-chevalier} (GEC in the following) considered
ionization and recombination processes as well as time-dependent
non-equilibrium cooling. They agreed on the onset of the instability
for shocks having velocities above 130~km/s -- 150~km/s. GEC found
this instability to be oscillatory which is in agreement with further
investigations presented by~\cite*{radshock:innes}. 

Recently, \cite*{radshock:plewa-1995} studied strong shock waves
($v_{S}>1000 $ km/s) running into dense material and found overstable
behavior also under these conditions.

In spherical symmetry, stability analyses of radiative shocks were
performed by~\cite*{radshock:imamura-wolff-stab1},
\cite*{radshock:houck-chevalier}, and~\cite*{hstab:dgani-soker-cadavid}.
Spherically symmetric cooling shells have significantly different
stability properties as long as their size is comparable to the
relevant radius. For small cooling shells,the stability
behavior approaches the planar limit behavior. Attempts to analyze the
thermal cooling instability of two-dimensional planar shocks have been
made by~\cite*{radshock:strickland-blondin-1995}.

The organization of the article is as follows: In
Sect.~\ref{sec:model} we briefly describe our physical model. In
Sect.~\ref{sec:code}, this is followed by some notes on the applied
numerical methods. In Sect.~\ref{sec:baspat} the basic flow patterns
are described and the sample of computations is presented. The
thermal cooling instability for the case of smooth flows is analyzed
in Sect.~\ref{sec:instab}, and in Sect.~\ref{sec:disturb} the same is
done for the case of disturbed flows. The evolution of the cold dense
layer is investigated in Sect.~\ref{sec:coldshell}. In
Sect.~\ref{sec:discussion} we discuss our results, and, finally, we
present the conclusions in Sect.~\ref{sec:conclusion}.

%
%
%
\section{The physical model}
\label{sec:model}
\subsection{The governing equations}
\label{subsec:model-equations}
Our physical model is based on the Euler equations and the use of
parameterized radiative loss functions (RLFs in the following). We
numerically solve the Euler equations in spherical symmetry:
\bea
\frac{ \pa \rho}{\pa t} +  \frac{\pa \rho v_r }{\pa r}
     & = & - \frac{2 v_r}{r}\rho ,
\label{eq:euler1} \\
\frac{\pa \left( \rho v_r \right) }{\pa t} 
+ \frac{\pa \left(\rho v_r^2 + p \right)}{\pa r}
& =  & - \frac{2 v_r}{r}\rho v_r ,
\label{eq:euler2} \\
\frac{\pa E}{\pa t} + 
  \frac{\pa \left[ v_r \left( E + p \right) \right]}{ \pa r}
& = & -  \frac{2 v_r}{r} \left(E + p \right)  \\
&   & -  \dot{\epsilon}(T,N) + q(T). \nonumber  
\label{eq:euler3} 
\eea
Here, $\rho$ denotes the mass density, $v_r$ the radial velocity, p
the pressure, $\dot{\epsilon}(T,N)$ the cooling term, and $ q(T) $ the
heating term. The last two are discussed below. The total energy
density $ E $ is the sum of the thermal energy density $ U $ and the
kinetic energy density:
\be
E = U + \rho v_{r}^{2} / 2 .
\ee
$\rho$ and the number density N are connected by
\be
\rho = N \mu m_{\rm H} ,
\ee
where $m_{\rm H}$ denotes the mass of the hydrogen atom and $\mu$ the
mean atomic weight. In our calculations we always use $\mu = 1.3$.

This set is closed by a polytropic equation of state (with $\gamma=5/3$
in all our calculations),
\be
U = \frac{p}{\gamma - 1} .
\label{eq:state}
\ee
The temperature is given by the ideal gas law as
\be
T = \frac{p \mu m_{\rm H}}{\rho k_B } . 
\ee
This model excludes heat conduction which is probably important for
strongly shocked astrophysical plasmae.
\subsection{The cooling model}
\label{subsec:model-cool}
We describe the radiated energy density per time, $\dot{\epsilon}$, by
a temperature dependent RLF $\Lambda(T)$ times the number density
squared,
\be
   \dot{\epsilon}(T,N) = N^{2} \cdot \Lambda(T) .
   \label{eq:erad}
\ee
For the RLF $ \Lambda (T) $ we use the approximation
\be 
   \Lambda(T) = \Lambda_{0_i}(T) T^{\beta_{i}} ,
   \hspace{0.5cm} T \in [T_{i},T_{i+1}].
   \label{eq:lambda}
\ee
RLFs have been published by several authors. The functions vary
significantly depending on the different physical processes which are
included. The assumed elemental abundances are of great importance for
the RLF. The first radiative loss function (RLF1) we used is a fit to
an optically thin, time independent RLF taken
from~\cite*{coolfunc:Cook} which uses solar photospheric abundances.
The second loss function (RLF2) is a fit to an optically thin, time
dependent radiative loss function taken from
\cite*{coolfunc:Schmutzler} which uses Allen's abundances. This RLF
has been obtained by a thermally self-consistent, time-dependent
computation of the cooling of a plasma from $10^{9}$~K to $10^{2}$~K.
Both RLFs are shown in Fig.~\ref{fig:cool-func} and their exact values
are given in Tables~\ref{tab:scal1} and~\ref{tab:scal2} in the
appendix.  In all our calculations $ \dot{\epsilon} $ is set to zero
for temperatures below a threshold temperature $ T_{C} $.

Our cooling model neglects processes which cool only linearly in the
density, e.g. cooling by inverse Compton scattering. The use of RLFs
also completely neglects ionization and recombination processes.
Therefore, we probably overestimate the post-shock temperature as the
kinetic energy converted by the shock is only used to heat the
matter. Moreover, in reality, the newly shocked matter is likely to be
underionized whereas gas on its cooling track tends to be overionized.
Consequently, the gas may radiate at a considerably different rate
from the equilibrium value used for RLF1. It probably also radiates at
a considerably different rate than computed by the time-dependent
RLF2, since the two cooling histories and thus the ionization
structures do not correspond. The consequences of these
simplifications are discussed in Sect.~\ref{sec:discussion}.
\bfi[htp]
\centerline{
      \psfig{figure=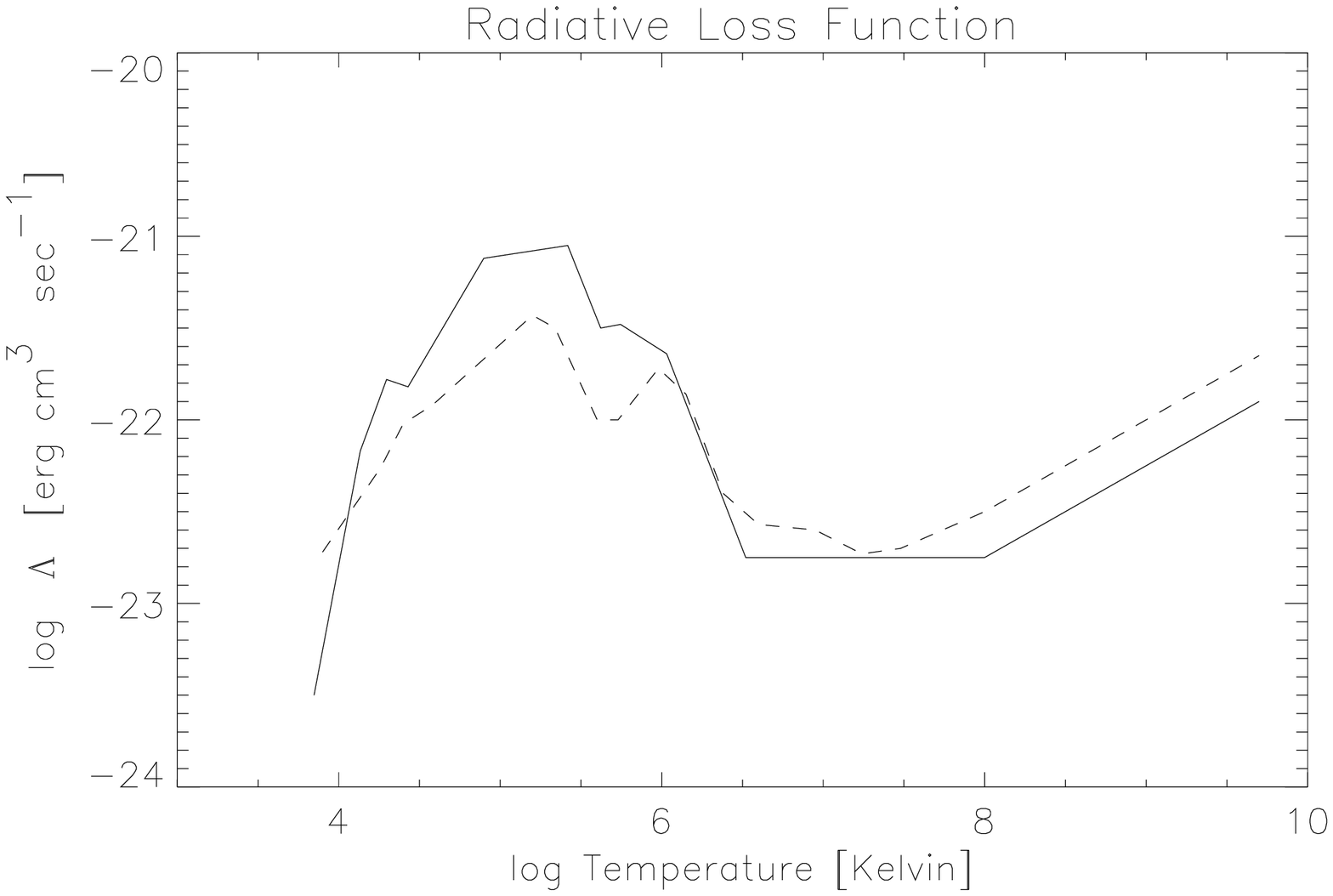,height=4.2cm,width=8.cm}
           }
\caption{The two RLFs used: Cook et al.~(1989) (solid line) and
         Schmutzler and Tscharnuter~(1993) (dashed line).}
\label{fig:cool-func}
\efi
\nocite{coolfunc:Cook}

Therefore, our physical model is more approximate than the models of
IGF and GEC. However, the use of an RLF has the big advantage to be
simple enough to allow an analysis of the connection between the RLF
and the response of the system. At the same time, it exhibits already
most of the features which are characteristic for radiative cooling
processes behind shocks. Moreover, the computational costs are
sufficiently low to allow a systematic study.

For later use we give the expression for the cooling time~$\tau $ (by
use of eqs.~\ref{eq:erad}, \ref{eq:lambda} and $U=3/2kT$),
\be
\tau(T,N) =  \frac{T}{\left | dT/ dt \right | } 
          =        \frac{3k}{2 \Lambda_{0_i}} \frac{T^{1-\beta_{i}}}{N}.
\ee
%
%
%
%
\subsection{The heating model}
\label{subsec:model-heat}
To account for radiative heating of the flow we have introduced a
heating term which, for temperatures below a certain threshold
temperature $ T_{H} \le T_{C} $, sets the gas temperature to
T$_{\mbox{H}}$.

%
%
%
%
\section{The numerical methods}
\label{sec:code}
\subsection{The AMRCART code}
Equations~\ref{eq:euler1}--\ref{eq:euler3} and~\ref{eq:state} are
solved by the spherically symmetric version of our AMRCART
code\footnote{A public version of the code will soon be
available. Requests please to walder@astro.phys.ethz.ch\,.} on the
basis of a Cartesian discretization. AMRCART combines a modern high
resolution Eulerian MUSCL scheme developed by~\cite*{meth:col-gl-real}
with the adaptive mesh refinement algorithm by~\cite*{berger:mesh1}.
It automatically refines the spatial grid and the time-step when the
local truncation error is estimated to be bigger than a given
threshold. This code was originally designed by 
\cite*{amr:be-rjl-amr-1} for pure gas-dynamics in two space
dimensions. It was adapted by~\cite*{walder-thesis} to compute complex
astrophysical problems, including reactive flows and arbitrary
physical source terms in one, two, and three space dimensions.
\bfi[tp]
\centerline{
  \psfig{figure=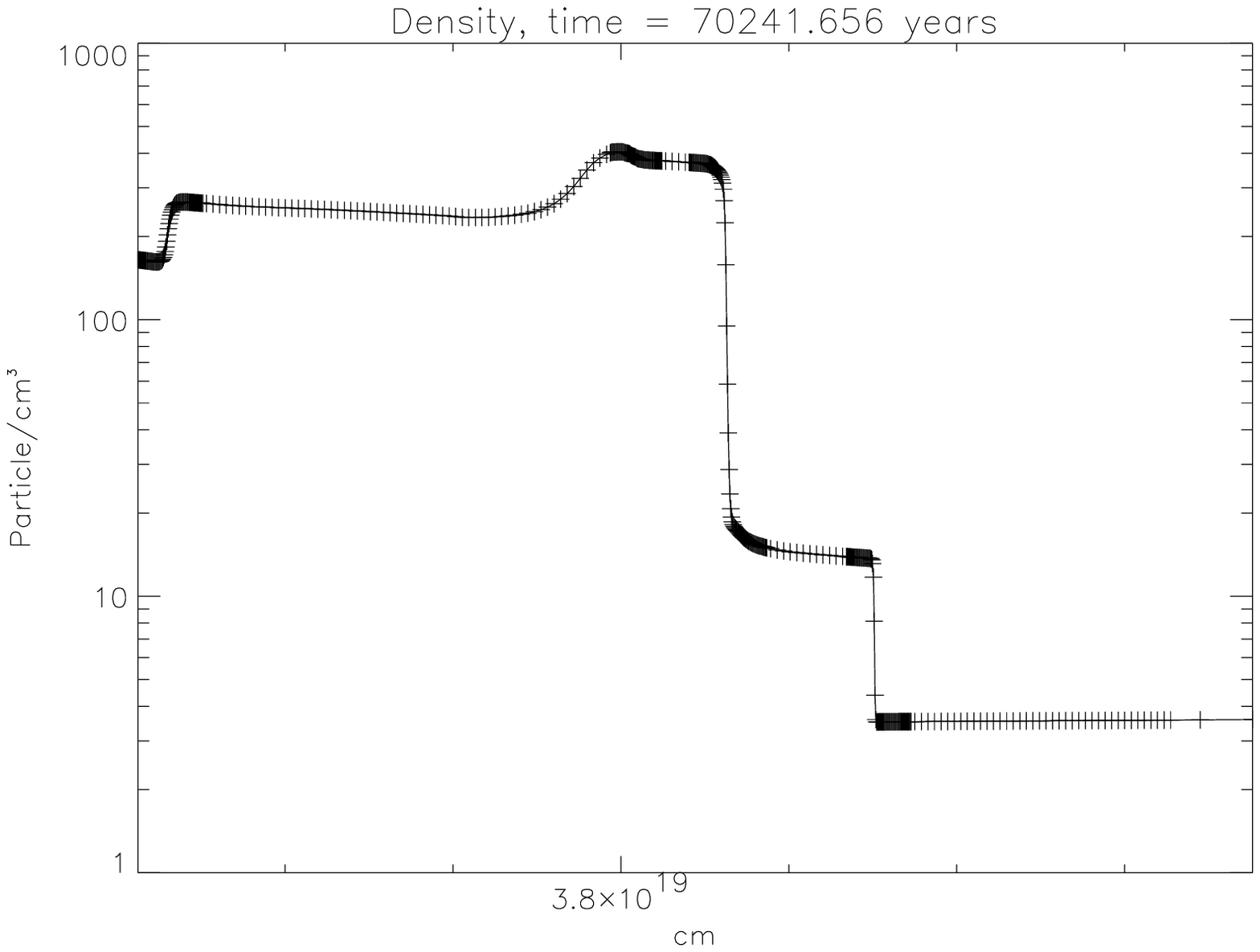,height=4.cm,width=8.4cm}
           }
\caption{Illustration of the mesh refinement algorithm.
         Shown are a computed density profile and the 
         numerical mesh. Three levels (out of five) of refinement 
         are shown. The mesh size is automatically refined by the code.}
\label{fig:code-grid}
\efi

The mesh refinement algorithm is crucial to the quality of our
simulations. Fig.~\ref{fig:code-grid} shows how finer meshes are
created to improve the spatial resolution where it is needed. In
regions with refined meshes the time-step is reduced by the same
factor as well. Thus, each time-scale which is dynamically important
can be resolved numerically. We emphasize that even in regions with
extreme high densities, and at temperatures at which the RLF
$\Lambda(T)$ is at its maximum, the radiative energy loss of each cell
during one time step $\Delta t$ is at most one percent of the current
thermal energy in this cell in all our calculations.
\subsection{Some technical details}
In all presented results the center of symmetry is located outside the
left boundary of the computational domain. At the left boundary we use
reflecting (supernova remnant) or inflow boundary conditions (wind
bubble and colliding winds). We use free outflow conditions at the
right boundary.
\subsection{A note on the numerical error}
\label{subsec:num_err}
\bfi[htp]
\centerline{
      \psfig{figure=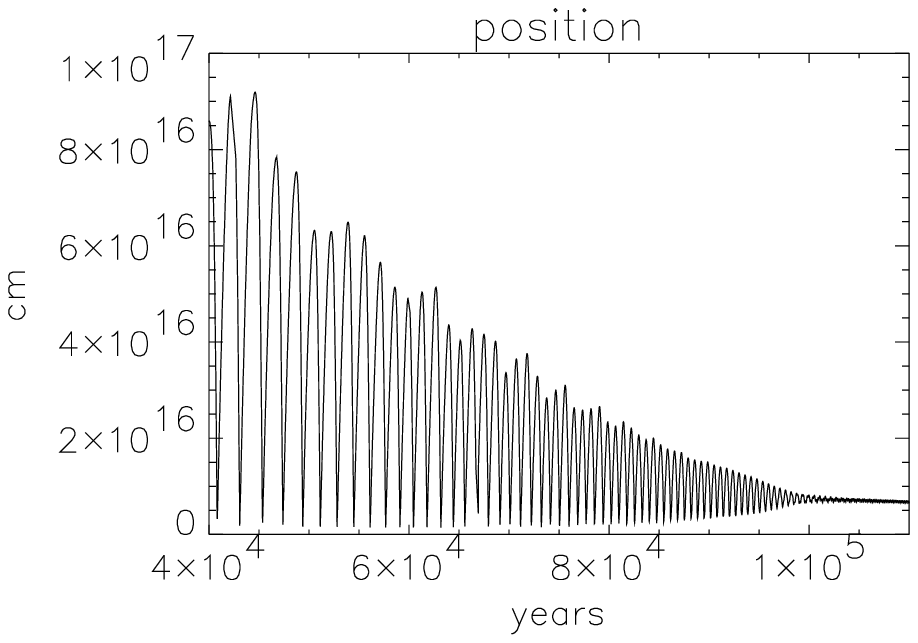,height=3.5cm,width=8.4cm}
           }
\centerline{
      \psfig{figure=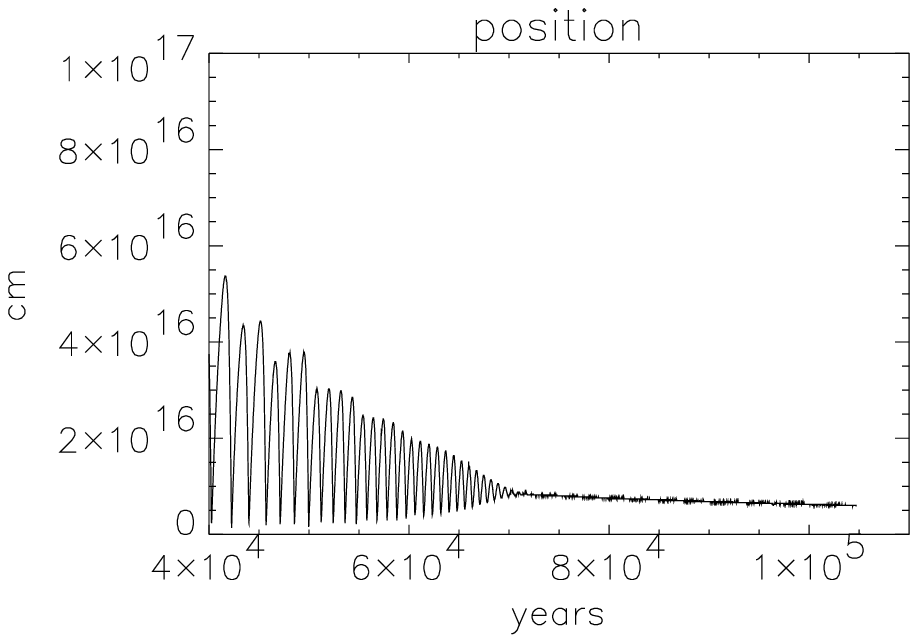,height=3.5cm,width=8.4cm}
           }
\caption{High- (top) and low-resolution (bottom) simulation of the
         wind bubble model WB1. Shown is the computed size of of
         the radiative shock. The high-resolution simulation shows 
         a bigger oscillation amplitude and the oscillation lasts
         nearly twice as long as in the low-resolution simulation.
         Note that in the high-resolution case the oscillation
         continues after 110'000~years but is substantially smaller
         after having undergone a mode-transition (see Fig.~7).}
\label{fig:rn_lev4_lev5}
\efi
Numerically obtained solutions are approximate solutions. Therefore,
one should always convince oneself that the numerical solution is
sufficiently close to the correct one.  Figure~\ref{fig:rn_lev4_lev5}
shows the significant, even qualitative difference between a {\it
high-} and a {\it low-resolution} numerical solution of an unstable
wind bubble (model WB1 of Table~\ref{tab:sample}). They have been
calculated with a finest spatial discretization of about
$2.4\cdot10^{13}$~cm and $1.92\cdot10^{14}$~cm respectively. The size
of the computational domain is $10^{20}$~cm. In both calculations the
cfl-number has been fixed to 0.2. Thus, the time-steps differ by a
factor of 8 as well. Note that even the low-resolution calculation has
still a remarkably good resolution compared to most computations given
in the literature.

This large error can be pinned down to the numerical smearing of the
boundary interface between the cooling layer and the cold compressed
layer. In connection with stiff source terms (radiative cooling in our
case), a too broad smearing of the interfaces can cause unphysical
wave speeds (see e.g.~\cite{meth:rjl-yee-stiffsources},
\cite{sources:colella-majda-Roytburd} or~\cite{source:klingenstein}). 
This problem is inherent to each numerical algorithm which is based on
interface capturing by finite volumes.

We estimate the errors of our calculated interface speeds to be at
most a few percent.
%
%
%
%
\section{Investigated flows}
\label{sec:baspat}
%
%
%
%
%
Simulations of wind bubbles (WB), supernova remnants (SNR) and
colliding winds (CW) are the basis for our study. These flows are well
documented in the literature and we give here only the characteristics
needed for this work.
\begin{figure}
%
\centerline{
  \psfig{figure=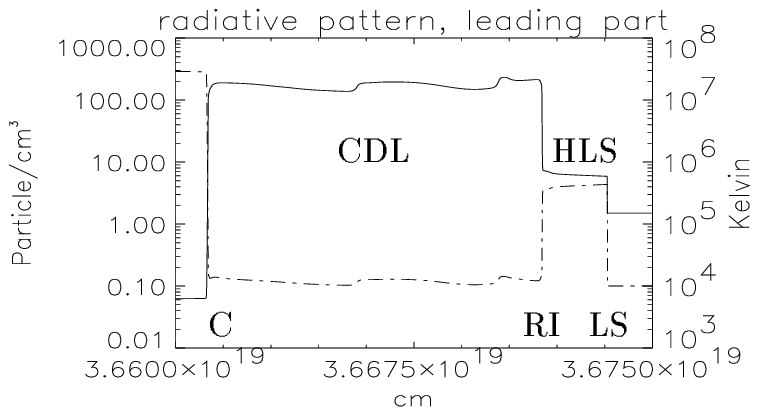,width=7.3cm,height=4.3cm}
            }
\vspace{-0.3cm}
\caption{Leading part of the interaction zone  in the radiative case. For
         details see text.}
\label{fig:adia-rad-pattern}
\end{figure}
A WB (see Fig.~\ref{fig:adia-rad-pattern}) is the structure which
results when a strong stellar wind blows into the uniform interstellar
medium. Similarly, a SNR is the result of a strong explosion driving
expelled gas against the uniform interstellar medium. In the case of
CW, a fast wind sweeps up a slower predecessor wind from the same
star. In the {\it adiabatic case}, all three flows are self-similar.
The leading shock decelerates in the WB-case according to $v_s\propto
t^{-2/5}$ (v$_{\mbox{s}}$ denotes the shock velocity, t the time) and
in the SNR-case according to $v_s\propto t^{-3/5}$. \cite*{sim:ryvi1}
give a nice overview of these flows. In the CW-case, the velocity of
the leading shock is constant (\cite{sim:ci1}). If {\it cooling is
included} the flows are no longer self-similar. However,
\cite*{radshock:bertsch-stab2} shows that under certain conditions
radiative SNR and WB can be described by piecewise self-similar
solutions.

We are interested in flows where the cooling layer is small against
the radius of the structure. In this case -- despite their different
driving mechanisms -- WB, SNR and CW have a similar leading part of
the interaction zone where the two flows collide. The leading part of
the interaction zone is illustrated in Fig.~\ref{fig:adia-rad-pattern}
at the example of the WB. In all three cases, it consists of a leading
shock (LS), a hot leading shell in which the shocked material cools
(HLS) and a 'cold' dense layer with compressed, already cooled shocked
material (CDL). The HLS is subject to the cooling instability, the CDL
can play an important role for the evolution of this instability. We
are interested in these two parts. The contact discontinuity (C) marks
the rear end of the leading part of the interaction zone. Notice that
in our study the reverse shock is nearly adiabatic since -- due to the
much lower density and the much higher temperature -- the cooling time
is considerably longer than the dynamical time. In the following we do
not discuss this reverse, driving part of the flow.

We list the exact settings of all our calculations in
Table~\ref{tab:scal1} in the appendix. We performed SNR and WB
simulations with different parameters up to the time where the thermal
instability ceased (see Fig.~\ref{fig:rn_lev4_lev5}). CW were
performed over a dynamical time which is equal to approximately
600~cooling times of the leading shell. All three flows have been
calculated with both RLFs given in Sect.~\ref{subsec:model-cool} and
in the appendix. The results presented below are the quintessence of
all these calculations. Of major importance is the WB
calculation. Since the structure moderately decelerates, it is
possible to scan a wide range of post-shock temperatures (in our case
from T $\approx 10^{6.5}$~K to T $\approx 10^5$~K) while the global
conditions remain nearly constant over one oscillation period of the
instability.

%
%
%
%
\section{The cooling instability of the leading shell for
                                   colliding smooth flows}
\label{sec:instab}
In this section we present the results for the case where two smooth
flows collide. In Sect.~\ref{sec:disturb} we investigate the
instability for disturbed flows. At present, smoothness means that
during one oscillation cycle the conditions for the radiative shock
remain nearly constant. In Sect.~\ref{sec:disturb} we give more
thorough constraints.
\begin{figure*}[htp]	
\centerline{\bf \large \hspace{3.75cm} 
snapshot 1 \hspace{3.cm} snapshot 2 \hspace{3.cm} snapshot 3 \hspace{3.75cm}}
%
%
%
\centerline{
 \psfig{figure=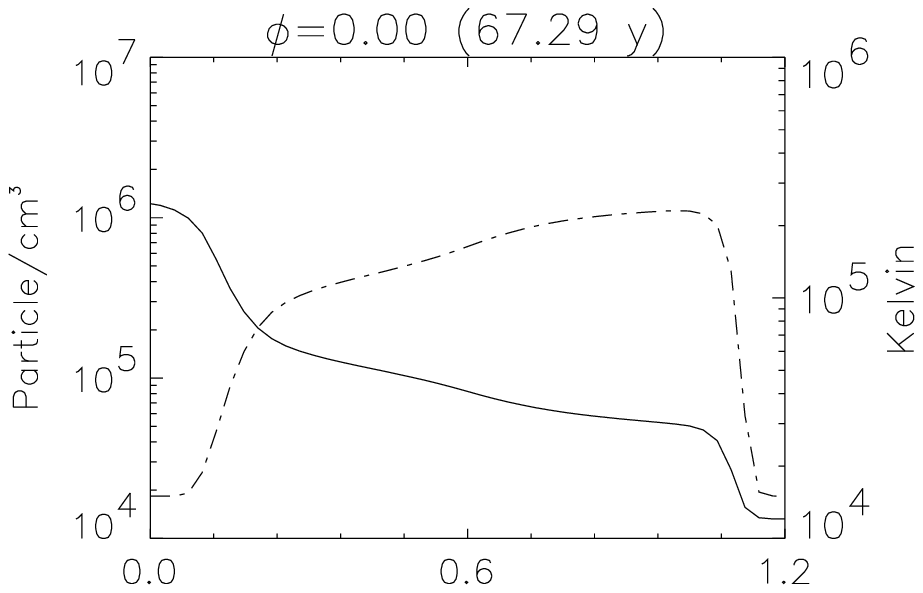,height=4cm,width=5.73cm}
 \psfig{figure=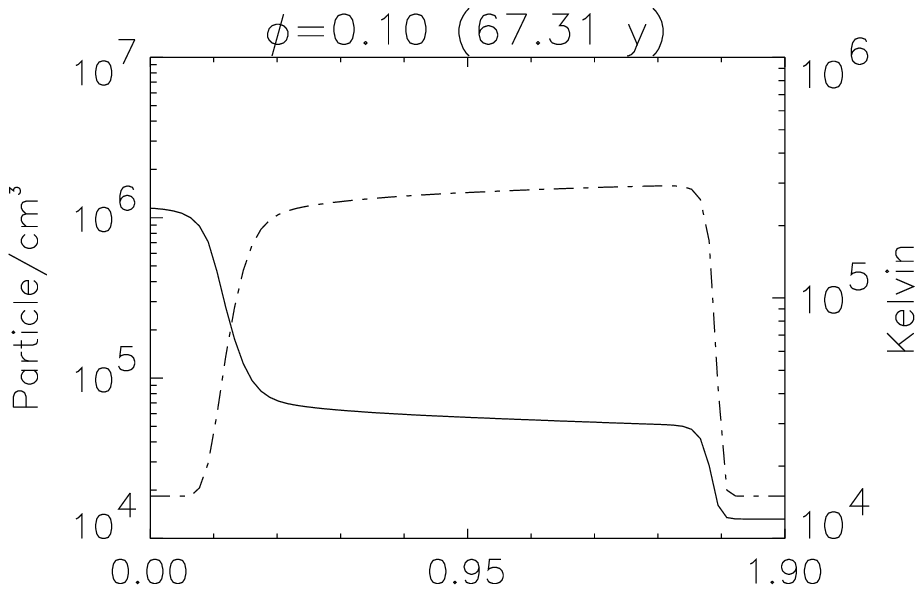,height=4cm,width=5.73cm}
 \psfig{figure=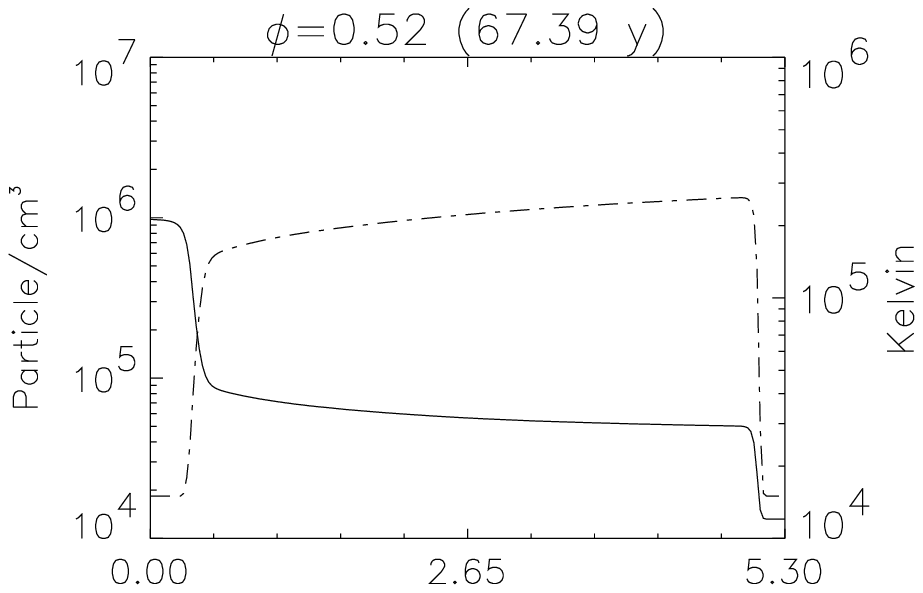,height=4cm,width=5.73cm}
           }    
%
%
\centerline{
\psfig{figure=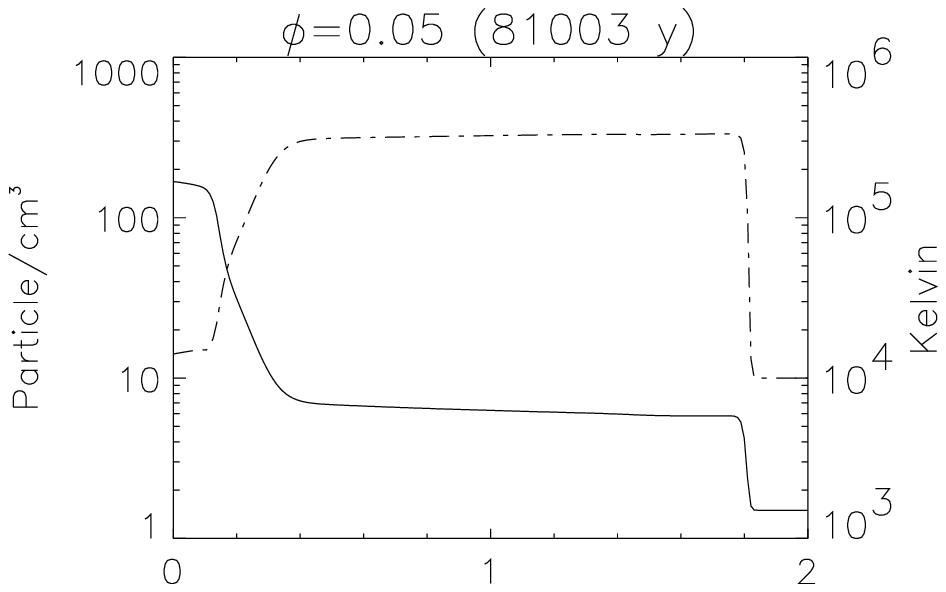,height=4cm,width=5.73cm}
\psfig{figure=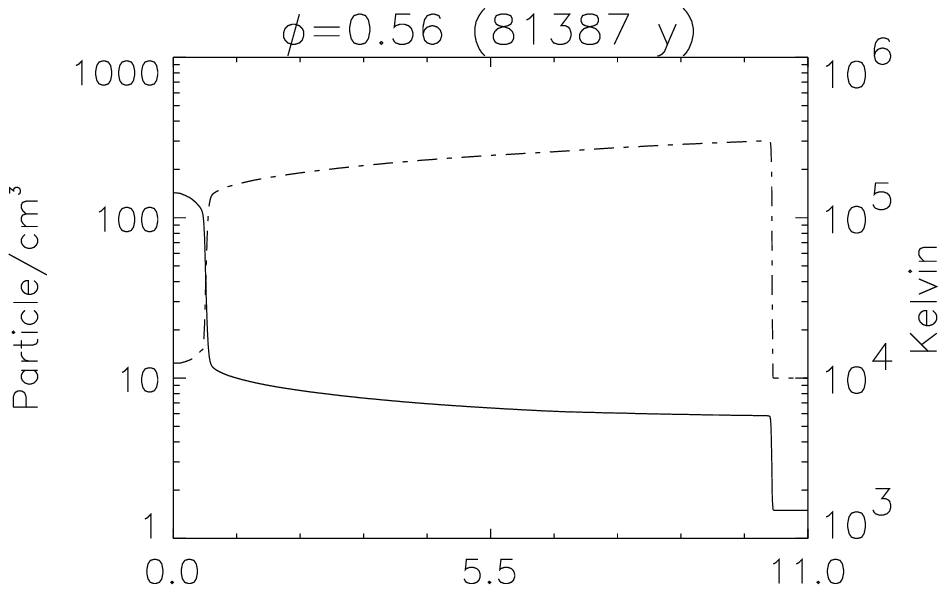,height=4cm,width=5.73cm}
\psfig{figure=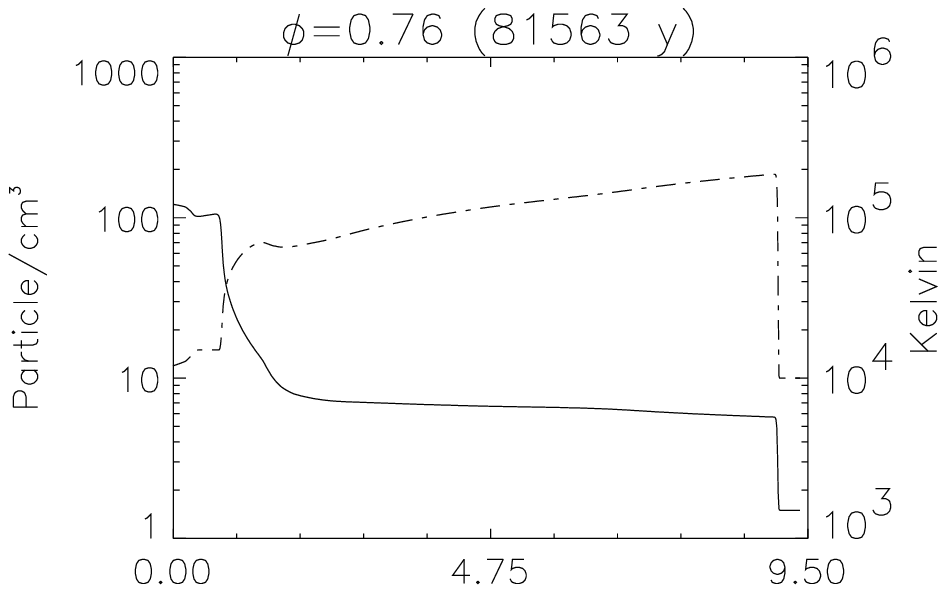,height=4cm,width=5.73cm}
           }
%
%
\centerline{
\psfig{figure=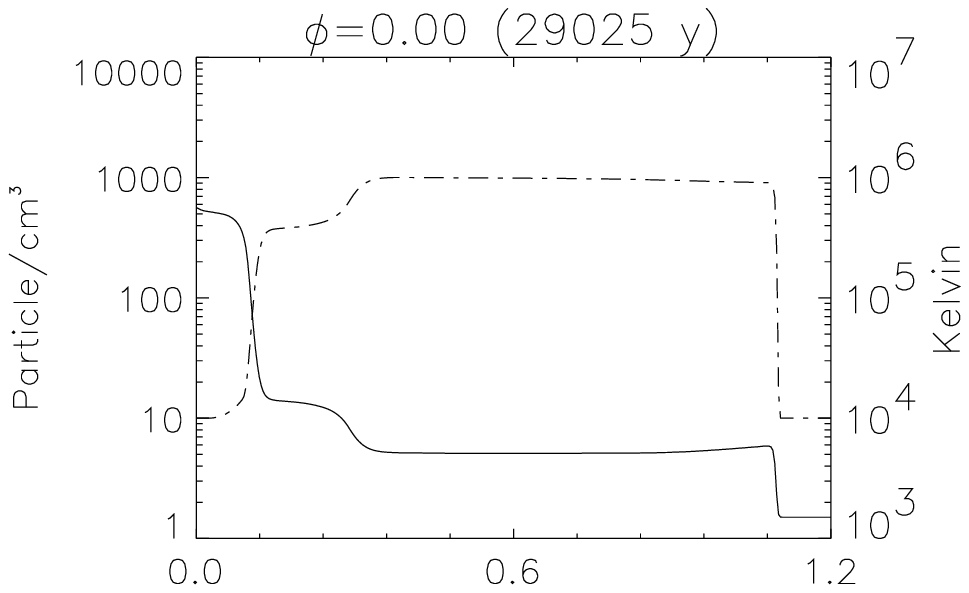,height=4cm,width=5.73cm}
\psfig{figure=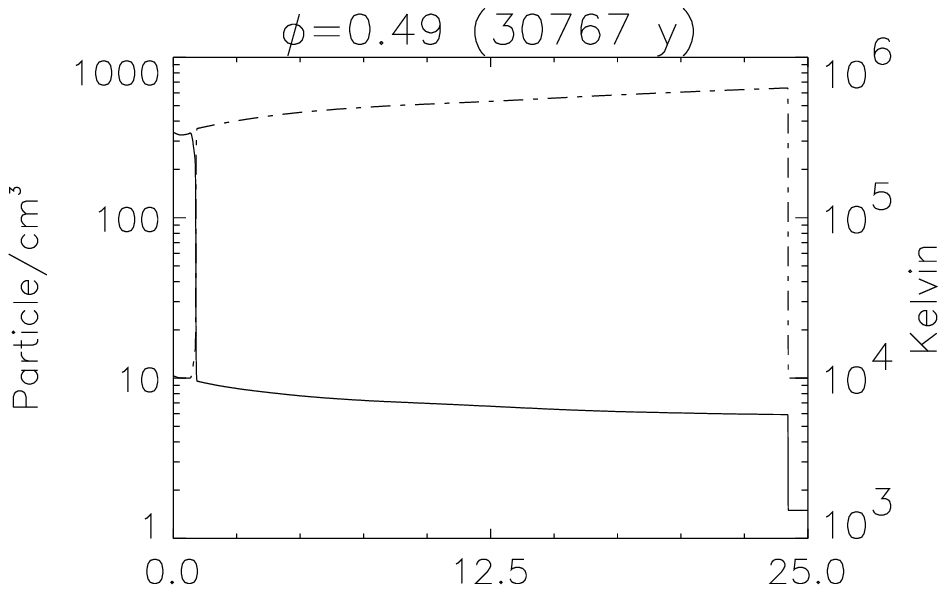,height=4cm,width=5.73cm}
\psfig{figure=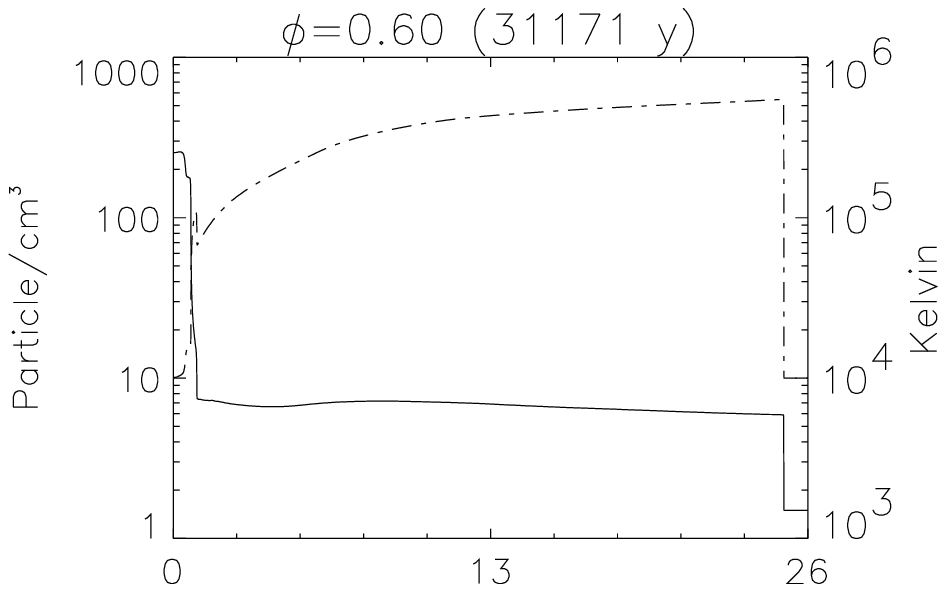,height=4cm,width=5.73cm}
           }
%
%
\centerline{
\psfig{figure=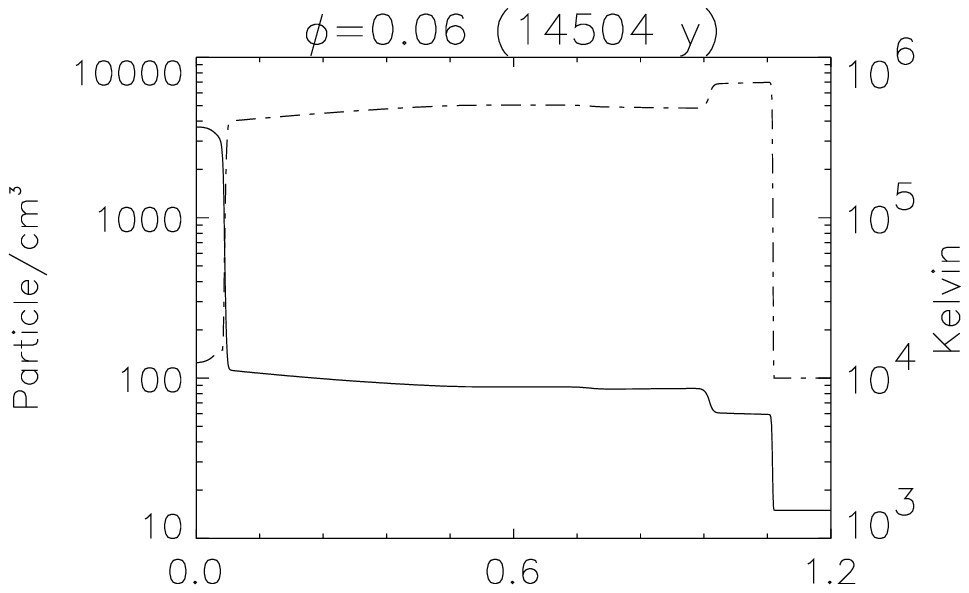,height=4cm,width=5.73cm}
\psfig{figure=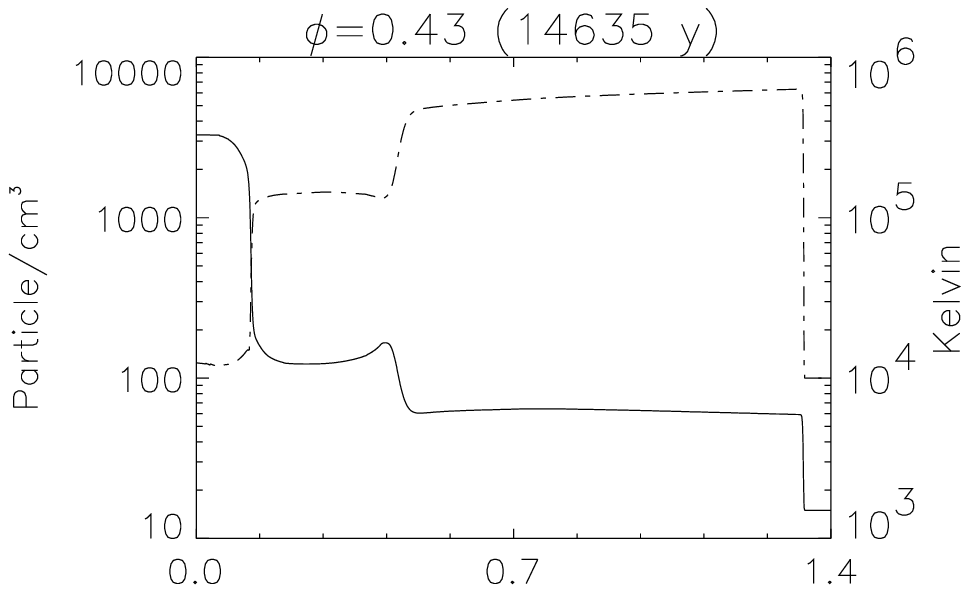,height=4cm,width=5.73cm}
\psfig{figure=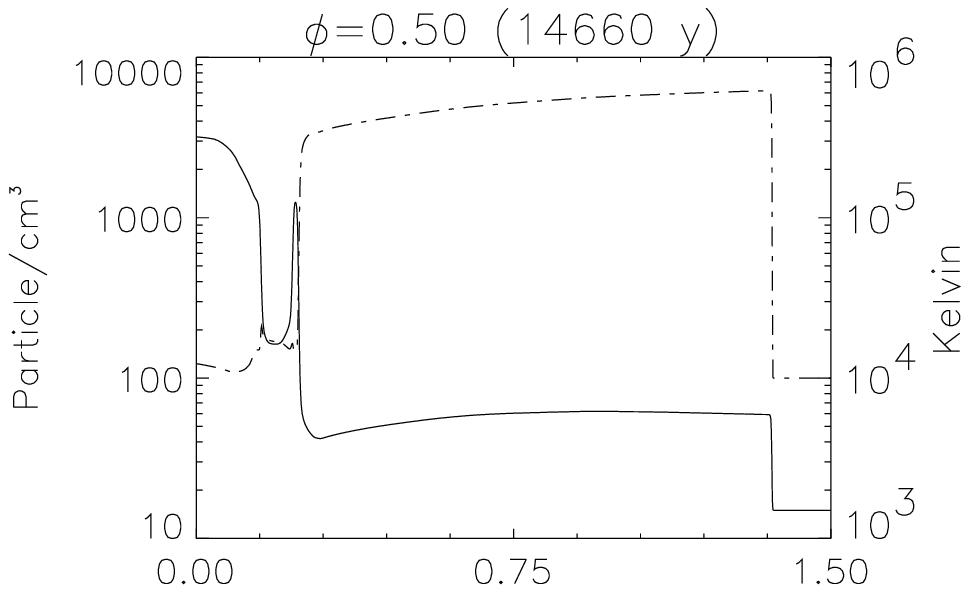,height=4cm,width=5.73cm}
           }
%
%
\centerline{
 \psfig{figure=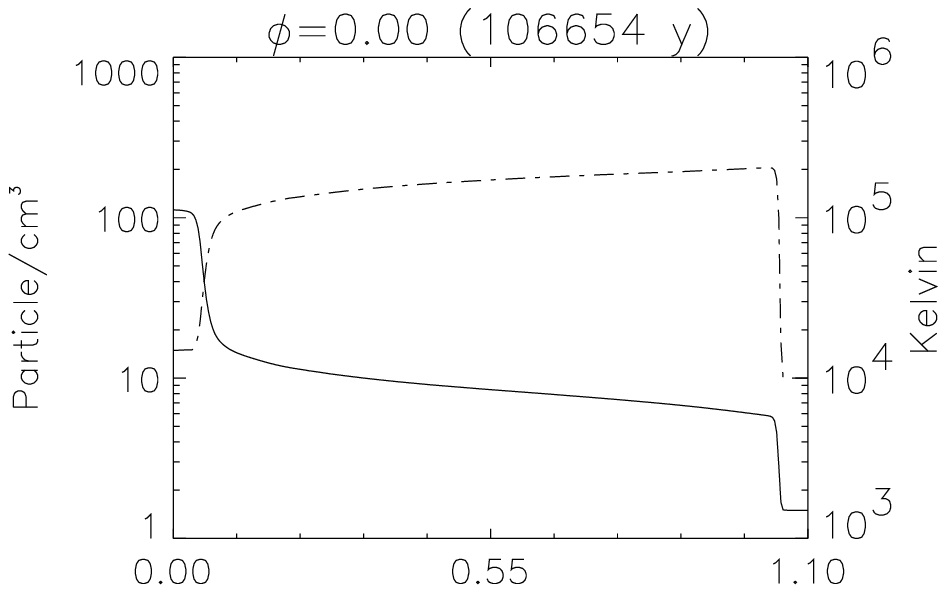,height=4cm,width=5.73cm}
 \psfig{figure=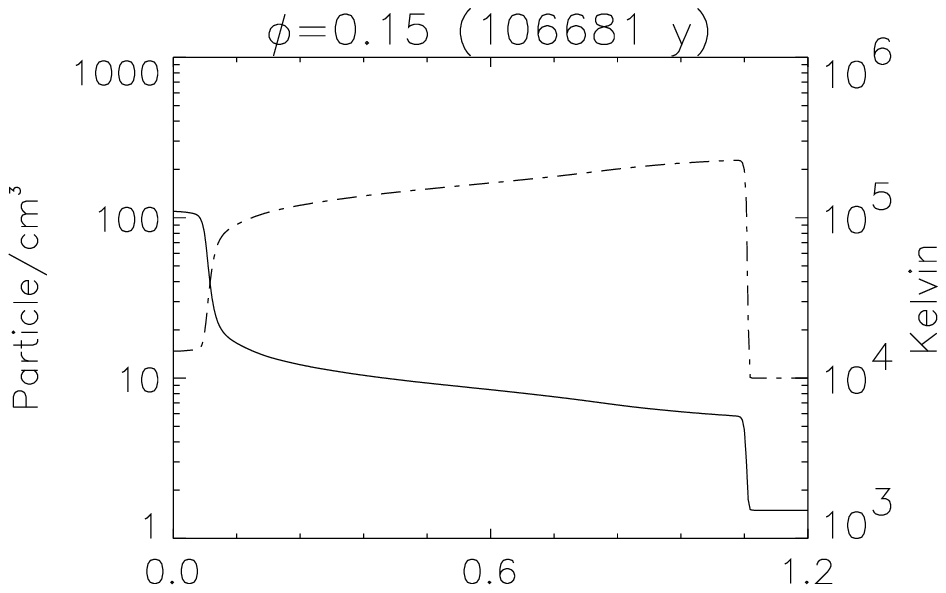,height=4cm,width=5.73cm}
 \psfig{figure=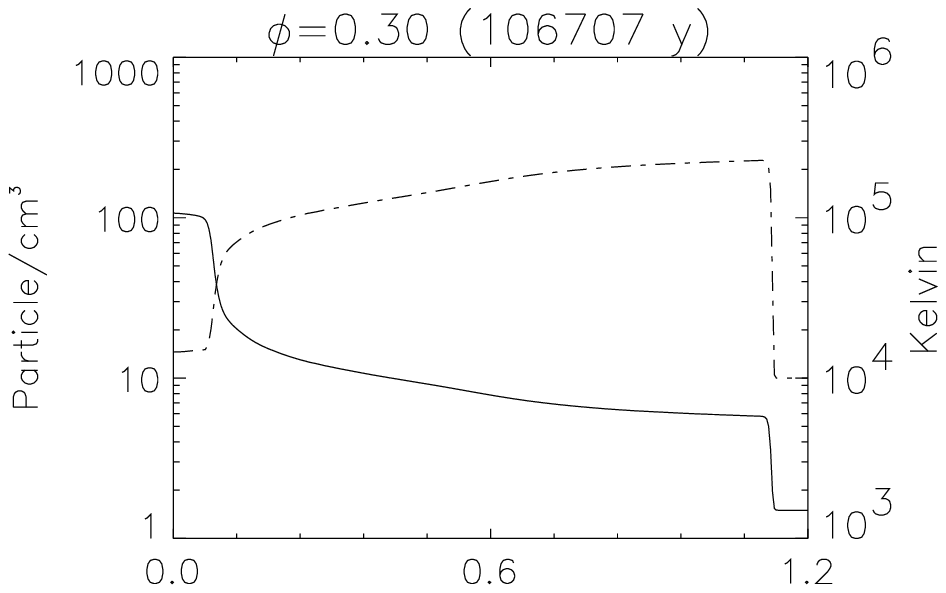,height=4cm,width=5.73cm}
}
\caption{{\bf~a)} 
         Time-series over one oscillation cycle (from left to right) for
         each different type of the overstability. The series continue
         on the next page. Shown are the density- 
         (solid line) and temperature-profiles (dashed line) of the
         cooling layer. From top to bottom:
         S1-type (F-mode with superimposed 1O-mode), I-type (F-mode), 
         C-type (F-mode), M-type (1O-mode), and S2-type (1O-mode).
         The series correspond to the cycles shown in Fig.~6. They start 
         approximately at the minimum extent of the shell (phase $\phi=0$) 
         and end at the next minimum ($\phi=1$). 
         (Caption continues next page.)}
\label{fig:cycle-evolution}
\end{figure*}
\begin{figure*}[htp]	
\centerline{\bf \large \hspace{3.75cm} 
snapshot 4 \hspace{3.cm} snapshot 5 \hspace{3.cm} snapshot 6 \hspace{3.75cm}}
%
%
\centerline{
 \psfig{figure=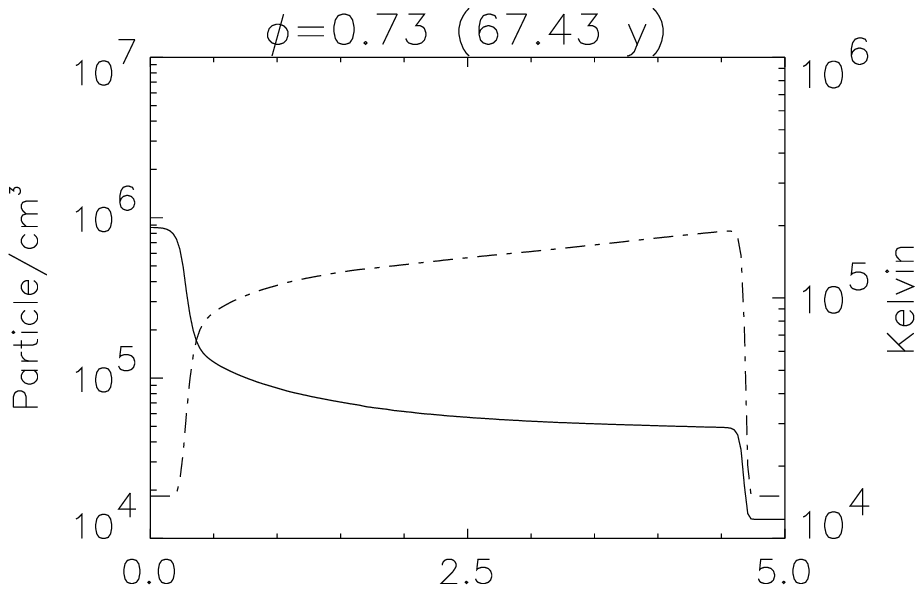,height=4cm,width=5.73cm}
 \psfig{figure=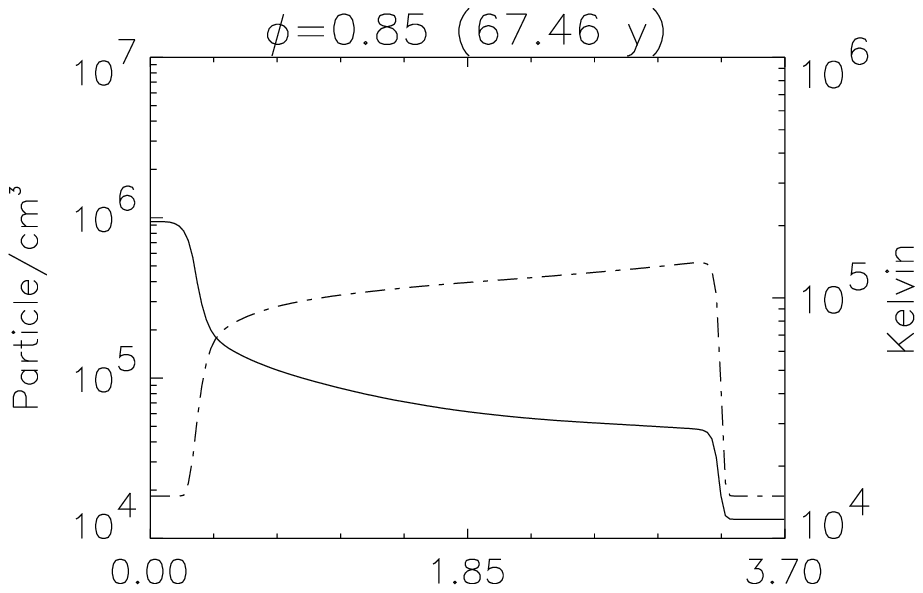,height=4cm,width=5.73cm}
 \psfig{figure=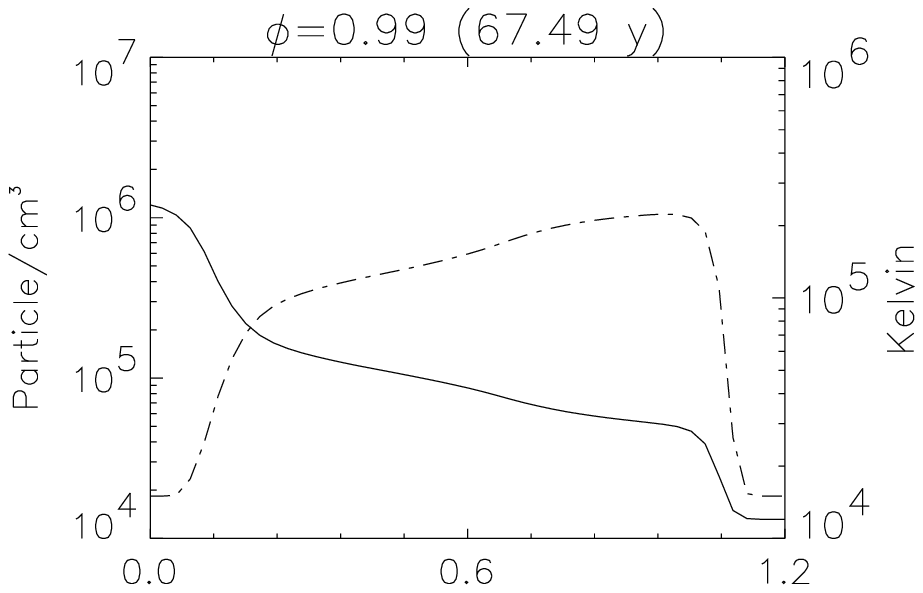,height=4cm,width=5.73cm}
           }
%
%
\centerline{
\psfig{figure=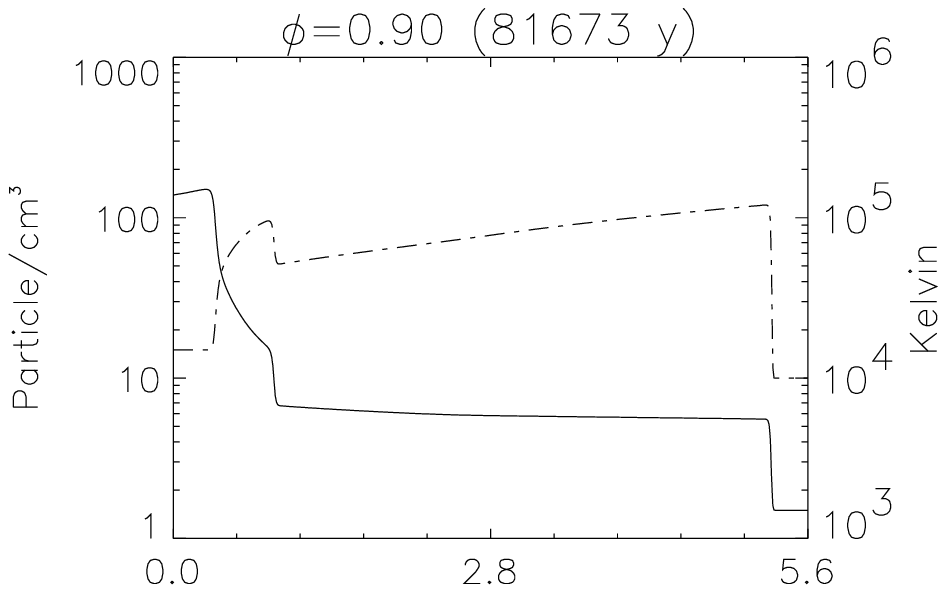,height=4cm,width=5.73cm}
\psfig{figure=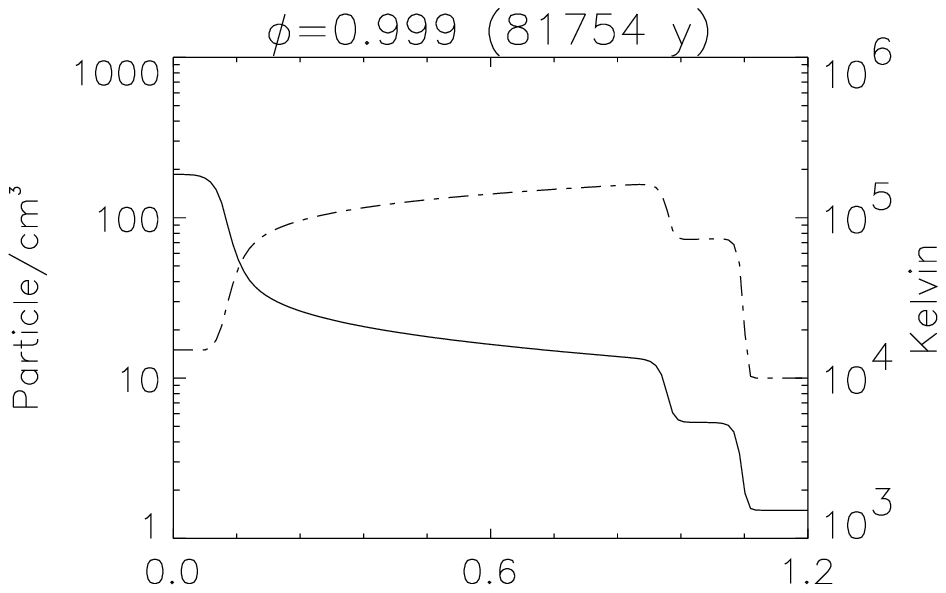,height=4cm,width=5.73cm}
\psfig{figure=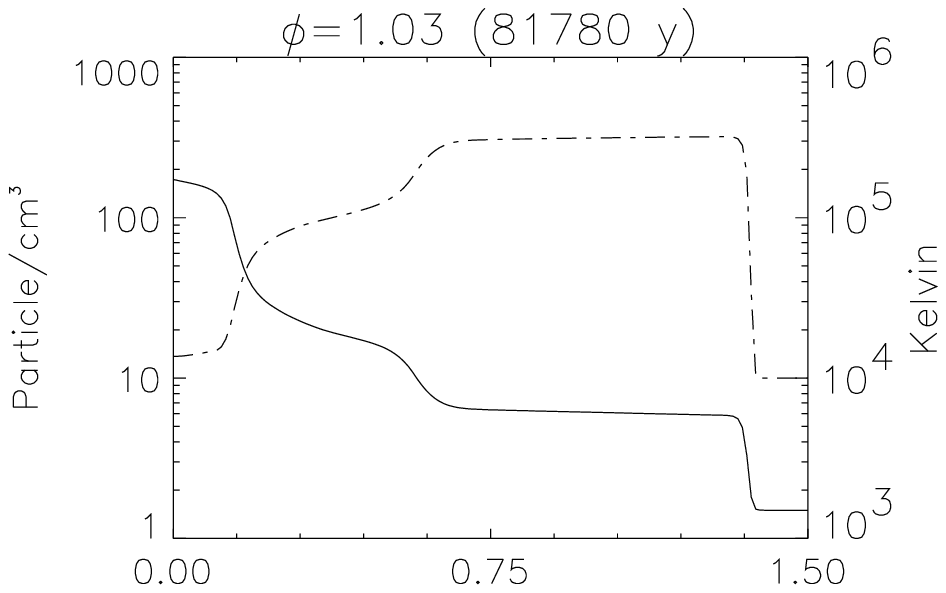,height=4cm,width=5.73cm}
           }
%
%
\centerline{
\psfig{figure=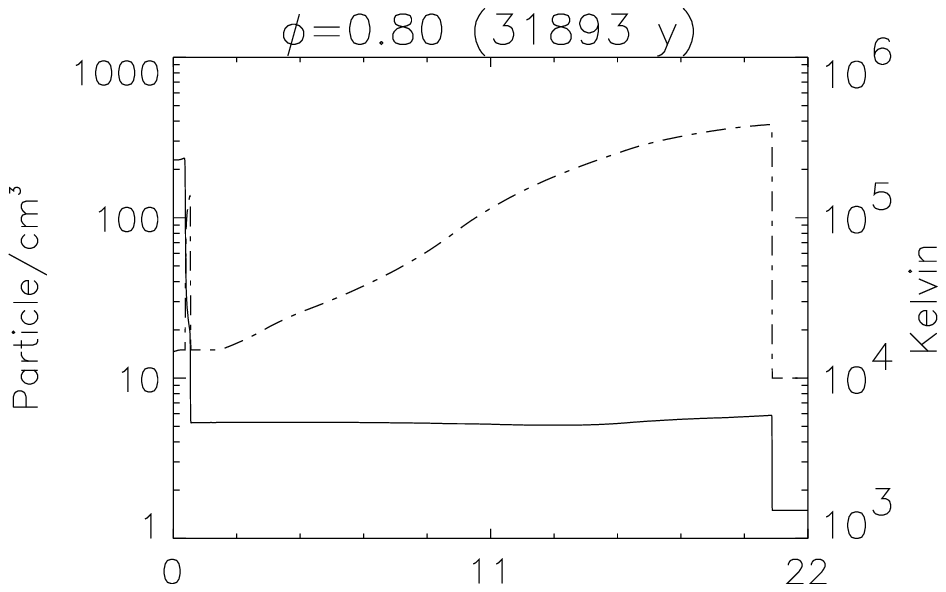,height=4cm,width=5.73cm}
\psfig{figure=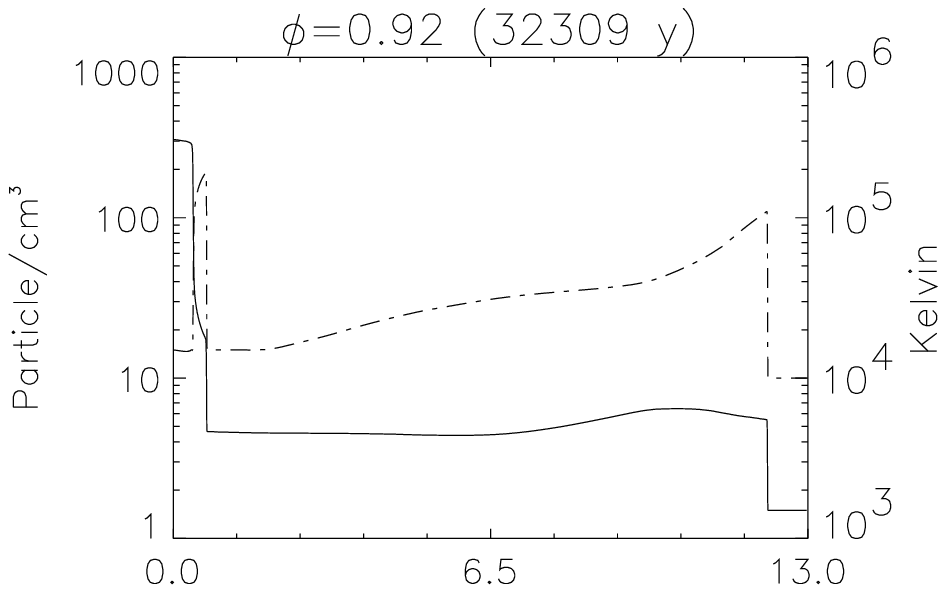,height=4cm,width=5.73cm}
\psfig{figure=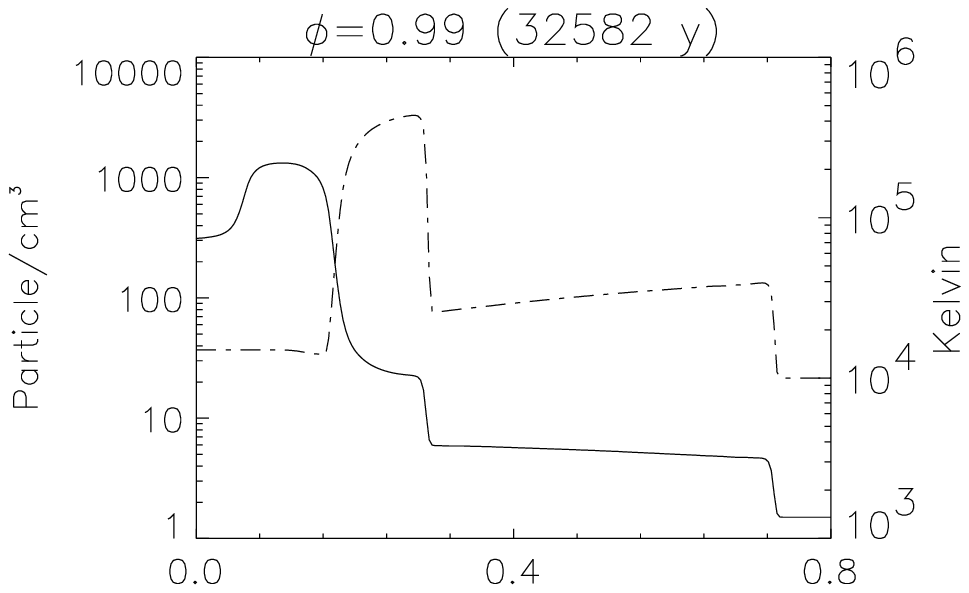,height=4cm,width=5.73cm}
           }
%
%
\centerline{
\psfig{figure=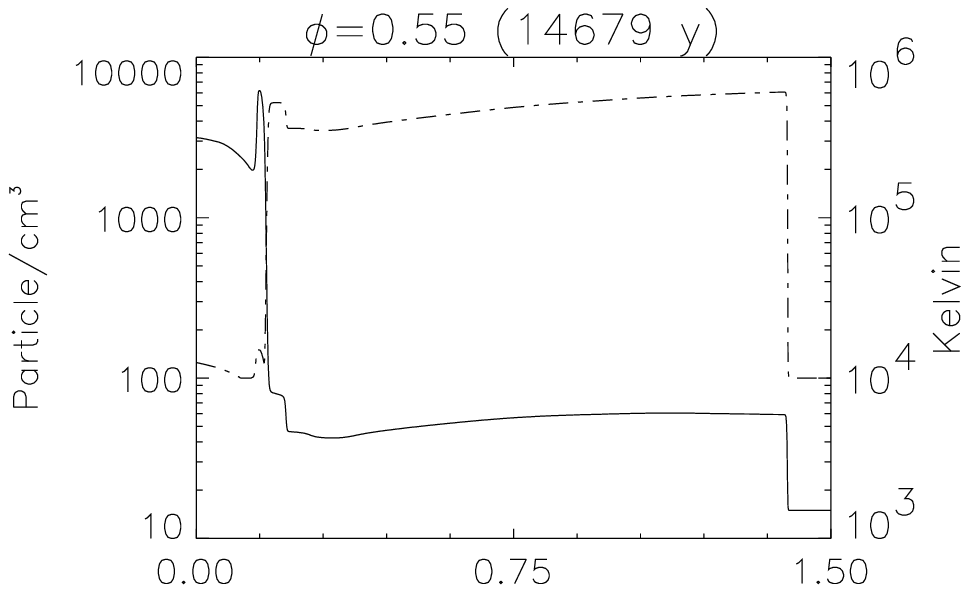,height=4cm,width=5.73cm}
\psfig{figure=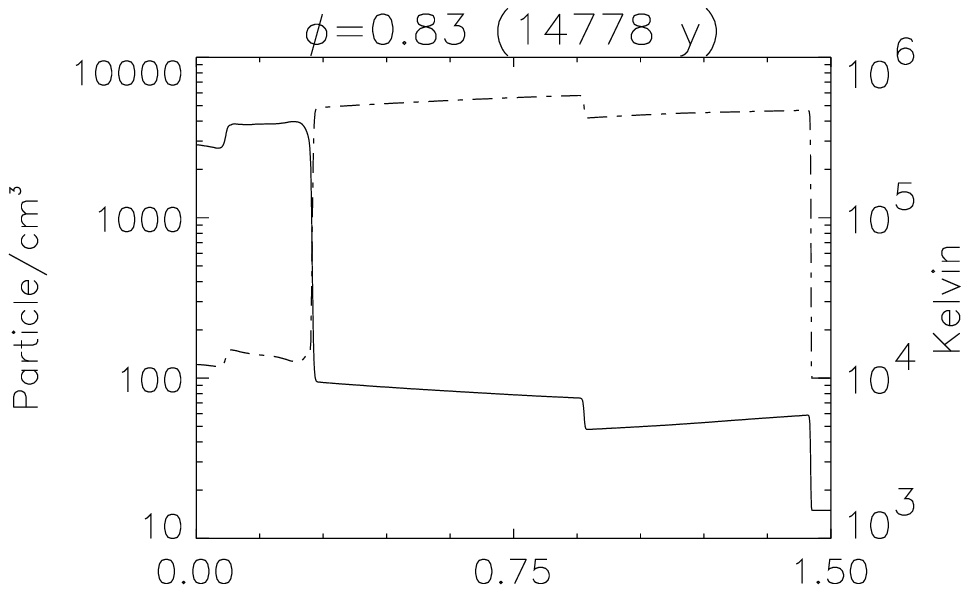,height=4cm,width=5.73cm}
\psfig{figure=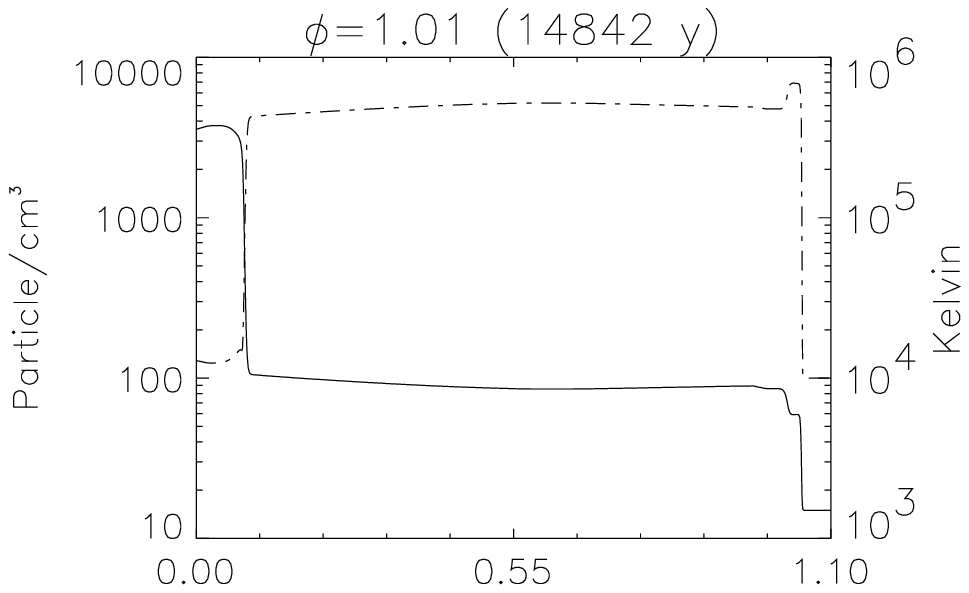,height=4cm,width=5.73cm}
    }
%
%
\centerline{
 \psfig{figure=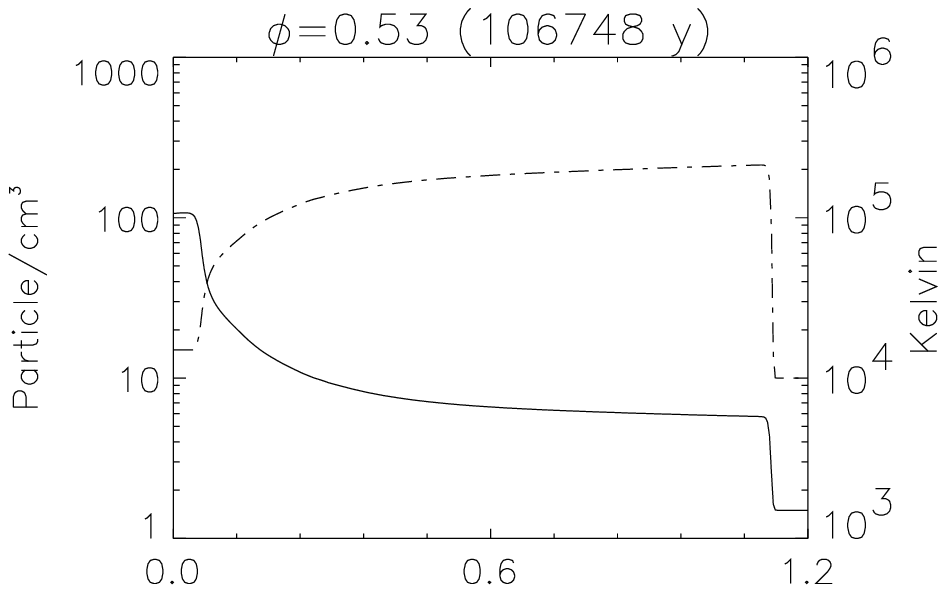,height=4cm,width=5.73cm}
 \psfig{figure=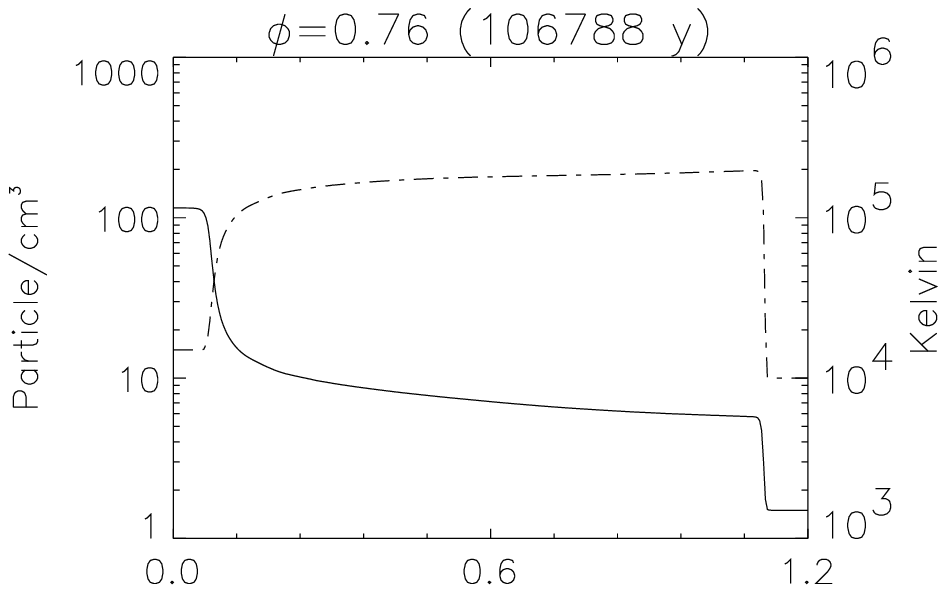,height=4cm,width=5.73cm}
 \psfig{figure=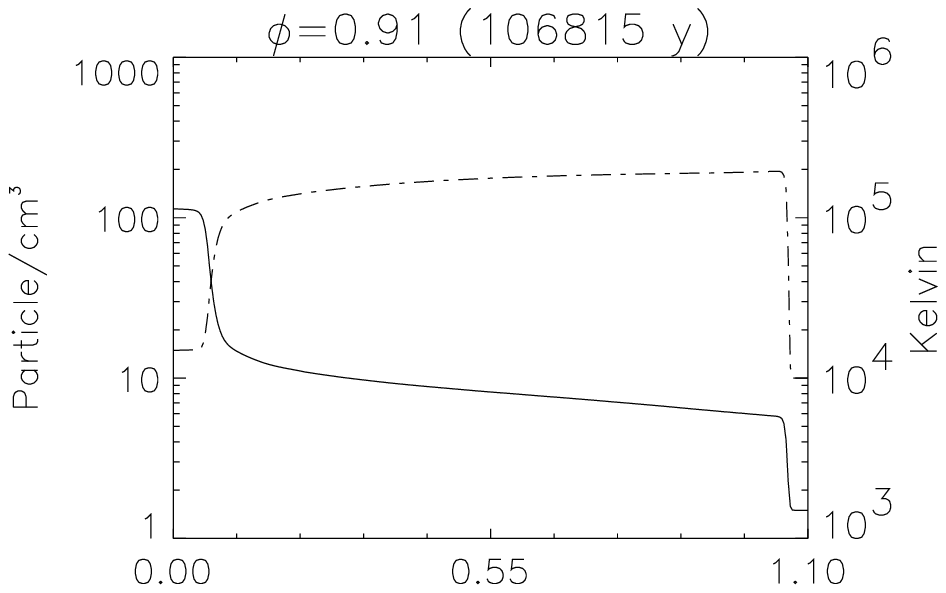,height=4cm,width=5.73cm}
            }
\setcounter{figure}{4}
\caption{{\bf~b)} Continuation of Fig.~5.~a) of previous page. 
         The length of the cooling layers are measured in units of the 
         lengths of their first minima ($\phi=0$). Notice the large
         difference in amplitude between the different types of
         oscillation for the F-mode. For the 1O-mode, the amplitude is 
         generally small. Note that the cooling layers may have different
         sizes at phases $\phi=0$ and  $\phi=1$. This is due to the
         influence of the dynamics in the cold dense layer (see 
         Sect.~7) and the general deceleration of the structure.
         Although the types are independent of the particular flows 
         we give in brackets the model from which we have taken the 
         example: S1-type (CW), 
         I-type (WB1), C-type (WB1), M-type (WB2f) and S2-type (WB1).}
%
%
%
\end{figure*}
\begin{figure*}[htp]
\centerline{\bf \large \hspace{0.4cm} 
S1-type \hspace{1.7cm} I-type \hspace{1.7cm} C-type \hspace{1.7cm} 
                       M-type \hspace{1.7cm}  S2-type \hspace{.1cm} }
\vspace{0.4cm}
\centerline{
 \psfig{figure=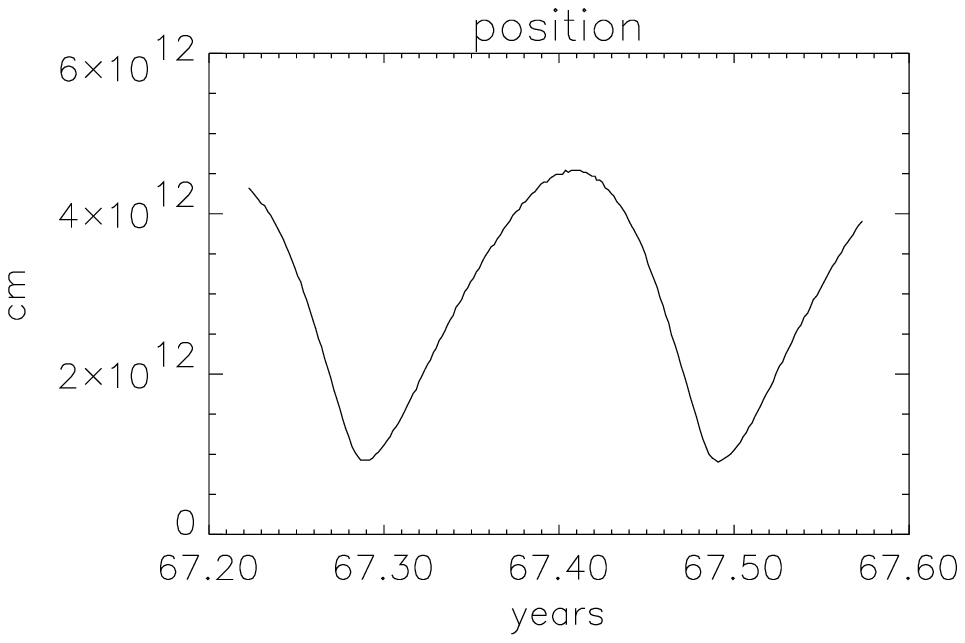,height=3cm,width=3.5cm}
 \psfig{figure=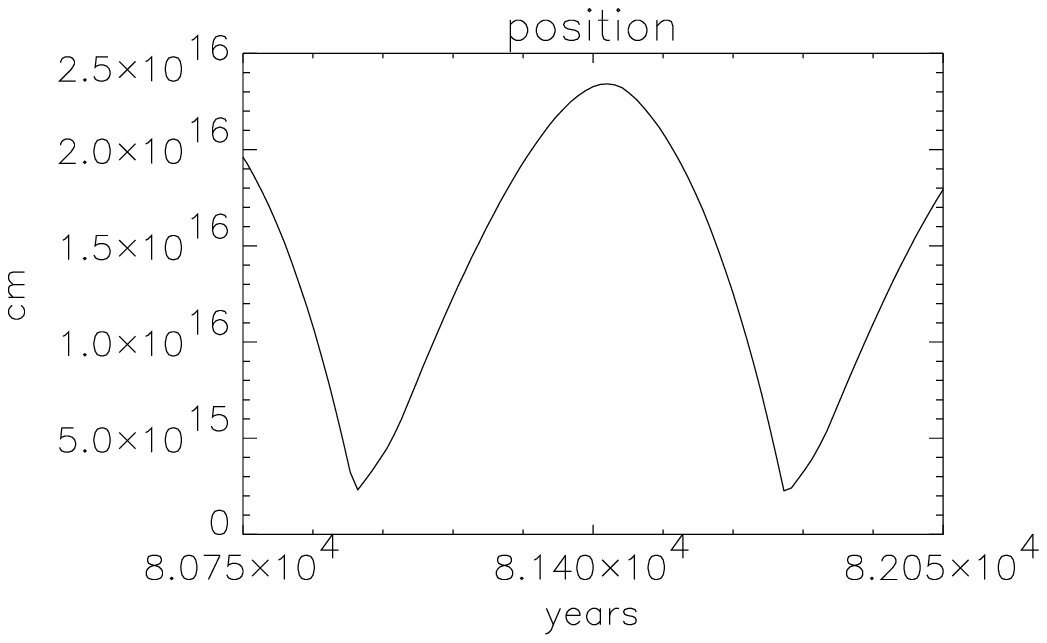,height=3cm,width=3.5cm}
 \psfig{figure=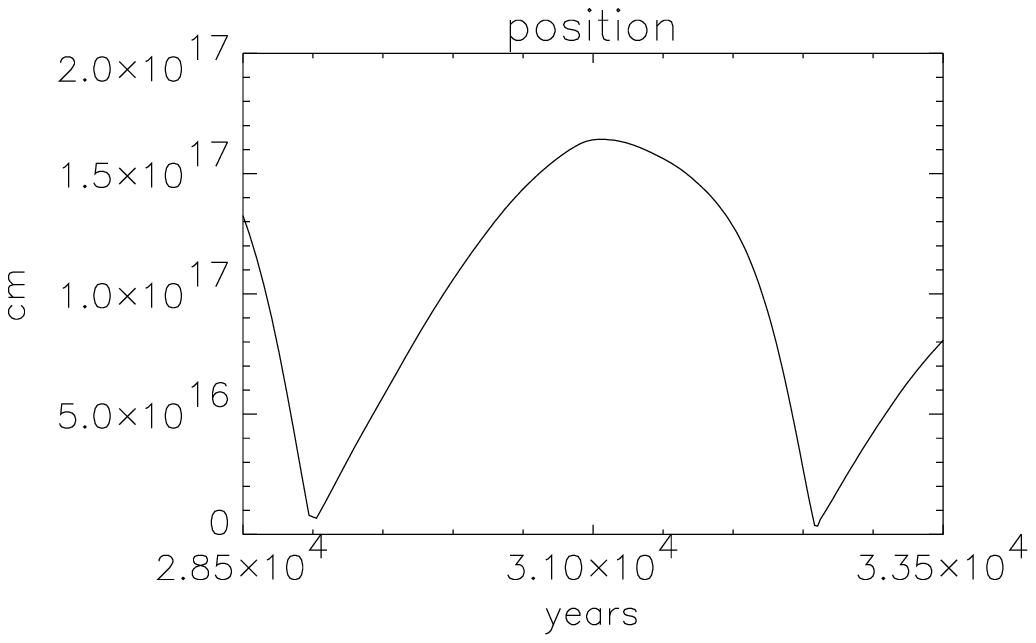,height=3cm,width=3.5cm}
 \psfig{figure=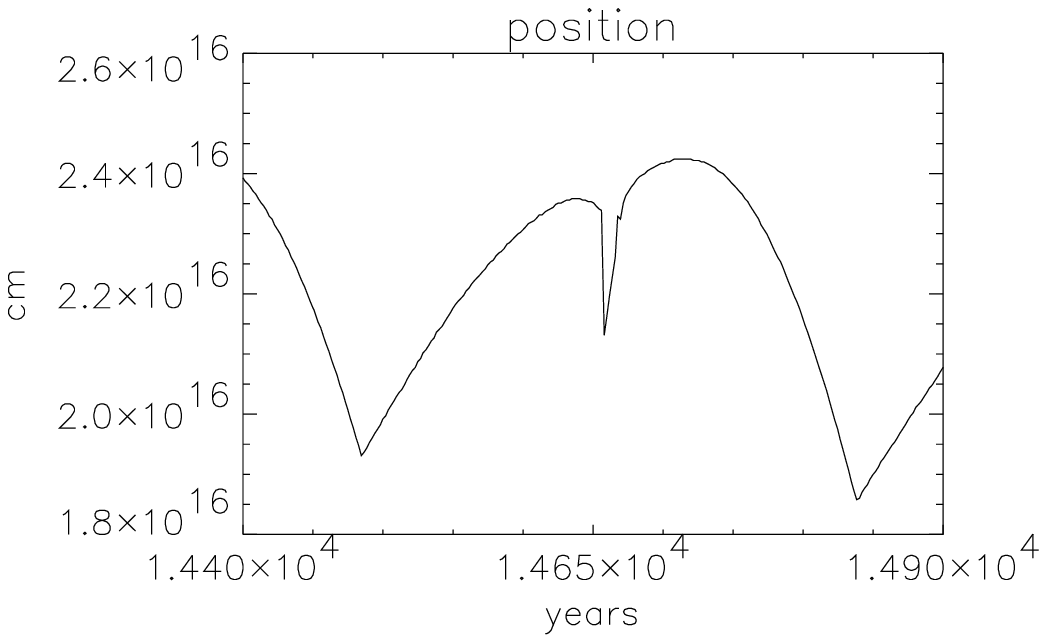,height=3cm,width=3.5cm}
 \psfig{figure=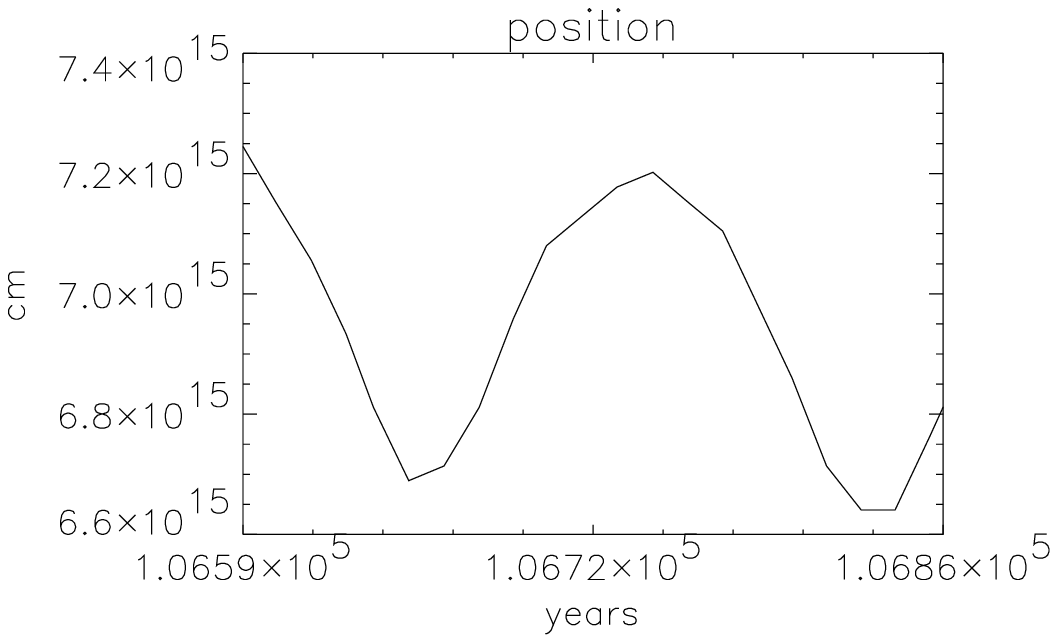,height=3cm,width=3.5cm}
            }
\centerline{
 \psfig{figure=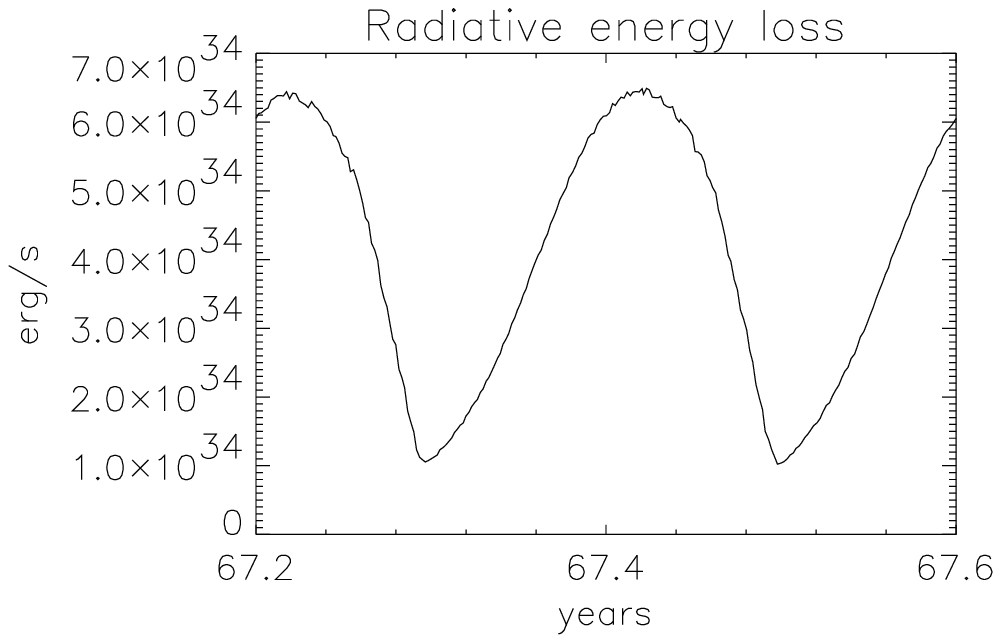,height=3cm,width=3.5cm}
 \psfig{figure=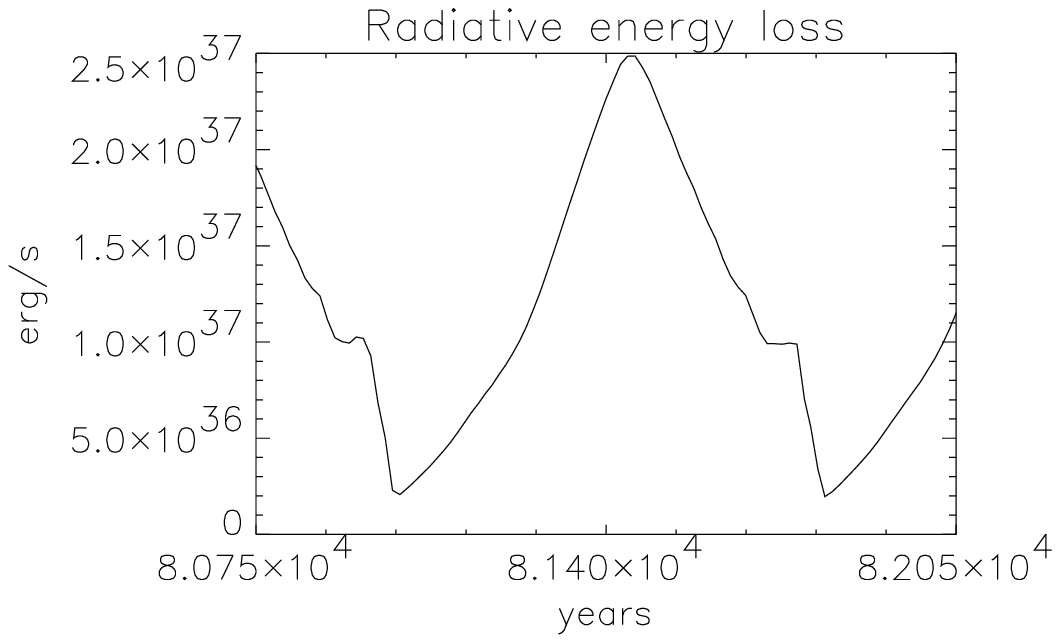,height=3cm,width=3.5cm}
 \psfig{figure=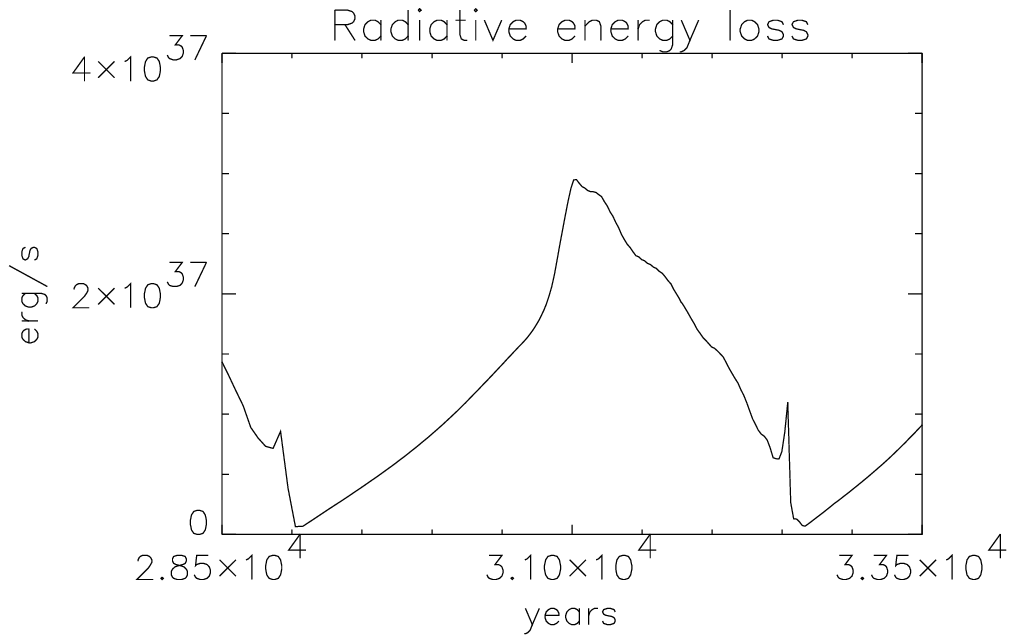,height=3cm,width=3.5cm}
 \psfig{figure=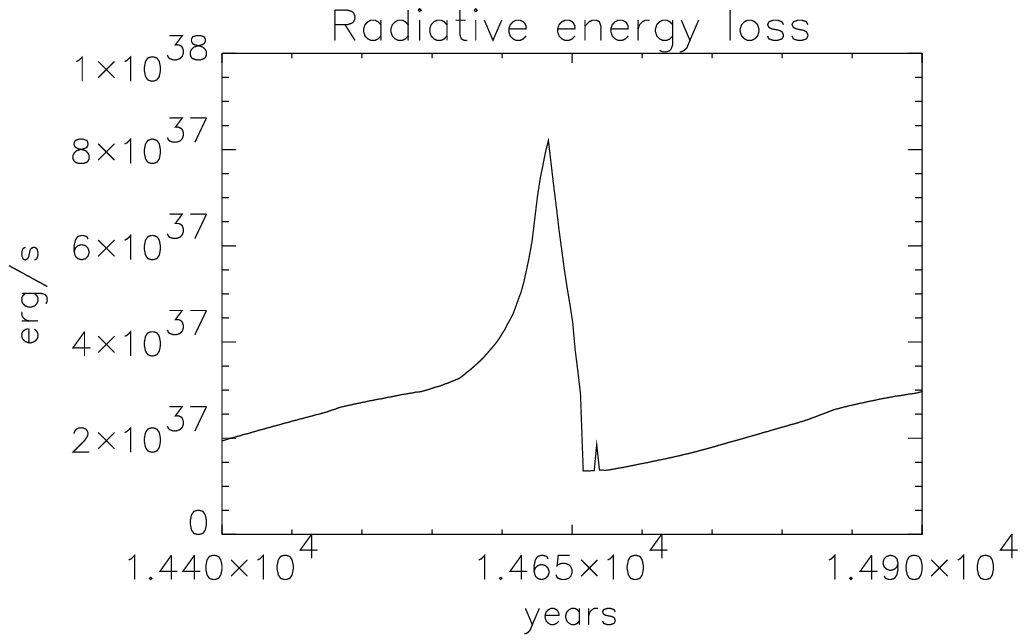,height=3cm,width=3.5cm}
 \psfig{figure=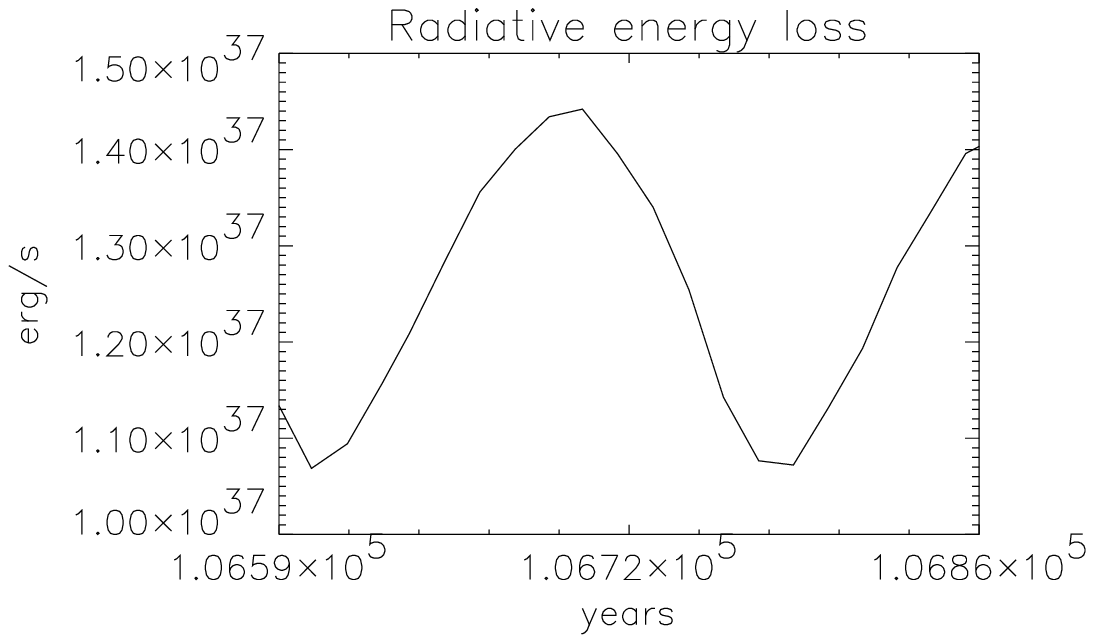,height=3cm,width=3.5cm}
           }    
\centerline{
 \psfig{figure=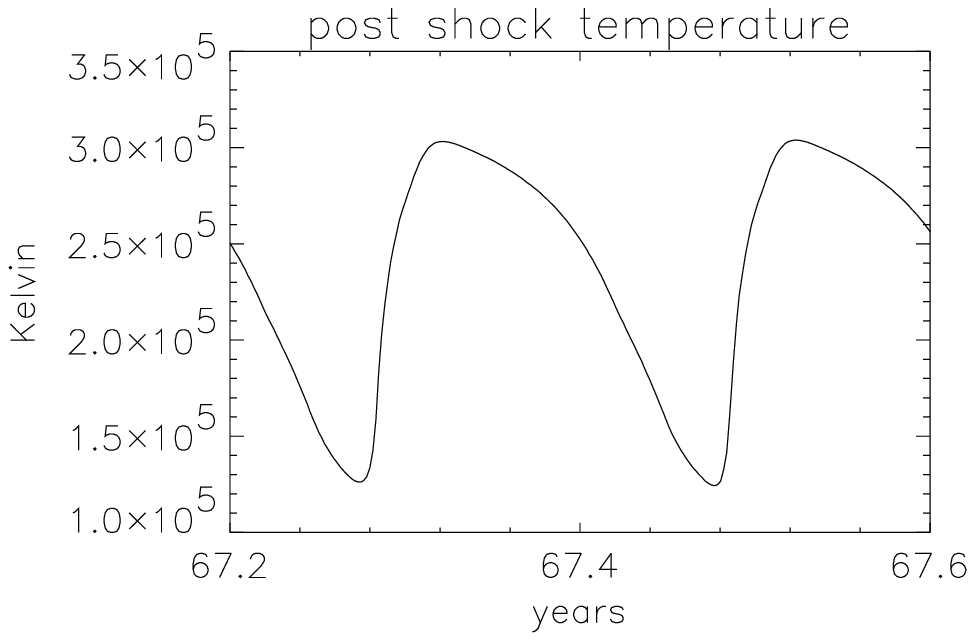,height=3cm,width=3.5cm}
 \psfig{figure=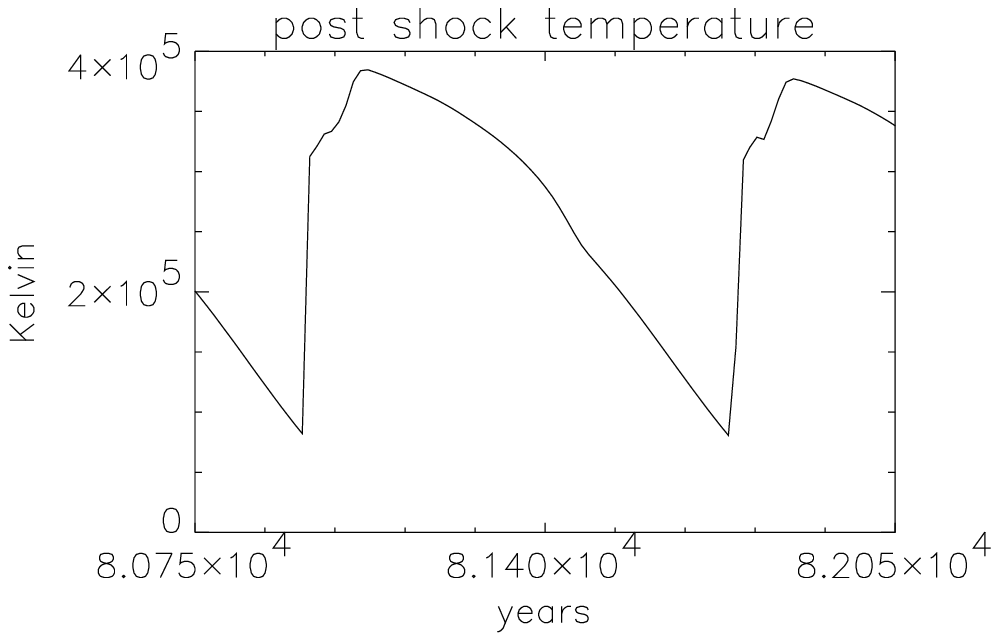,height=3cm,width=3.5cm}
 \psfig{figure=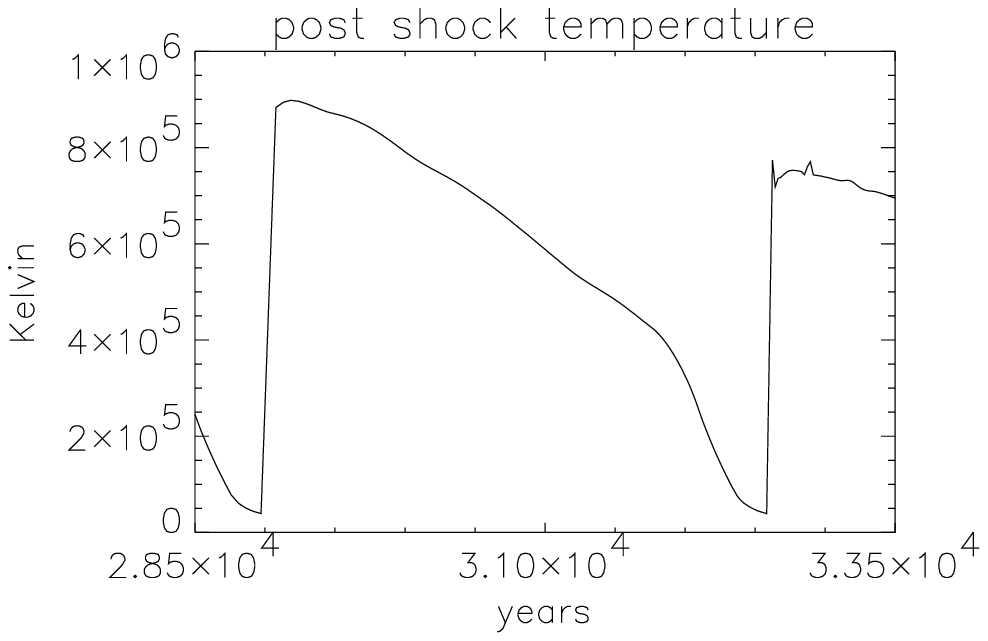,height=3cm,width=3.5cm}
 \psfig{figure=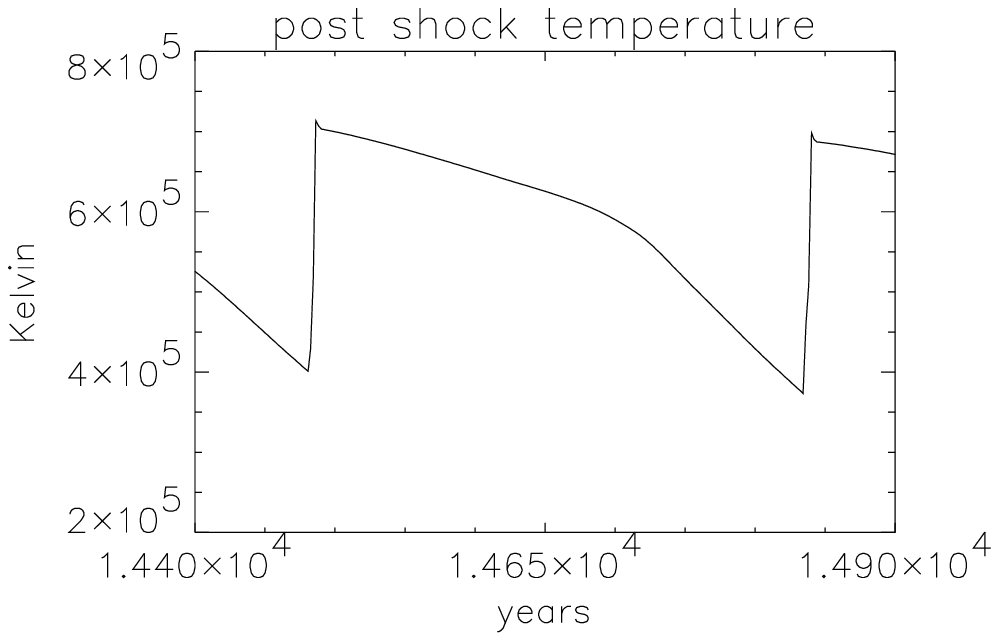,height=3cm,width=3.5cm}
 \psfig{figure=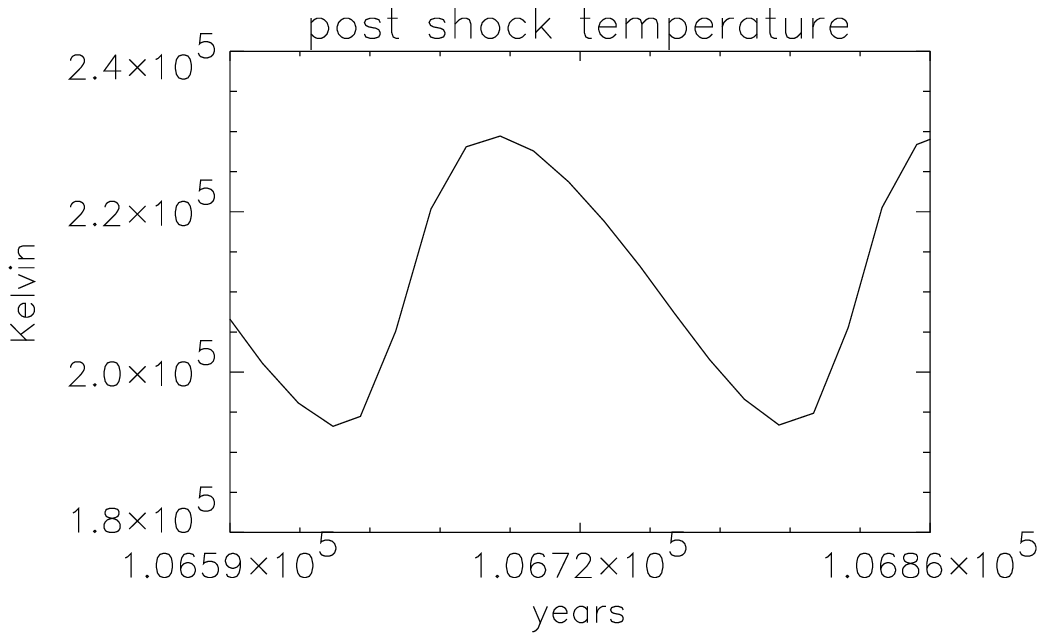,height=3cm,width=3.5cm}
           }
\centerline{
 \psfig{figure=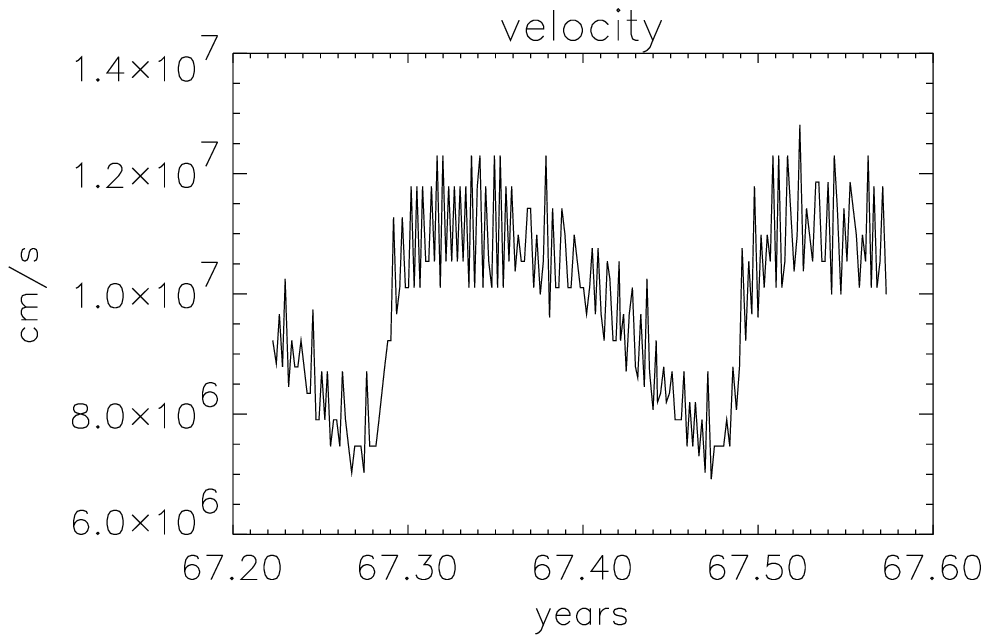,height=3cm,width=3.5cm}
 \psfig{figure=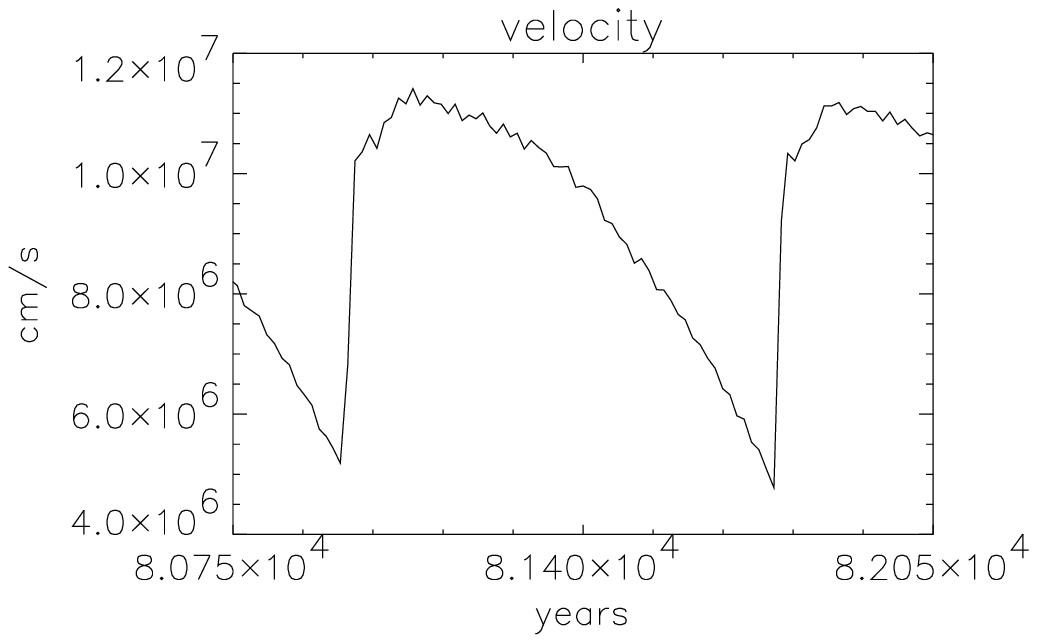,height=3cm,width=3.5cm}
 \psfig{figure=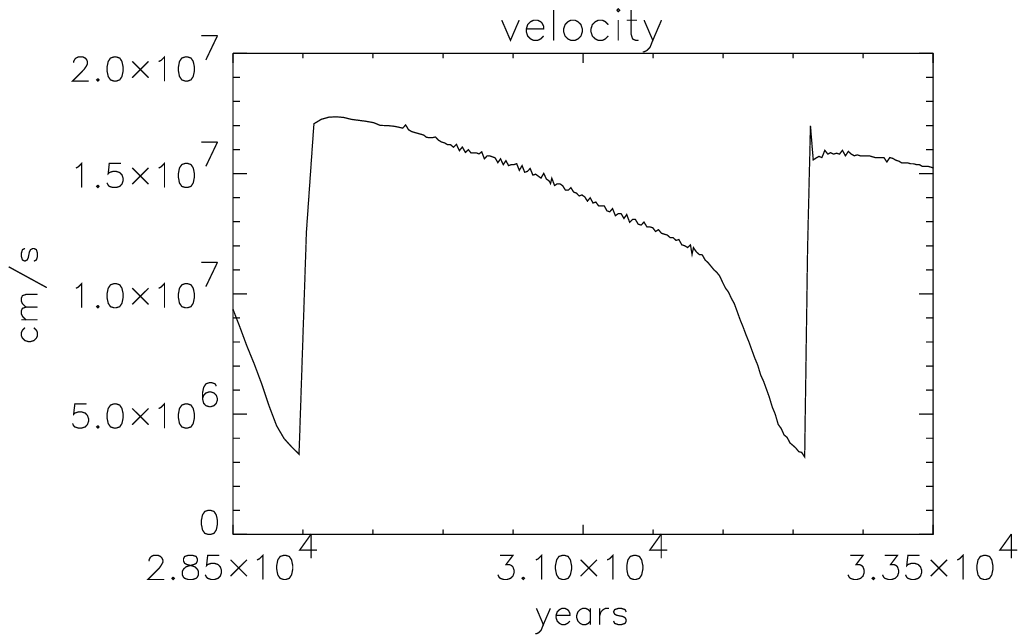,height=3cm,width=3.5cm}
 \psfig{figure=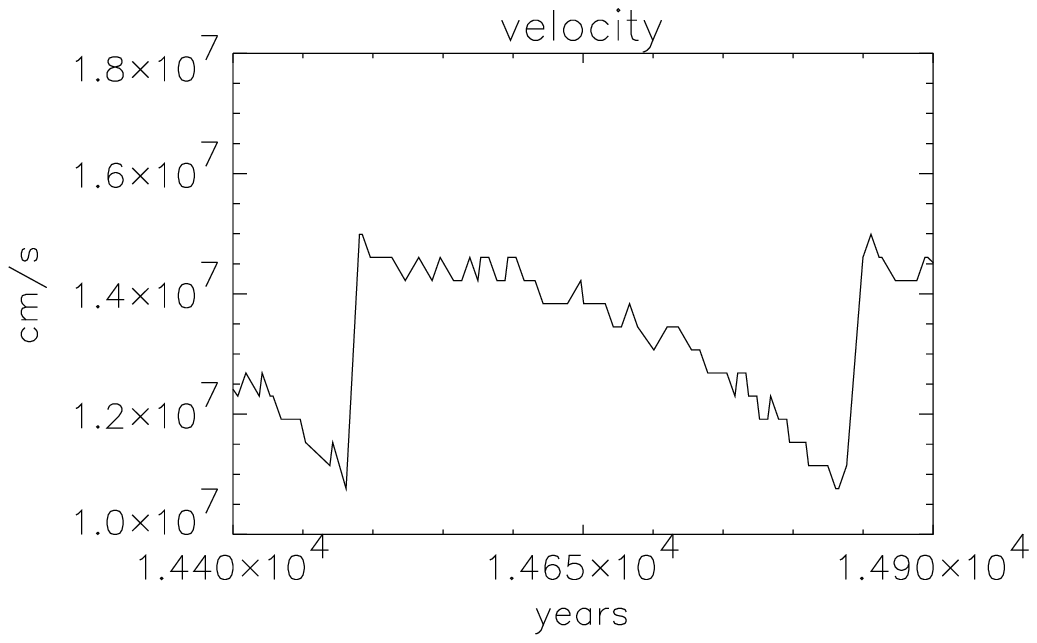,height=3cm,width=3.5cm}
 \psfig{figure=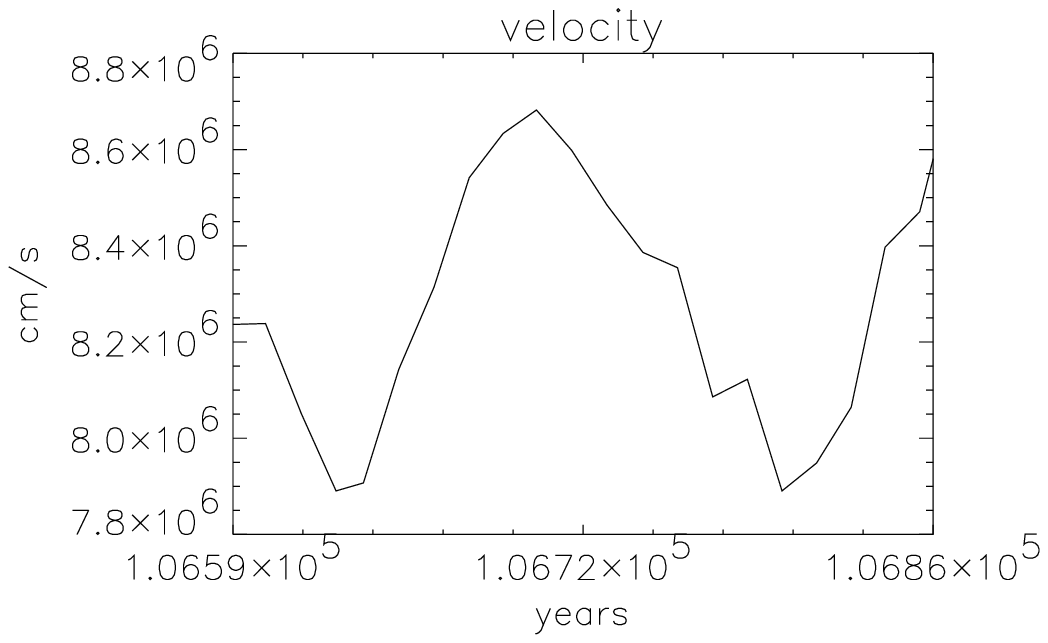,height=3cm,width=3.5cm}
           } 
\centerline{
 \psfig{figure=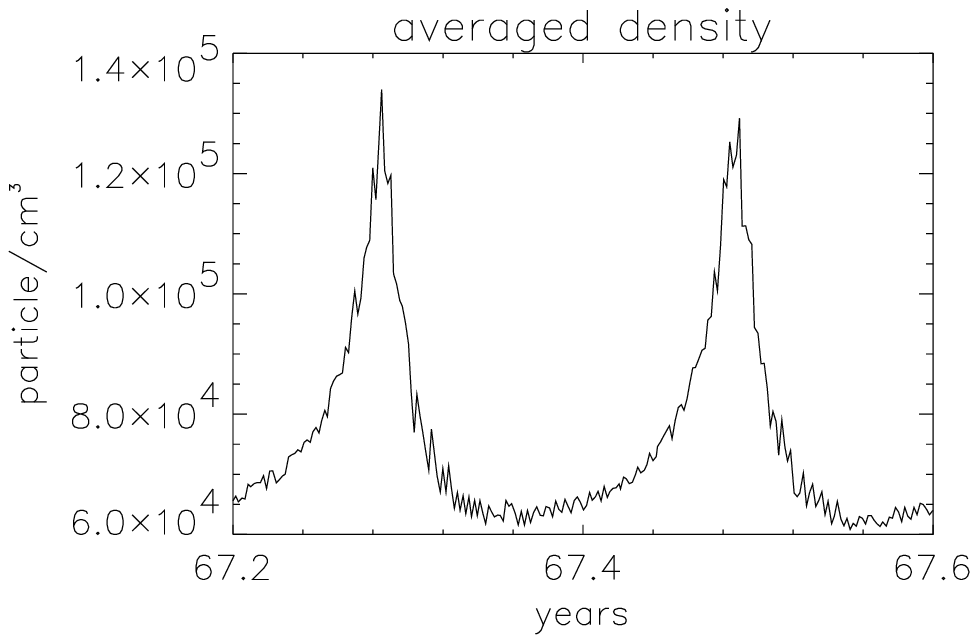,height=3cm,width=3.5cm}
 \psfig{figure=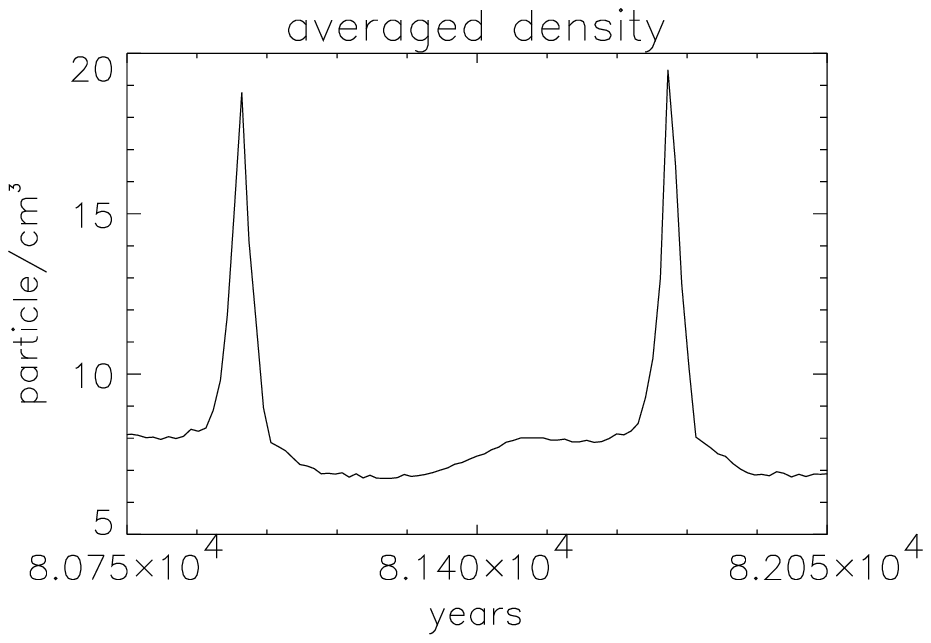,height=3cm,width=3.5cm}
 \psfig{figure=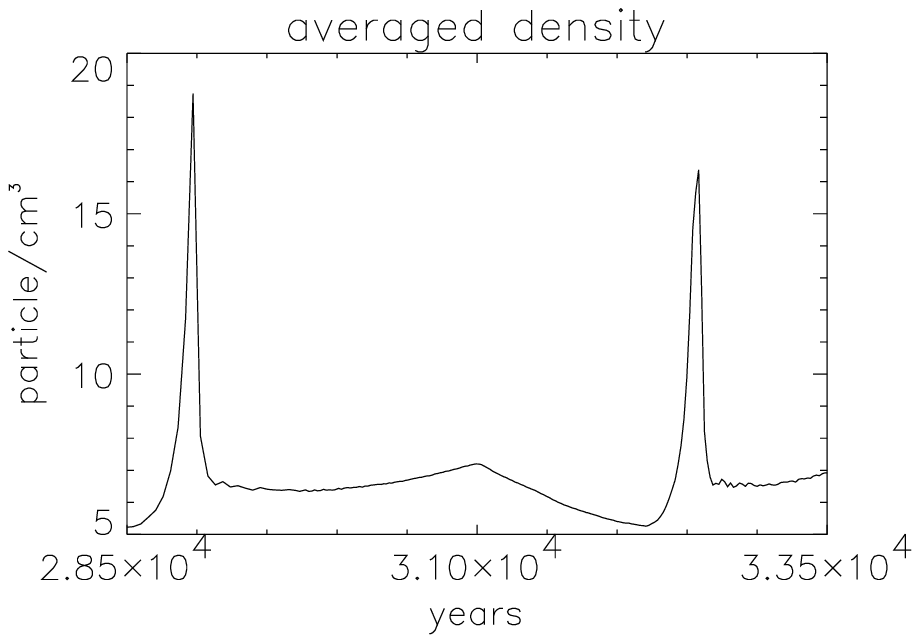,height=3cm,width=3.5cm}
 \psfig{figure=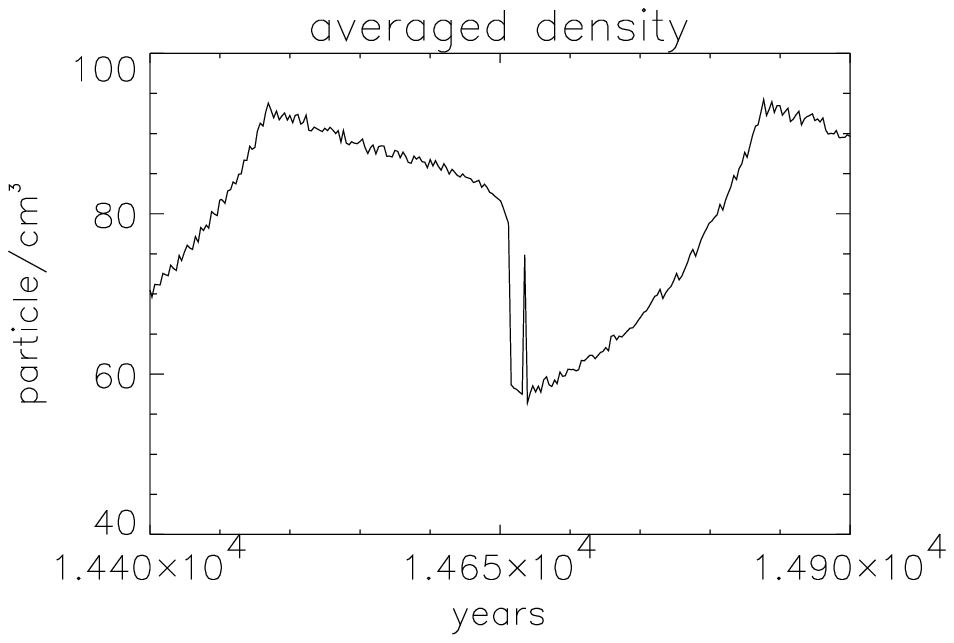,height=3cm,width=3.5cm}
 \psfig{figure=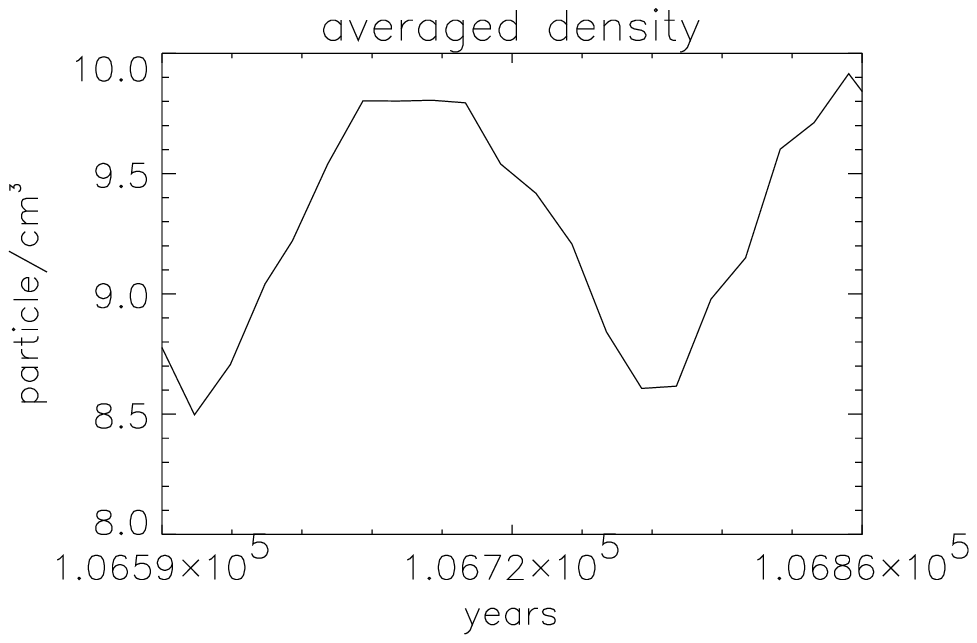,height=3cm,width=3.5cm}
           }
\centerline{
 \psfig{figure=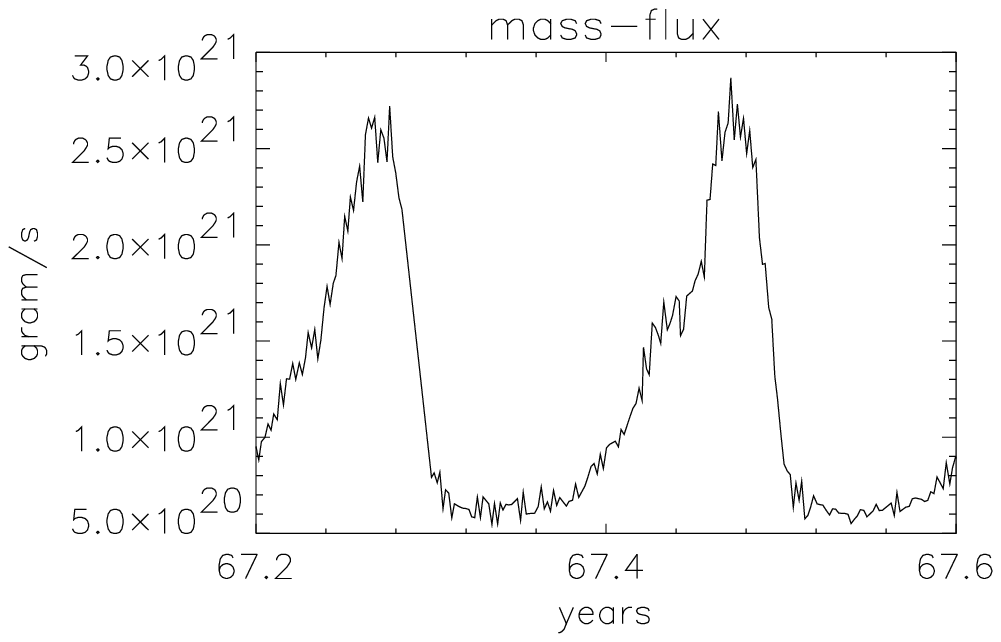,height=3cm,width=3.5cm}
 \psfig{figure=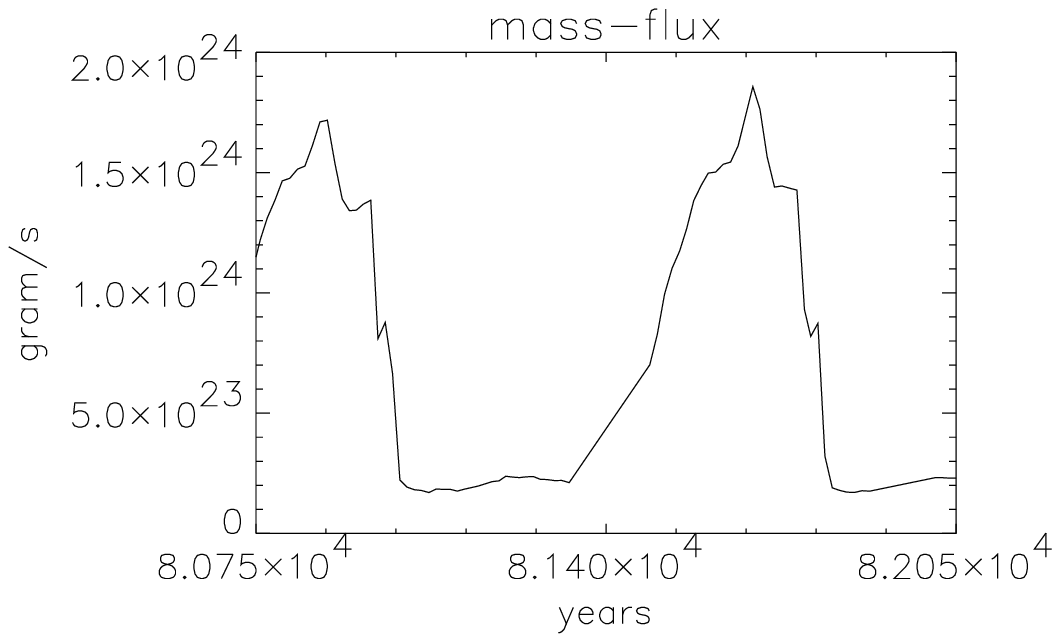,height=3cm,width=3.5cm}
 \psfig{figure=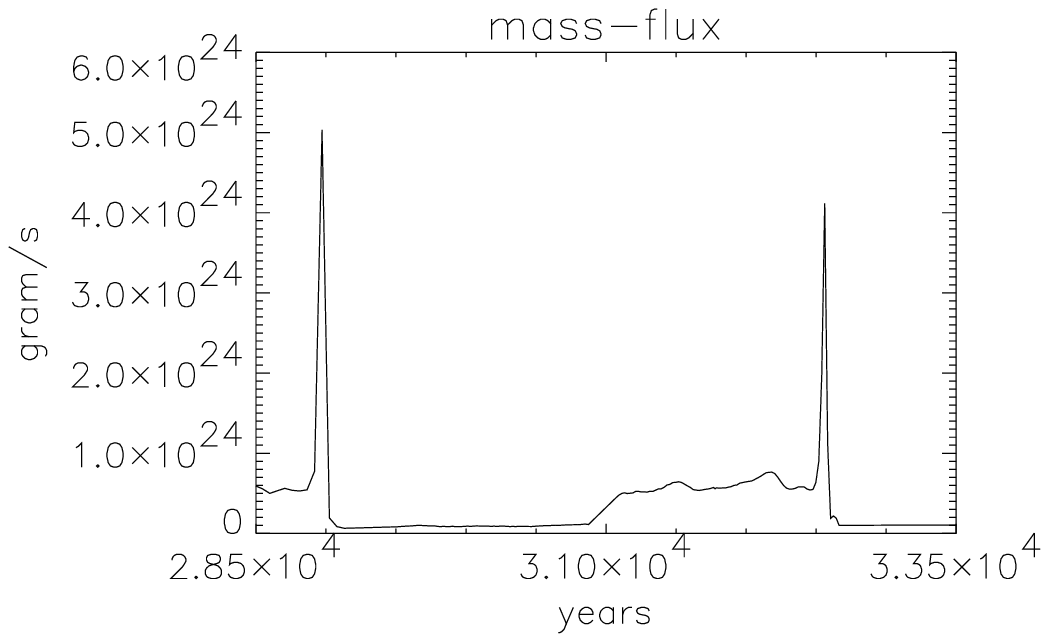,height=3cm,width=3.5cm}
 \psfig{figure=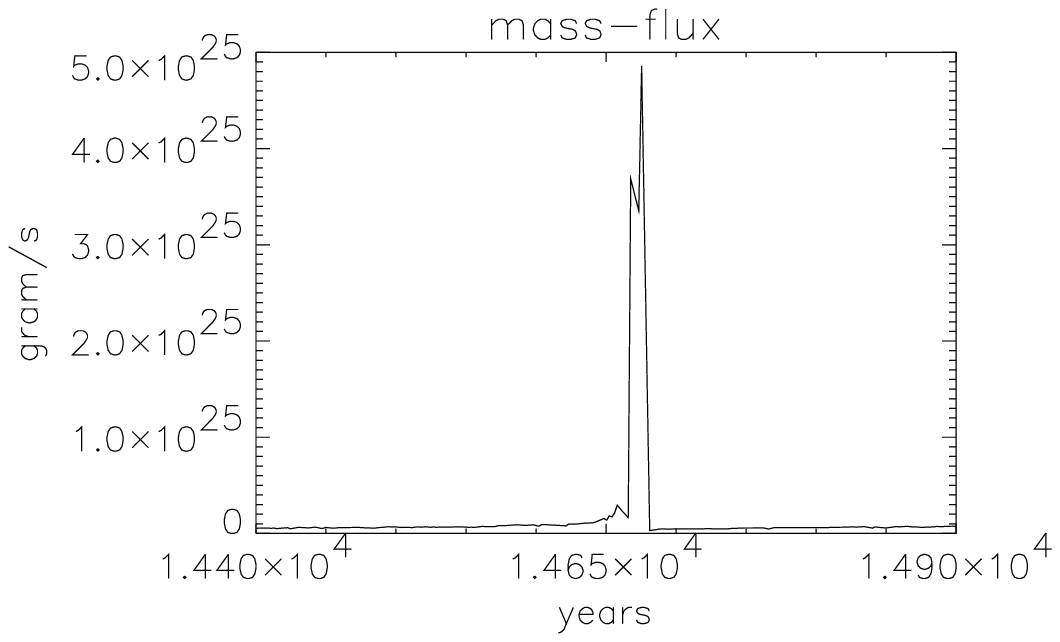,height=3cm,width=3.5cm}
 \psfig{figure=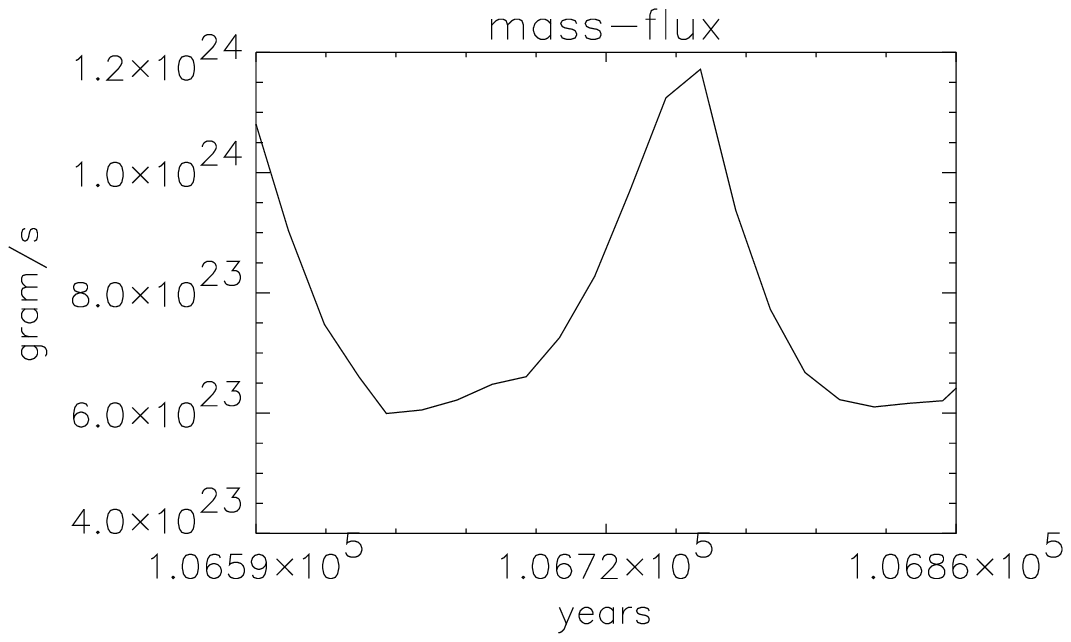,height=3cm,width=3.5cm}
           }    
\centerline{\bf \large \hspace{2.cm} F-mode \hspace{1.5cm} F-mode \hspace{1.5cm} F-mode \hspace{1.5cm} 1O-mode \hspace{1.5cm}  1O-mode \hspace{1.5cm} }
\caption{The different types and modes of the cooling overstability shown
         in some characteristic quantities. The graphs correspond to the
         cycles shown in Fig.~5.
         From top to bottom: Size of the leading shell,
         radiated power (of the entire leading shell), 
         post shock temperature, shock-velocity, 
         average density of the leading shell and mass-flux from the
         leading shell into the cold compressed gas layer. 
         Notice the highly different amplitudes of the oscillations
         in the different types.}
\label{fig:instab-type}
\end{figure*}

In agreement with earlier investigations, our systematic study shows
that the cooling instability for smooth flows -- if present --
manifests itself as an overstability: All quantities, e.g. the size of
the cooling layer or the radiated energy, evolve in a periodic way
with fixed period and amplitude. The linear stability analyses of CI
and \cite*{radshock:bertsch-stab2} show that the period of the
overstability of the radiative shock is determined by dimensionless
quantities $ \delta_i$ ($i=0,1,2,,... $) which are eigenvalues of the
boundary value problem describing the flow between the radiative shock
and the rear end of its cooling layer. One usually speaks of the
fundamental mode (F-mode, $i=0$) and of overtone modes (1O-mode,
2O-mode, etc). The governing eigenvalue $\delta$ is related to the
steady state position of the shock, $x_{0}$, the velocity of the
inflowing matter relative to the shock, u$_{in}$, and the oscillation
period, $T$, by the relation $\delta = x_0/u_{in} \cdot 2 \pi /T $.
Previous investigations have shown that even in the non-linear regime
the period is in good agreement with the fundamental (F) and the first
overtone (1O) mode determined by linear stability analysis.

However, our analysis shows -- also in agreement with earlier
investigations -- that the overstability can show different
phenomenological manifestation when oscillating in one particular
mode. Therefore, we propose to introduce an additional classification
of the overstability, namely the notation of types.
\subsection{Different types of oscillations}
\label{subsec:instab-type}
Following phenomenological criteria, we propose five different
oscillation types. The most striking differences between the different
types are: the oscillation amplitude, which we express as ${\cal A}$
=(q$_{max}$-q$_{min}$)/q$_{min}$ where q is a characteristic quantity of
the cooling layer, whether the oscillations are sinusoidal or not,
and the oscillation mechanism. Fig.~\ref{fig:cycle-evolution} and
Fig.~\ref{fig:instab-type} illustrate the characteristics of the
different types. In the following, we give only a brief description of
some aspects of these types and refer to earlier authors (see
Sect.~\ref{subsec:basics-therm_inst}) for more detailed descriptions.

The {\it catastrophic C-type} is characterized by a dramatic collapse
of the temperature at a certain moment of the oscillation cycle. In
wide ranges of the cooling layer the temperature drops nearly to
nebular temperature on a very short time-scale. The density is hardly
affected despite the rapid change in pressure due to the strong
cooling (snapshots~4 and 5). A huge pressure cavity evolves and a
secondary shock forms at the boundary to the cold layer as the
pressure equilibrium starts to be re-established (visible in
snapshots~3 --~6). The renewed growth sets in when the secondary shock
in front of the cold layer hits the undisturbed medium again
(snapshot~1).  The amplitude ${\cal A}$ is large and can reach up to
two orders of magnitude. None of the quantities oscillates
sinusoidally. In particular, the radiated energy is rather peaked.
For the RLF1 we find a C-type oscillation for stationary post shock
temperatures $T_{st}^{ps} > 4\cdot 10^{5}$~K. The upper limit probably
lies higher than $T_{st}^{ps}\approx 4\cdot 10^{6}$~K, the highest
limit post shock temperature we have investigated for RLF1. For the
RLF2 the C-type is present between $3\cdot 10^{5}$~K
$<T_{st}^{ps}<5\cdot 10^{5}$~K. We observe the C-type only in the
F-mode.

As another extreme, we observe the {\it smooth S-types}. There, the
shell remains hot during the entire oscillation cycle and no pressure
cavity evolves. No secondary shocks are observed. All quantities
oscillate rather sinusoidally. The S1-type can be associated with the
F-mode, the S2-type with the 1O-mode. $\cal{A}$ is always of order
one, but is generally smaller for the S2-type. In the case of the
S1-type, as the shell shrinks, a pressure gradient evolves between the
cold layer and the leading shock. The force it exerts leads to the
renewed acceleration of the leading shock.  In the case of the
S2-type, on the other hand, a pressure wave evolves at the boundary to
the cold layer which later hits the leading shock and accelerates
it. This leads to the phase-shift between the energy loss and the size
of the shell (see Fig.~\ref{fig:instab-type}). The differences between
the S1-type (F-mode) and the S2-type (1O-mode) correspond to the
results of the linear analysis of CI which predict knots and
phase-shifts for overtone modes. In the RLF1 we find the S2-type for
temperatures
$1.7\cdot10^{5}$~K~$<~T_{st}^{ps}~<~2.5\cdot10^{5}$~K. Note that the
stability limit is determined by the insufficient spatial resolution
in the simulation and may be somewhat lower in reality. We find no
S1-type in the RLF1. For the RLF2, we find the S1-type for
temperatures between $1.8\cdot 10^{5}$~K $ < T_{st}^{ps} <
2.2\cdot10^{5}$~K and the S2-type for temperatures $6.6\cdot 10^{5}$~K
$ < T_{st}^{ps} < 8.5\cdot10^{5}$~K.

The {\it intermediate I-type} occurs during the transition from C-type
to S-types.  The leading shell always remains rather hot, but -- as in
the C-type -- the shell vanishes apart from the secondary shock. The
amplitude of the oscillation is still rather big. We find the I-type
for temperatures in the range of $2.5 \cdot 10^{5}$~K $ < T_{st}^{ps}
< 4\cdot 10^{5}$~K for the RLF1 and in the range of $2.2\cdot
10^{5}$~K $ < T_{st}^{ps} < 3\cdot 10^{5}$~K for the RLF2. Also the
I-type is observed only in connection with the F-mode.

In the {\it mixed M-type} the leading shell is divided into two parts,
namely a high density part near the cold layer and a low density part
behind the leading shock. While the low density part remains hot the
high density part strongly cools (snapshots~2 and~3). When this cooled
dense matter is swallowed by the cold dense layer a pressure wave or a
weak shock is generated (snapshot~4) which travels through the leading
shell towards the leading shock (snapshots~4, 5 and~6). On its way it
enhances the density of the areas it passes. Finally, it hits the
leading shock and accelerates it (snapshot~6). By now the shell is
again divided into a high density and a low density part (snapshot~6
and~1). The oscillation amplitude is of order one. Most of the energy
is released during a very short time. We find the M-type for the
temperature range between $5\cdot10^{5}$~K $ < T_{st}^{ps} <
6.6\cdot10^{5}$~K of the RLF2, we do not observe it for the RLF1. The
{M-type} oscillates in the 1O-mode. Like the S2-type, the M-type shows
knots and phase-shifts.

The phenomenology of the overstability and, in particular, the
amplitudes of the oscillations are essentially determined by the
non-linear terms. Each of the modes lives mainly in two different
forms, a weak and a strong one. The {\it weak form} of the F-mode
corresponds to the S1-type, the {\it strong form} to the C-type, while
the {\it weak form} of the 1O-mode to the S2-type and the {\it strong
form} of the 1O-mode to the M-type.
\subsection{What determines the modes and the types?}
\label{subsec:instab_mode_det}
We do not yet have a final answer to this question. However, we
present some observations. They are mostly based on mode-transitions
which occur in WB-simulations when the post-shock temperature slowly
decreases due to the deceleration of the entire interaction region. We
begin with two general observations.

$\bullet$ 1) According to our experience, for a given radiative loss
function, the type of the overstability is fixed by the post-shock
temperature $T_{st}^{ps}$ of the (unstable) stationary solution alone:
Different forms of excitation of the instability at the same
$T_{st}^{ps}$ result -- after a certain relaxation time -- in the same
mode and the same type. {\it We detect no sign of multiple solutions
or a bifurcation of the solution.}

$\bullet$ 2) Linear theory predicts that whenever the F-mode is
present the 1O-mode is present as well. However, {\it in the strong
forms of the F-mode we do not observe any sign of a superimposed
1O-mode}. Only at the transition from the I-type to the S-type a
slight trace of a superimposed 1O-mode shows up.  At this point, we
\begin{figure}[hp]
\centerline{
\psfig{figure=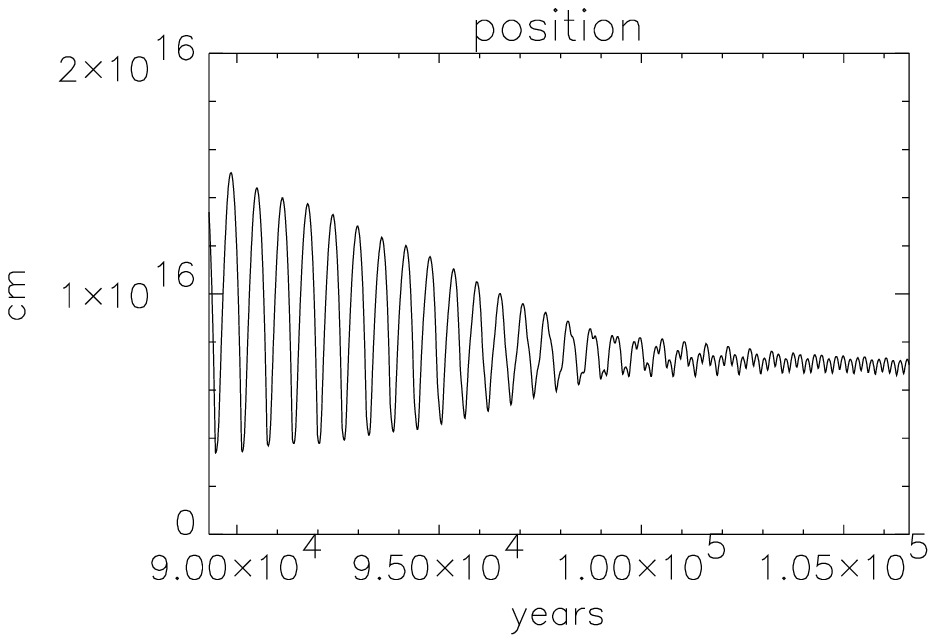,height=3.4cm,width=8.8cm}
           }
\centerline{
\psfig{figure=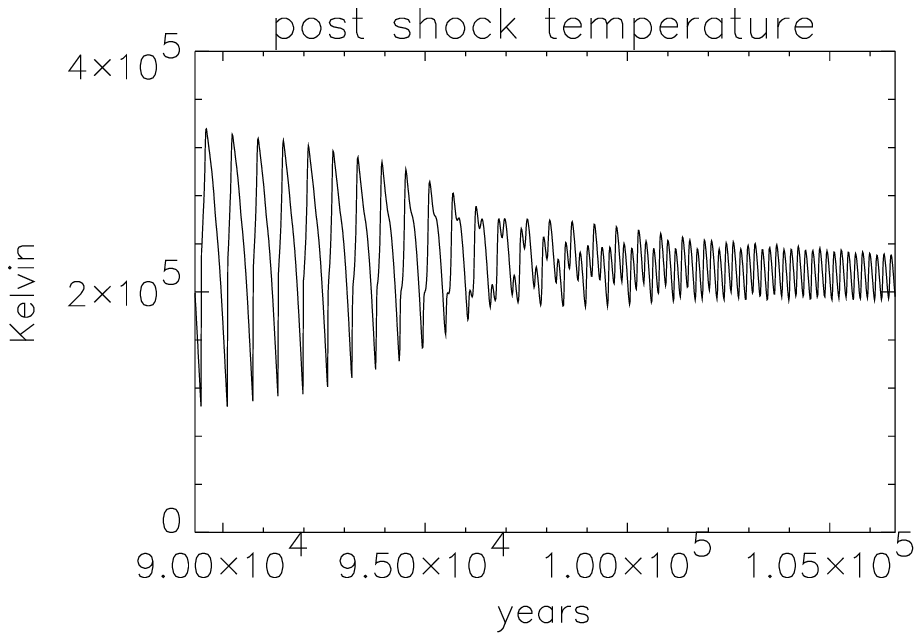,height=3.4cm,width=8.8cm}
           }    
\vspace{-0.3cm}
\caption{Mode-transition in WB1 from the F-mode (I-type) to 
         the 1O-mode (S2-type) for the RLF1. Shown are the 
         time-evolution of the size of the leading shell (top) 
         and the post-shock temperature (bottom).}
\label{fig:smooth-mode-transition}
\end{figure}
observe no phase shift between the modes due to the non-rational ratio
of the eigenfrequencies. The phase shift only sets in when the
overstability is already very smooth
(Fig.~\ref{fig:smooth-mode-transition}). The physical reason for the
absence of overtone modes in the strong forms may be the total
destruction of the leading shell and, therefore, the history after each
oscillation cycle.
\begin{figure}[tp]
\centerline{
\psfig{figure=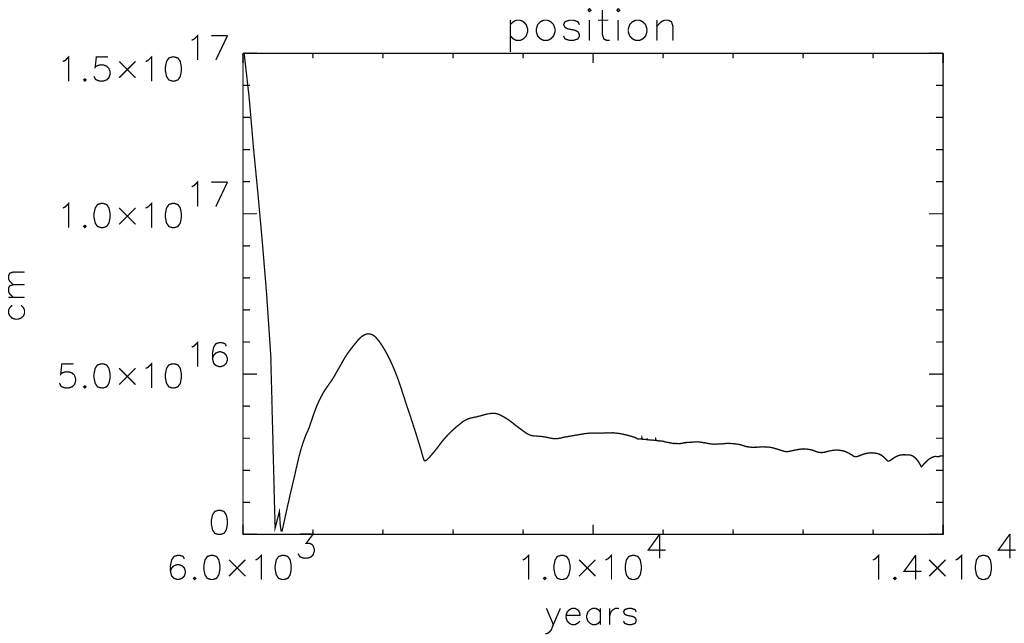,height=3.3cm,width=8.8cm}
           }    
\centerline{
\psfig{figure=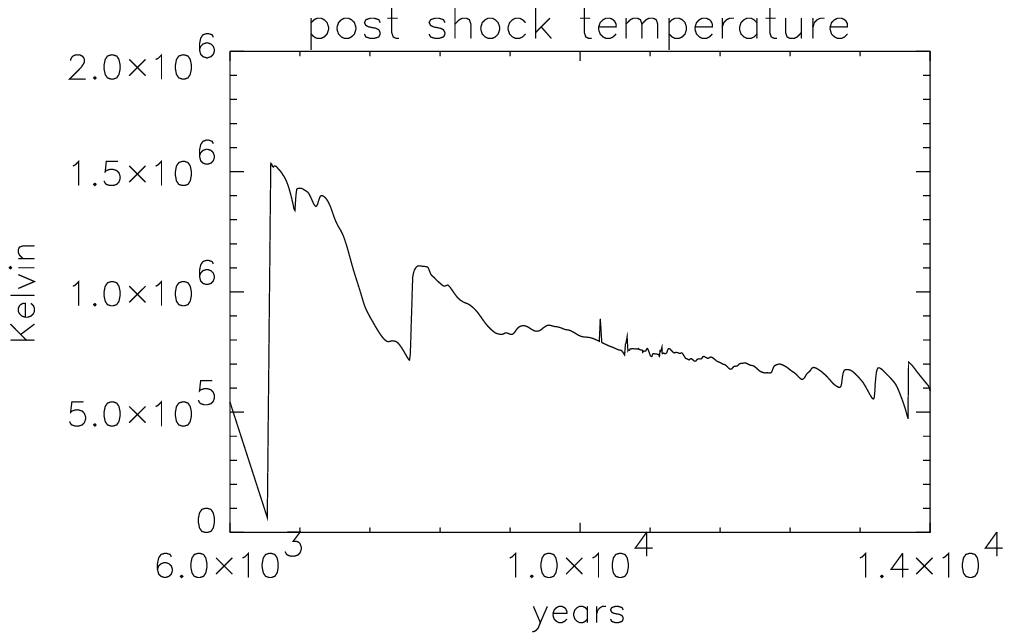,height=3.3cm,width=8.8cm}
           }  
\vspace{-0.3cm}
\caption{WB2f: Stability behavior when applying RLF2 for $1\cdot10^6 >
         T^{ps}_{st} > 5\cdot10^5$. Shown is the size of the leading 
         shell (top) and the post-shock temperature (bottom).
         Above T$_{st}^{ps} \approx 7\cdot10^5$
         the (artificially excited) F-mode (C-type) is strongly damped,
         below this temperature a 1O-mode is growing.
         The further evolution is shown in the next figure.}
\vspace{-0.3cm}
\label{fig:rlf2-mode-transition_1}
\end{figure}

The next two points are dealing with the transition of types and modes
when scanning RLF1 in the temperature range where the logarithmic
slope changes from a negative to a positive value (see
Fig.~\ref{fig:smooth-mode-transition}).

$\bullet$ 3) We observe a mode and type-transition from the F-mode
(I-type) to the 1O-mode (S2-type) at T$_{st}^{ps}\approx
2.5\cdot10^{5}$~K when applying RLF1. That is approximately the
temperature where $\beta$ jumps from a strongly negative value to a
moderately positive value. {\it For this transition, the change of
$\beta$ associated with the post-shock temperature seems to be
critical.} 
\begin{figure}[tp]
\centerline{
\psfig{figure=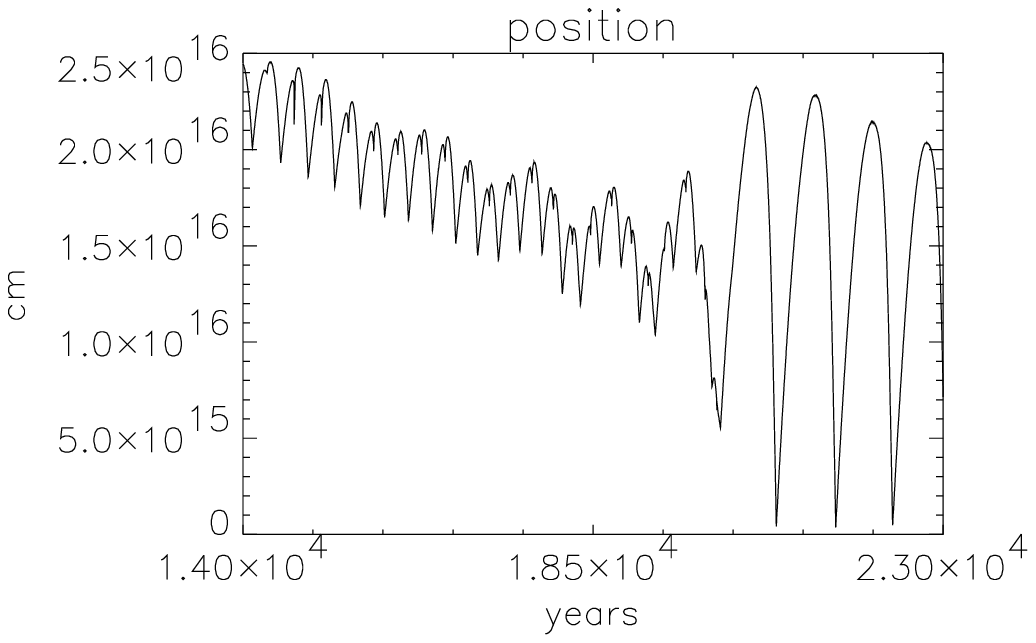,height=3.3cm,width=8.8cm}
           }    
\centerline{
\psfig{figure=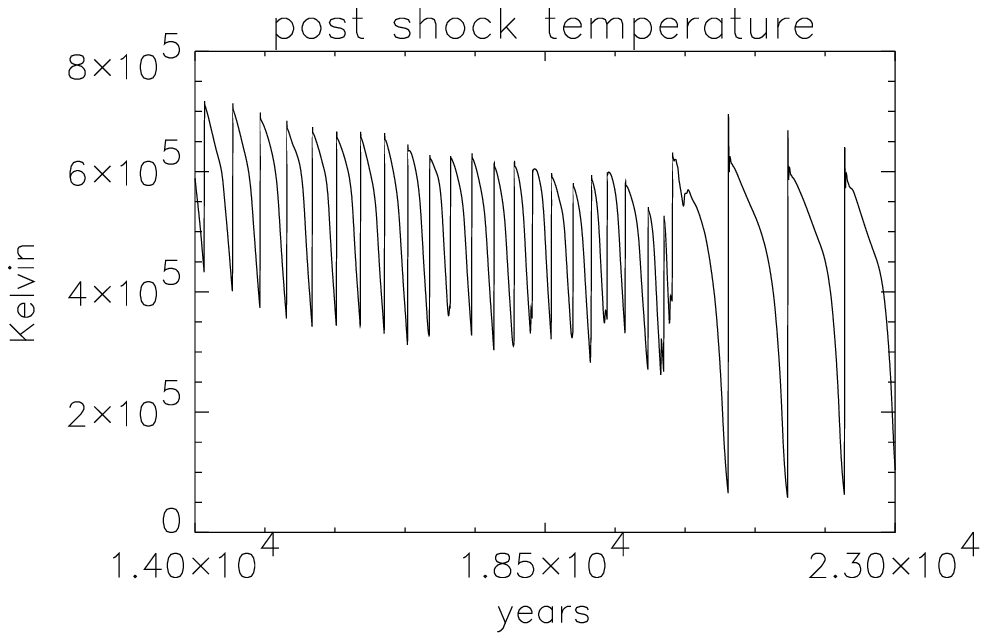,height=3.3cm,width=8.8cm}
           }    
\vspace{-0.3cm}
\caption{WB2f: Mode-transition from 1O-mode (M-type) to F-mode (C-type)
         at $T^{ps}_{st} < 5\cdot10^5$ when applying RLF2. 
         Shown is the size of the leading shell (top) and the post-shock 
         temperature (bottom).(The evolution at
         earlier times is shown in the previous figure.)}
\vspace{-0.3cm}
\label{fig:rlf2-mode-transition_2}
\end{figure}

$\bullet$ 4) The absence of the F-mode below T$_{st}^{ps} \approx
2.5\cdot10^{5}$~K for RLF1 is surprising. The post-shock temperature
varies over the entire oscillation cycle between $1.9\cdot10^5$~K and
$2.3\cdot10^5$~K and even most parts of the shell have always
temperatures above $7.9\cdot10^4$~K. For these temperatures, $\beta$
is equal to 0.15 while below $7.9\cdot10^4$~K it becomes strongly
positive. Linear analysis predicts for $\beta=0.15$ the F- {\bf and}
the 1O-mode to be present. Indeed, when using a RLF assuming a
constant $\beta = 0.15$ we observe both modes. We conclude that {\it
the mode is not governed by the $\beta$ which is associated with the
stationary post-shock temperature alone}. The entire cooling layer
and, in particular, the rear boundary interface may be of similar
importance. The strongly positive $\beta$ for T~$ \le 7.9\cdot10^4$~K
reduces the cooling efficiency as compared to $\beta=0.15$ and in this
way seems to damp the F-mode.

The next three points summarize the transition of modes and types when
scanning RLF2 in the temperature range $1\cdot10^{6} \gapprox
$T$_{st}^{ps} \gapprox 4\cdot 10^{5}$~K (see
Figs.~\ref{fig:rlf2-mode-transition_1}
and~\ref{fig:rlf2-mode-transition_2}). This is exactly the range where
RLF2 forms a 'valley'. For $T < 4\cdot 10^{5}$~K the logarithmic slope
of the RLF2 is equal to -1.92, for $4\cdot10^{5}< T <
5.37\cdot10^{5}$~K it is zero and above $5.37\cdot10^{5}$~K it is
equal to 1.12.  The planar models, CPF\_55 to CPF\_25, prove that the
different modes and types observed in this 'valley region' of the RLF2
are not merely transition phenomena but persist if the structure does
not slow down.

$\bullet$ 5) The excited F-mode (C-type) at
T$_{st}^{ps}\gapprox9\cdot10^{5}$~K is very strongly damped when RLF2
is applied. This suggest a stable situation for this temperature, and
indeed the planar model CPF\_55 with T$_{st}^{ps}=9\cdot10^5$~K is
stable. At this temperature we have $\beta=1.12$ which is in the
stable range according to linear theory. However, large parts of the
cooling layer have temperatures associated with an unstable
$\beta$. Despite this fact, {\it in this case the $\beta$ associated
with the post-shock temperature seems to determine the stability.}

$\bullet$ 6) Below T$_{st}^{ps}\approx8.5\cdot10^{5}$~K a 1O-mode of
the S2-type grows for RLF2 (CPF\_45 for T$_{st}^{ps}\approx
8.25\cdot10^5$~K shows S2-type behavior).  The post shock temperature
does not leave the region with $\beta=1.12$ over the entire
oscillation cycle. Interestingly, nearly in the entire cooling layer
the temperature is in the range where $\beta=1.12$! With decreasing
T$_{st}^{ps}$, the type of the overstability slowly changes to the
strong form of the 1O-mode (M-type) (CPF\_35 for T$_{st}^{ps}\approx
6.1\cdot10^5$~K is clearly of the M-type). $\beta$ associated with
T$_{st}^{ps}$ is still 1.12! {\it This again indicates that the RLF
for temperatures below T$_{st}^{ps}$ plays a significant role.}

$\bullet$ 7) As can be seen from
Fig.~\ref{fig:rlf2-mode-transition_2}, at T$_{st}^{ps}\approx 5\cdot
10^{5}$~K another mode-transition takes place for RLF2, this time from
the 1O- (M-type) to the F-mode (C-type). This temperature corresponds
approximately to a change in $\beta$ from the stable value 1.12 to
zero, a value at which -- according to linear theory -- both modes
should be present. {\it The change of $\beta$ corresponding to the
post-shock temperature seems to be important for the mode-transition.}

The emerging picture is not homogeneous. Sometimes, the $\beta$
associated with T$_{st}^{ps}$ seems to play an outstanding role. More
often, however, the presence, the mode, and the type of the
overstability cannot be determined on the basis of the local
characteristics of the RLF at this temperature. We come back to these
points in Sect.~\ref{sec:discussion}

%
%
%
\section{The cooling instability of the leading shell for disturbed flows}
\label{sec:disturb}
\subsection{The set of disturbed flows}
We now discuss the cooling instability for disturbed flows. We disturb
the flow upstream of the interaction zone between position r$_0$ and
r$_0 + \lambda_0$ by setting the density to
\be
\rho(r) = \rho_0 \left[ 
      1 + \xi \sin \left( \pi \frac{r - r_0}{ \lambda_0}  \right) \right] 
\mbox{if} \hspace{0.2cm}  r_0 \le r \le r_0 + \lambda_0  
\label{eq:disturb}
\ee
where $\rho_0$ is the density of the undisturbed flow.
The disturbance is applied after the system has evolved into a limit
cycle. The leading shock can hit it at an arbitrary phase $0 \le
\phi_0 \le 1$ of the oscillation cycle (where $\phi=0,1$ are the
phases of minimum extent of the leading shell). Note the difference to
the setting of IGF who let the shock in its stationary position run
against a, however similar, density disturbance.

We performed a systematic investigation of disturbed I-type systems on
the example of WB1.  All disturbances were applied around
69'000~years. The exact settings of these models can be taken from
Table~\ref{tab:disturb-i} of the appendix. Except for two models
(WB1\_d0 and WB1\_d1) the amplitude of the disturbance, $\xi$, is
never bigger than 1. The ratio of the mass of the disturbance,
M$_{\mbox{d}}$, to the mean mass (over a period) of the hot leading
shell, M$_{\mbox{ls}}$, varies from 0.017 to 73, the ratio of
M$_{\mbox{d}}$ to the mass of the subsequent cold dense layer,
M$_{\mbox{cdl}}$, from $1.4\cdot 10^{-4}$ to 0.6155. The ratio of the
length of the disturbance, $\lambda_0$, to the characteristic length
of the limit cycle, $\ell=T \cdot v^{sh}_{st}$, varies from 0.05 to
8.8 ($T$ denotes the oscillation period and $v^{sh}_{st}$ the
stationary shock velocity). The shock hits the disturbance at phases
between $\phi_0=0.07$ and 0.52. The results presented in this section
are a summary of the analysis of this set.
\subsection{Noisy, original periodic, aperiodic, and modulated 
                                          periodic evolution}
We found five clearly different responses of the I-type system to
disturbances. The time-evolutions of three of them are shown in
Fig.~\ref{fig:disturb-i-type-ap-periodic} and a schematic
representation of all possible responses is given in
Fig.~\ref{fig:disturb-i-type-phase}. If the disturbance is small (in
\bfi[htp]
\centerline{
  \psfig{figure=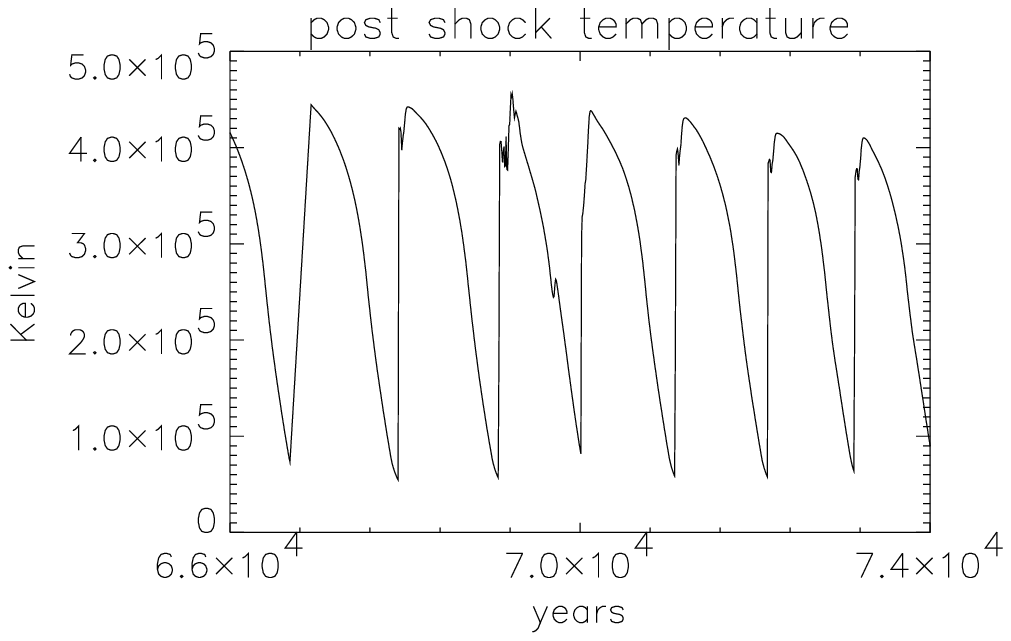,height=3.cm,width=8.2cm}
            }
\centerline{
  \psfig{figure=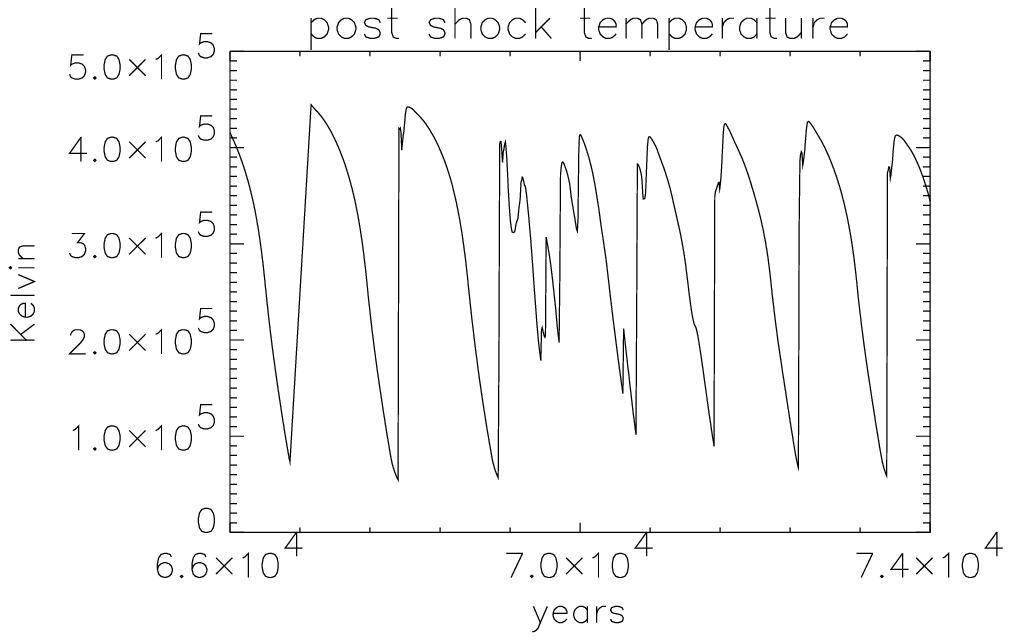,height=3.0cm,width=8.2cm}
            }
\centerline{
  \psfig{figure=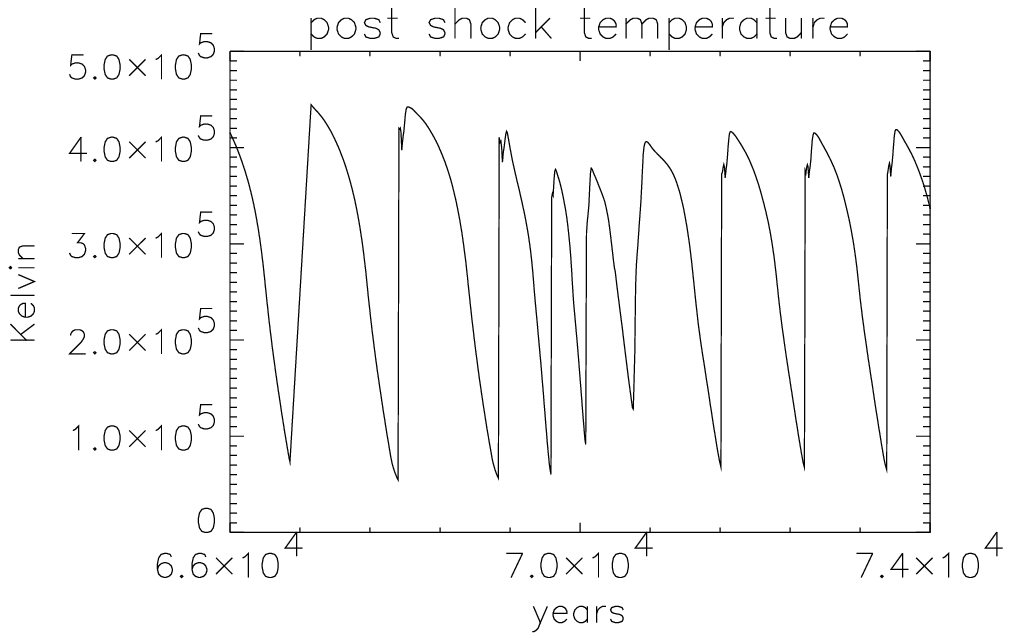,height=3.cm,width=8.2cm}
           }
\caption{Noisy original periodic, aperiodic and modulated 
         periodic evolution of the radiative shock after being 
         disturbed at time 68'893~years.
         Shown are post-shock temperatures of models WB1\_d4.0.1,
         WB1\_d4.1 and WB1\_d6 (from top to bottom).}
\label{fig:disturb-i-type-ap-periodic}
\efi
which sense we see below), one observes a rapid, aperiodic, much
smaller scaled variation of only one cycle of the original
oscillation. We call this a {\it noisy, original periodic evolution
(np)}. If the disturbance is appropriately scaled, the periodicity of
the limit cycle is completely destroyed and we have an {\it aperiodic
evolution of the radiative shock (ap).} However, after a certain time,
the system relaxes to the original, periodic evolution. Disturbances
on a larger length scale and not too big amplitude $\xi$ result in a
modulation of the limit cycle, governed by the slowly changing
pre-shock conditions. In this case, we speak of a {\it modulated
periodic evolution (mp)}. On the other hand, if we have a very massive
\bfi[htp]
\centerline{\psfig{figure=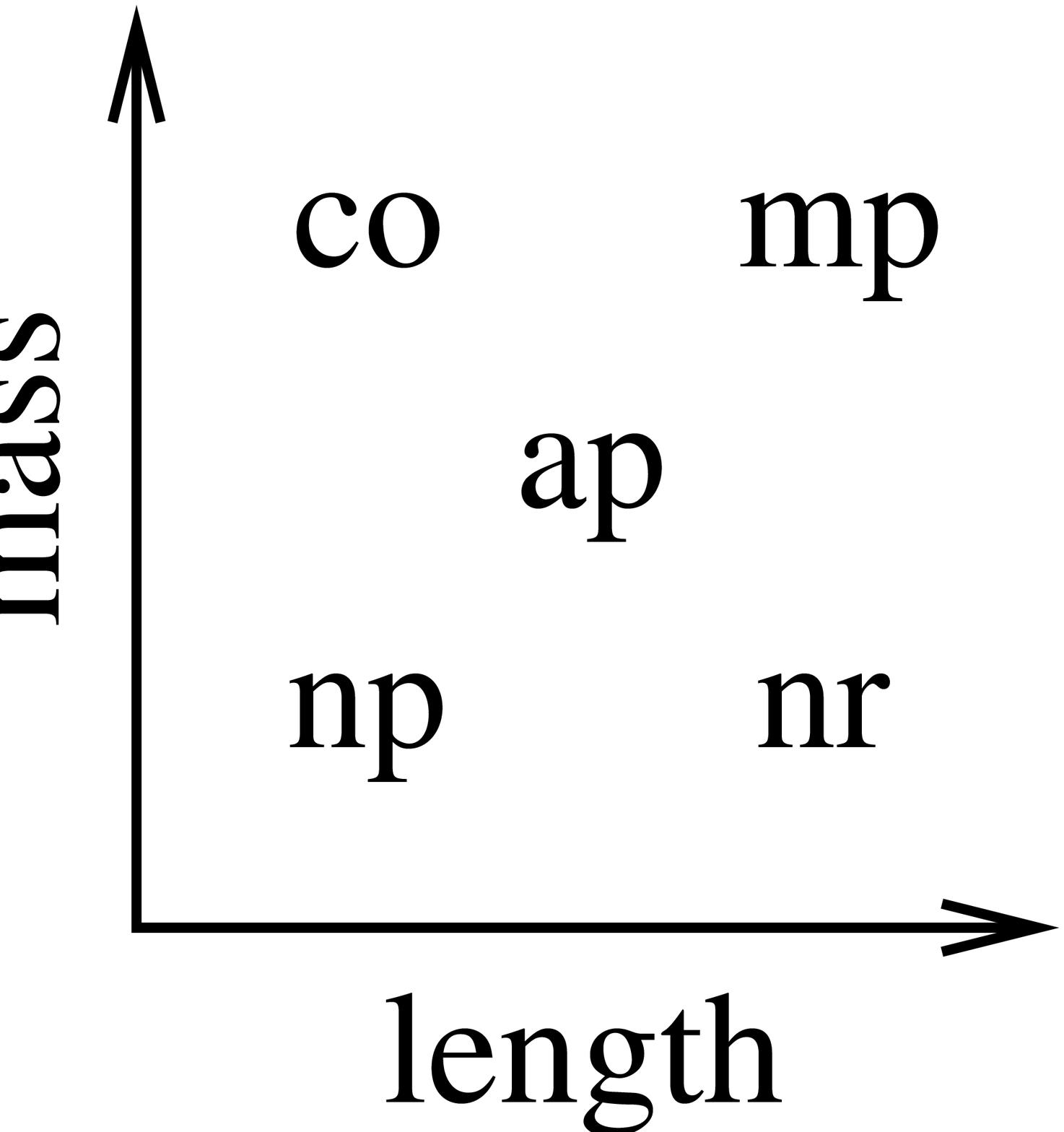,width=3.7cm}}
\caption{Schematic representation of the different responses
         versus mass and length of the density disturbance.}
\label{fig:disturb-i-type-phase}
\efi
disturbance on a relatively short length-scale, the leading {\it shell
collapses nearly immediately (co)}. 
The history of the leading shell
is destroyed and the shell behaves according to the new global
conditions. For example, in model WB\_d0 the accumulation of the mass
of the disturbance by the cold dense layer leads to a slow down of the
entire interaction zone and the leading shock becomes stable. Finally,
if we encounter a small mass disturbance on a large length scale, the
system shows nearly {\it no reaction (nr)}. We note already here, that
big disturbances may significantly disturb the cold dense layer
downstream of the radiative shock wave (see
Sect.~\ref{sec:coldshell}).
\begin{figure*}[htp]
\hbox{
\psfig{figure=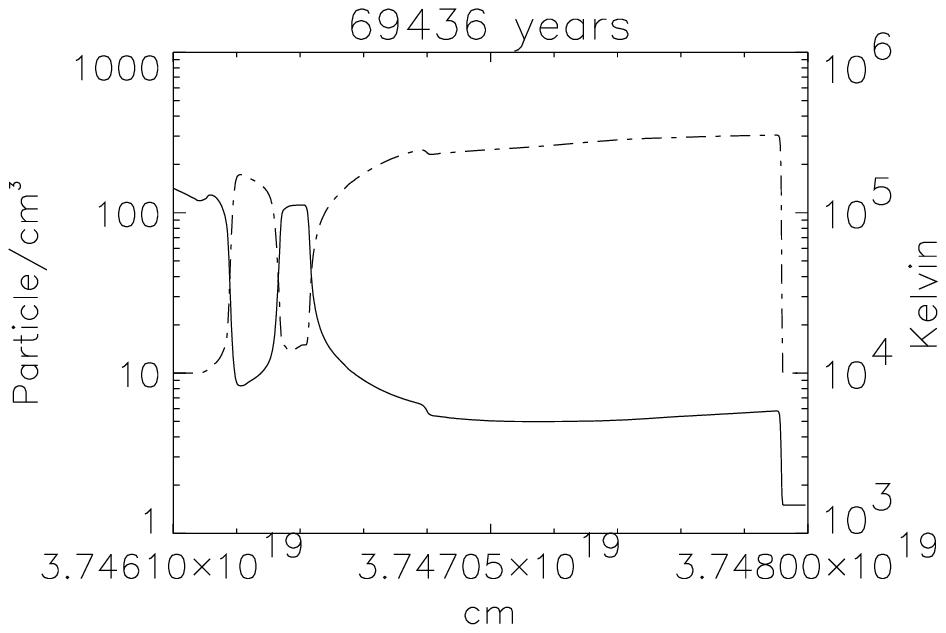,height=3.8cm,width=5.9cm}
\psfig{figure=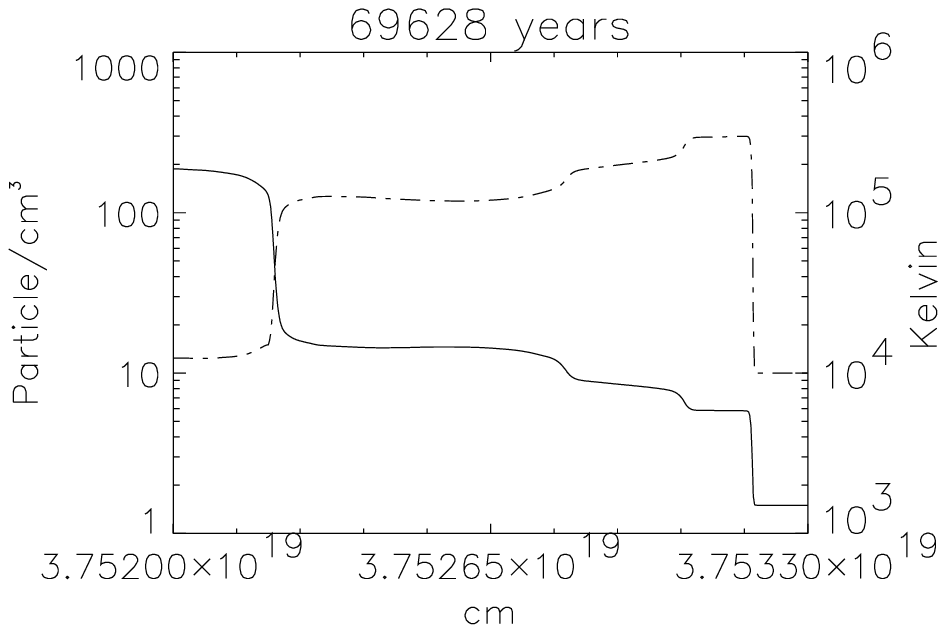,height=3.8cm,width=5.9cm}
\psfig{figure=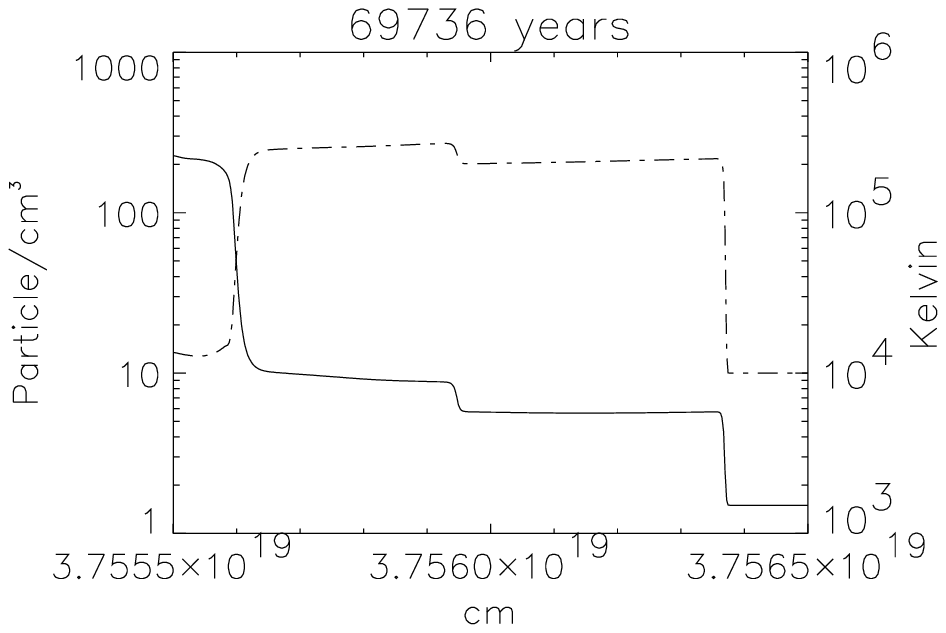,height=3.8cm,width=5.9cm}
     }
\vspace{-0.2cm}
\caption{The aperiodic evolution of the radiative shock and its
         cooling layer from example of WB1\_d4.1. Shown are
         density- (solid line) and temperature-profiles (dashed line)
         of three successive times. The evolution is governed by the
         fast partial cooling and the waves which are created
         during this process. Details can be taken from Sect.~6.3.}
\label{fig:caseB.disturbance.example.1}
\end{figure*}
\subsection{The aperiodic evolution}
\label{subsec:disturb_aperiodic}
The oscillation mechanisms of the np- and the mp-cases are similar to
the mechanisms described in Sect.~\ref{sec:instab}. Here we describe
the aperiodic response of the radiative shock at the example of model
WB1\_d4.1. It evolves very similar to what was described by IGF.
We concentrate on some aspects important to the understanding of
the constraints for the evolution of disturbed shocks.

The size of the initial density disturbance is crucial for the time
evolution illustrated in Figs.~\ref{fig:disturb-i-type-ap-periodic}
and \ref{fig:caseB.disturbance.example.1}.  As the density disturbance
passes through the leading shock it shrinks in size while being
compressed.  Cooling of the hot shell starts in the region with
enhanced density.  In fact, cooling is so fast that the flow cannot
maintain pressure balance and a pressure cavity forms in the cooled
region.  A sheet of cold and initially over-compressed matter is then
created within the hot leading shell (snapshot~1 of
Fig.~\ref{fig:caseB.disturbance.example.1}). The relaxation of the
over-compressed sheet generates two pressure waves or weak shocks
traveling forwards and backwards in the hot shell. Their strengths are
a direct measure of the violence of formation of this second dense
sheet. The dynamical consequences of these waves are two\-fold.
Firstly, temperature and density increase again in those regions where
the waves pass. The consequence for the cooling scale cannot be
estimated easily, since in this region of the radiative loss function
the increase in temperature and density are working in opposite
directions. For RLF1 and in this temperature range the cooling scale
likely is to decrease behind weak waves, it is likely to increase
behind strong waves. In our example, the first possibility applies. A
second important consequence of these waves is encountered when they
eventually hit the leading shock. The pushing wave transfers part of
its energy to the leading shock which is accelerated. As a consequence
of the shock jump conditions, the post-shock temperature is higher and
the density is lower in the freshly shocked part than in the older
parts of the shell (compare snapshot~2). The further development can
hardly be followed in detail. The older, denser parts cool faster than
the newer parts. Further cold, compressed sheets are formed, usually
near the boundary to the cold layer. Their formation creates again new
waves and the process continues (snapshot~3).

Our investigation proves -- for the I-type range and for parameterized
cooling -- that this process does not last forever and that the limit
cycle finally recovers.
\subsection{The constraints for the evolution and relaxation times}
\label{subsubsec:disturb-constraints}
As can be seen from Table~\ref{tab:disturb-i} in the appendix, models
having a length of the disturbance between 5 and 20 percent of the
characteristic length of the limit cycle and having masses between 10
and 40 percent of the mass of the leading shell react with an
aperiodic evolution to the disturbance (case ap).

To achieve the most efficient aperiodic evolution, the region of
enhanced density has to be {\it moderately smaller} than the hot
shell, since otherwise the entire leading shell would be affected
nearly similarly and we would encounter a modulation of the original
oscillation (case mp) or a collapse of the shell (case co). On the
other hand, the region with enhanced density has to be large enough to
be able to create {\it secondary waves} as powerful as possible, since
these waves are driving the aperiodic evolution.

For the aperiodic evolution, we define the {\it relaxation time},
$\tau_{rel}$, as the difference between the time when the aperiodic
evolution first sets in and the time when the regular periodic
evolution starts again. $\tau_{rel}$ can be taken from
Table~\ref{tab:disturb-i}. The biggest ratio of $\tau_{rel} /
\tau_{0}$ ($\tau_{0}$ being the period of the limit cycle) we measured
is 3 (model WB1\_d4.1). In this case, the ratio of the characteristic
lengths, $\lambda_0 / \ell_{c} $, is 0.2, the mass ratio,
M$_{\mbox{d}}$/M$_{\mbox{hls}}$, is 0.4. Comparing models WB1\_d3,
WB1\_d2.2 and WB1\_d4.1 we notice the influence of the different
masses. A larger mass of the disturbance leads to the generation of
stronger pressure waves, and the relaxation time increases.
%
%
%
\section{The cold dense layer (CDL) and the hot leading shell (HLS): 
                                        An interacting system}
\label{sec:coldshell}
Downstream of the radiative shock and its cooling layer, a thin layer
of cold, compressed gas is established (see
Fig.~\ref{fig:adia-rad-pattern}). In this section we study the
dynamics of this layer, in particular its interaction with the
thermally unstable radiative shock.
\subsection{Mass feeding of the CDL}
\label{subsubsec:coldshe-mass_feeding}
We define the boundary between the cold dense layer and the hot
leading shell to be at the surface where cooling is balanced by
heating. Through this boundary interface, there is a net mass-flux from
the cooling layer of the radiative shock into the cold dense layer
which is shown in the last row of Fig.~\ref{fig:instab-type}.

It is interesting that for smooth flows the five different types of
the overstability manifest themselves also in these mass-fluxes
(Fig.~\ref{fig:instab-type}). Most remarkable, there is a very
significant difference between the strong and the weak form of the
overstability. For S-type instabilities the mass-fluxes are smooth and
vary only moderately over one cycle. In the strong forms of the
overstability, most of the mass enters the CDL during a very short
time-interval of the cycle. This is a direct consequence of the sudden
collapse of large amounts of the hot shell. Again we note a
significant difference between the modes: In the 1O-mode (M-type), the
peak of the mass-flux is correlated with the largest, in the F-mode
(I- and C-types) with the smallest extension of the cooling
layer. Note also that in the I- and C-type overstability the onset of
the mass-flux at approximately one third of the period is accompanied
by the formation of the secondary shock which is attached to the CDL.
Obviously, the irregularity of the mass flux is enhanced in disturbed
flows.
\subsection{Response of the CDL to the cooling overstability of
            the leading shell}
\label{subsubsec:coldshell-response}
The non-steady mass-flux through the interface between the HLS and the
CDL introduces disturbances into the CDL, even when smooth flows
collide. Assuming that the size of the CDL is much bigger than the
HLS, \cite*{radshock:bertsch-stab2} has shown by linear stability
analysis that the disturbances damp on a scale which is much smaller
than the size of the CDL. When we encounter the same conditions, our
numerical study shows indeed this behavior. However, there are many
cases where the size of the CDL and the HLS are comparable, e.g. in
the period after the transition of the shock from the adiabatic to the
radiative regime. Moreover, the analysis in
Sect.~\ref{subsubsec:coldshe-mass_feeding} shows that in strong forms
of the overstability the mass feeding into the CDL -- and, therefore,
the introduced disturbance -- is extremely pulsed with a large pulse
height. When disturbed flows collide, these disturbances can become
even larger. Such disturbances are clearly beyond the scope of linear
analysis. Therefore, it is not surprising that under these different
circumstances the disturbances can survive for a significant time.
\bfi[tp]
\centerline{
      \psfig{figure=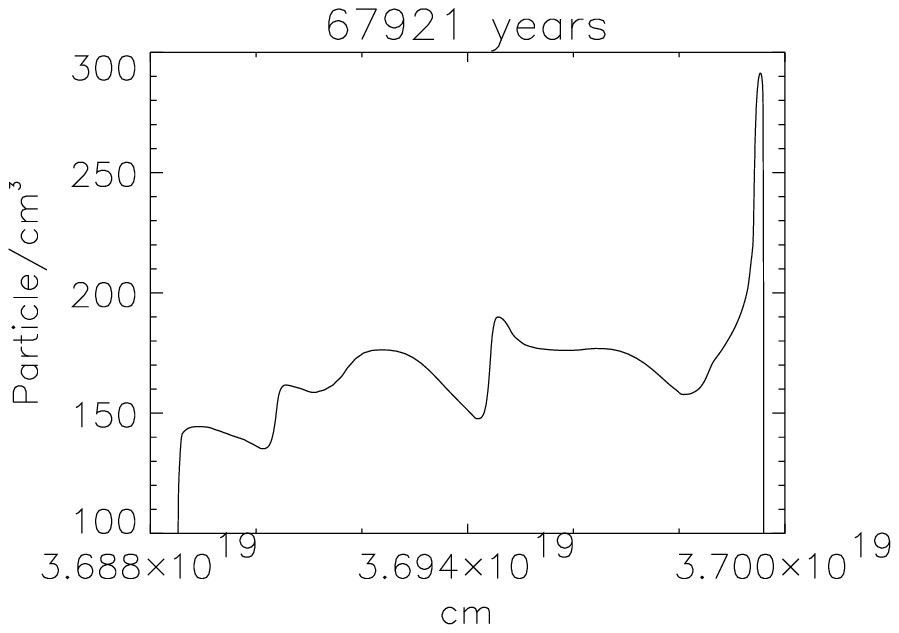,width=4.3cm}
      \psfig{figure=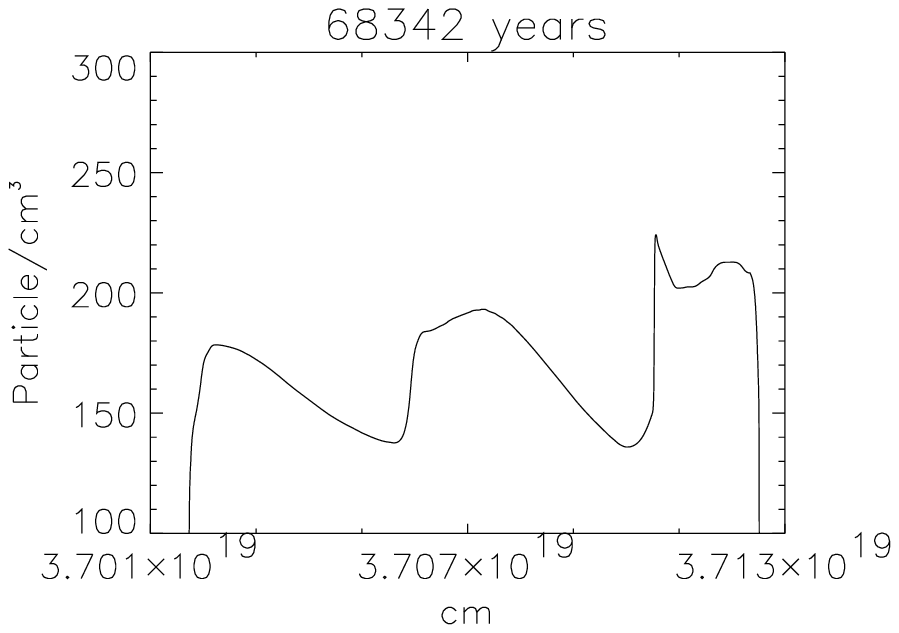,width=4.3cm}
      }
\caption{Density distribution of the CDL of WB1 (I-type instability)
         immediately after the 
         leading shell has collapsed and the overcompressed sheet of
         freshly cooled matter is swallowed by the CDL ({\bf left}) and
         half a period later in the limit cycle ({\bf right}).}
\label{fig:cold-layer-rn-dens-diff_times}
\efi
\subsubsection{Traveling waves and density variation}
As described in Sect.~\ref{subsec:disturb_aperiodic} the sudden
cooling of large parts of the hot leading shell leads to the formation
of an initially overcompressed sheet of cold matter. Apparently, the
same mechanism also works in the strong forms of the limit cycle of
the overstability when smooth flows collide, with the difference that
the newly cooled, overcompressed sheet is in contact with the already
\bfi[htp]
\centerline{
 \psfig{figure=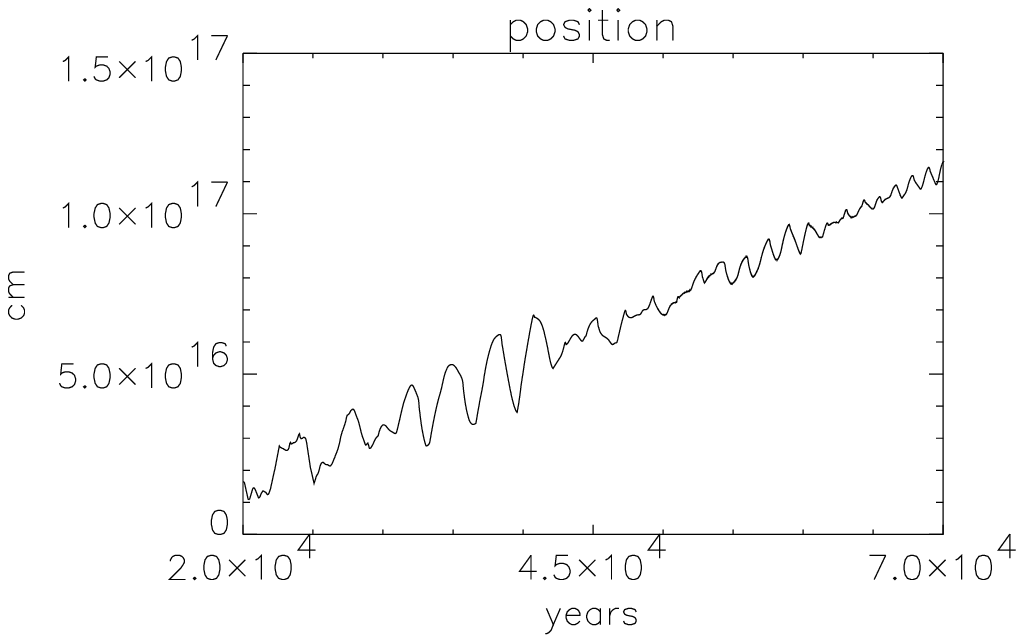,height=3.1cm,width=8.4cm}
           }
\centerline{
 \psfig{figure=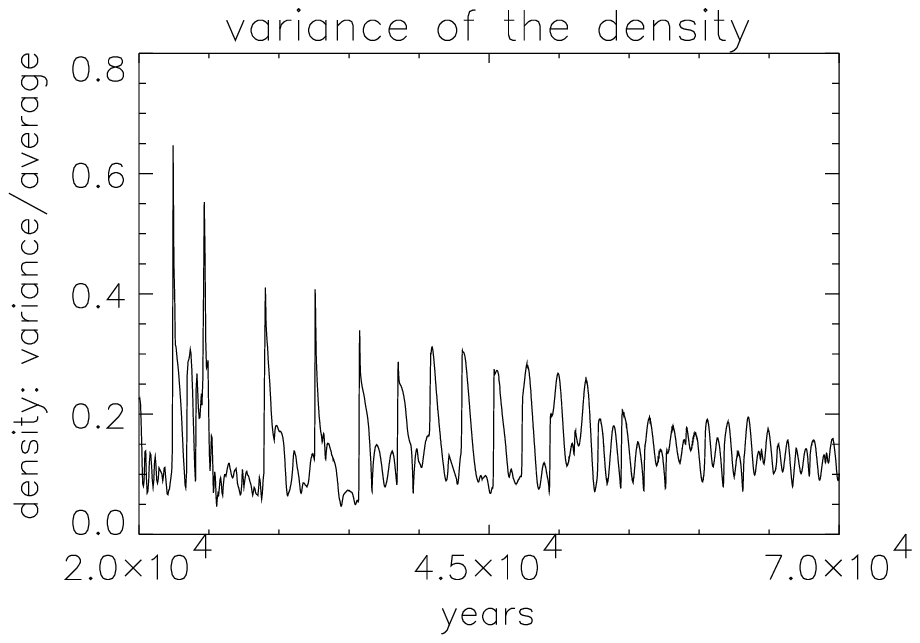,height=3.1cm,width=8.4cm}
           }
\caption{Oscillating size ({\bf top}) and density-variance $\sigma$
         ({\bf bottom}) of the CDL. Shown are 18~cycles after the 
          transition from the adiabatic to the radiative regime of model WB1.
         The overstability of the leading shell is of the C- and I-type.}
\label{fig:cdl-variations}
\efi
present CDL. The swallowing of such an overcompressed sheet by the CDL
is shown on the left of Fig.~\ref{fig:cold-layer-rn-dens-diff_times}.
The thin sheet at the very right boundary of the CDL, having
approximately twice the density of the rest, contains the freshly
cooled matter. At this time, the density variance in the shell is
greatest (see Fig.~\ref{fig:cdl-variations}). In the following,
relatively quiescent phase of the limit cycle, the overcompression
relaxes by a wave traveling through the CDL (right graph of
Fig.~\ref{fig:cold-layer-rn-dens-diff_times}). Eventually, the
propagating wave is reflected at the rear interface of the CDL and
interacts with the other waves. As a result, the density structure of
the CDL is rather complicated in the presence of strong forms of the
overstability of the radiative shock. The evolution of the density
variance shown in Fig.~\ref{fig:cdl-variations} exactly mirrors the
cycle of the overstability of the leading shell. It shows peaks every
time a new, overcompressed sheet is incorporated into the CDL. The
figure demonstrates that the density variance is rather high as long
as we encounter I-type instability. Not visible in the figure, the CDL
rapidly smoothes after the instability changes to S-type. Note also,
that we do not find a significant difference between the density
variance of an isothermal CDL (model WB1c) and an adiabatic CDL (model
WB1).
\subsubsection{Oscillations of the CDL}
The CDL is expected to become compressed when it pushes the shock
after the collapse. This should lead to a considerable shrinking of
the size of the CDL, at least as long as there is not too much mass in
the CDL. Indeed, we observe such oscillations (see
Fig.~\ref{fig:cdl-variations}). However, the traveling, reflecting,
and interacting waves disturb the basic oscillation cycle
remarkably. Periods of nearly no oscillation are followed by periods
of nearly regular oscillation and vice-versa. When smooth flows
collide, we find changes in the size of the CDL up to 15 percent as
long as we encounter I-type instability in the WB and SNR simulations.
Much larger oscillation of the CDL can be excited in disturbed flows
as will be shown in the next paragraph.
\subsection{The back-coupling to the cooling layer}
\label{subsec:coldshell-backcoupling}
We now investigate how the dynamics of the CDL affect the radiative
cooling overstability of the hot leading shell. Of course, the
oscillation of the CDL changes the rear boundary conditions of the
cooling layer with time, which influences the evolution of the
radiative shock. This back-coupling can be strong in disturbed flows
as will be demonstrated below. When smooth flows collide, the effect
is present but never large (See Fig.~\ref{fig:rn_lev4_lev5}). Only
immediately after the transition from the adiabatic to the radiative
phase it is considerable. In this phase the amplitude of the
overstable oscillation can change up to 15 percent from one period to
the next. Although weaker, the effect can be observed as long as the
overstability is of the I-type.
\subsubsection{The evolution of model WB1\_d7}
WB1\_d7 is a disturbed model with a modulated periodic behavior (see
Table~\ref{tab:disturb-i}). The disturbing mass is 7.3 times the
average mass of the hot leading shell and 6.18 percent of the mass of
the CDL. The encounter of the interaction zone with this density
disturbance excites an oscillation of the CDL with an amplitude of
about 15 percent. The period of the oscillation approximately
corresponds to the length of the density disturbance. As demonstrated
in Fig.~\ref{fig:cdl-mass-disturb}, this oscillation of the CDL
significantly influences the further evolution of the cooling
overstability.

Note also the direct consequence of the density disturbance for the
overstability (see again Fig.~\ref{fig:cdl-mass-disturb}). The leading
shock slows down. The oscillation period becomes smaller due to the
decreasing cooling time. However, the radiative energy release
increases due to the density enhancement which overcomes the decrease
of the emitting volume due to the smaller size of the shell. When the
\bfi[htp]
\centerline{
      \psfig{figure=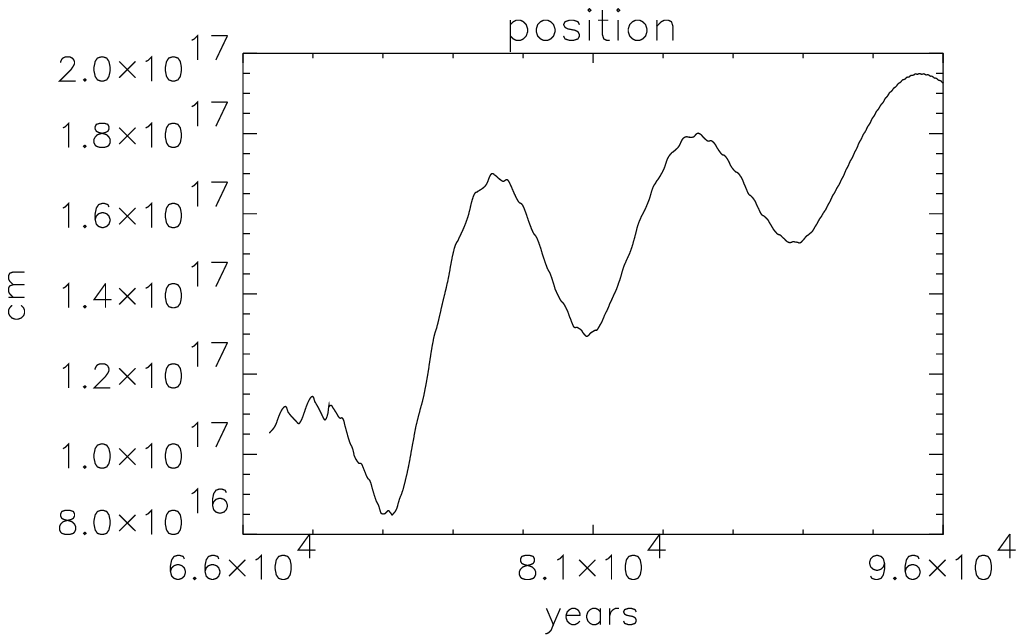,height=3.1cm,width=8.4cm}
           }
\centerline{
      \psfig{figure=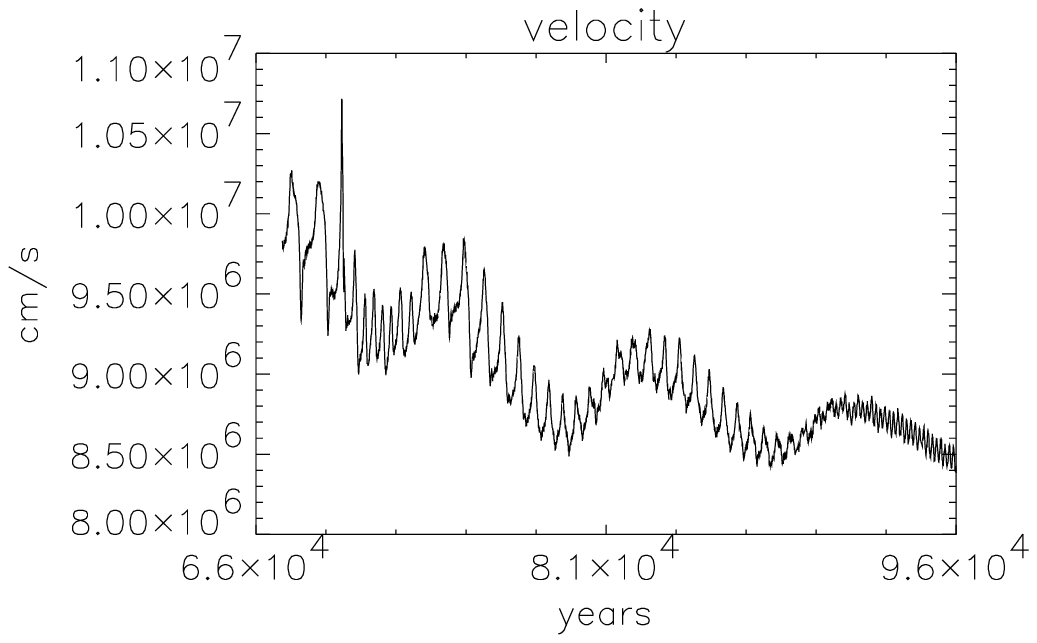,height=3.1cm,width=8.4cm}
           }
\centerline{
      \psfig{figure=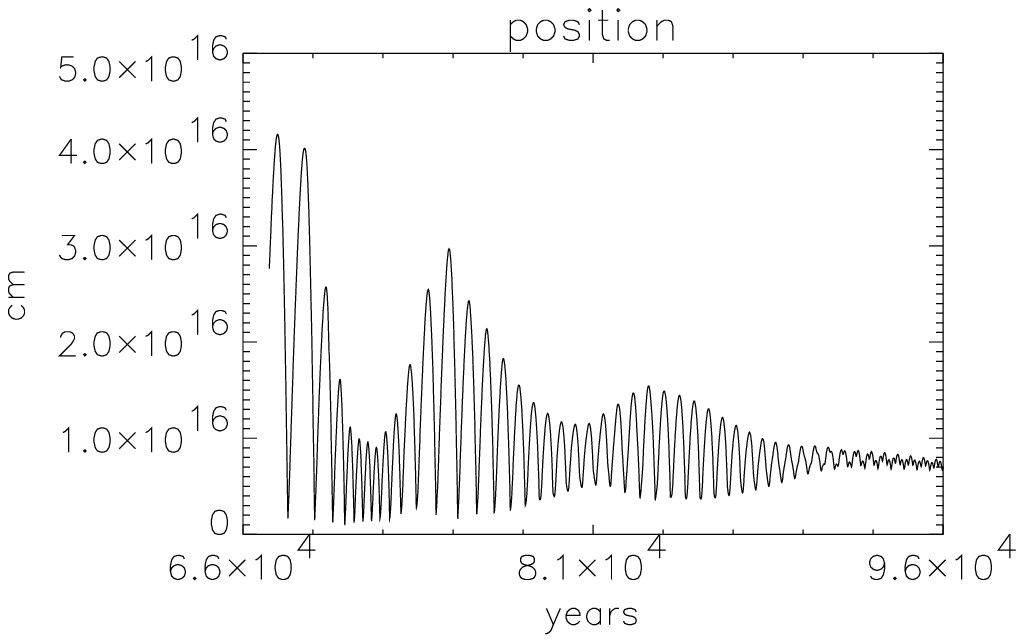,height=3.1cm,width=8.4cm}
           }
\centerline{
      \psfig{figure=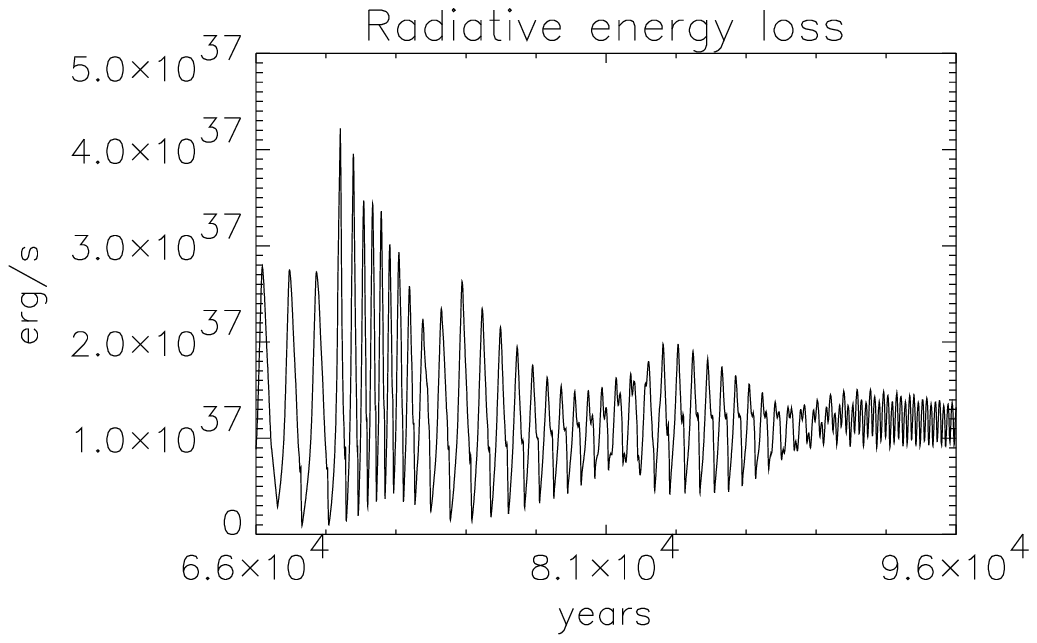,height=3.1cm,width=8.4cm}
           }
\caption{Time evolution of the model WB1\_d7 after running into a 
         disturbance having 6.18 percent of the mass of the CDL.
         {\bf 1. line:} Size of the CDL,
         {\bf 2. line:} Velocity of interface between CDL and HLS,
         {\bf 3. line:} Size of HLS,
         {\bf 4. line:} Radiated energy per time from the HLS.
         }
\label{fig:cdl-mass-disturb}
\efi
interaction zone has passed the density peak, it accelerates again. At
73'500 years the oscillation of the leading shell has nearly recovered
the state it had before the disturbance. However, the absorbed mass
has led to a general slow down of the interaction zone and to the
oscillation of the CDL.
\subsubsection{Switch on, switch off system}
\label{subsubsec:switch_on_off}
The onset of the cooling instability is very sensitive to the
post-shock temperature. Thus, the instability can switch on and off
under the influence of an oscillatory cold dense layer as we will
demonstrate with a CW calculation.  At the beginning of the evolution
shown in Fig.~\ref{fig:system-pn} a
strong wave is traveling backwards and forwards in the CDL. Due to the
oscillation of the CDL, the velocity of the rear boundary interface of
the cooling layer is trans-critical: If the velocity falls below the
critical value, the cooling instability stops, if the interface is
again pushed over the critical value, the instability reappears. The
overstability is of the S1-type. The instability immediately becomes
periodic if the critical post-shock temperature is exceeded.

The third instable time period is most interesting. The instability is
excited by a very small overshooting of the shell over the stationary
value.  Although the excitation is small, the amplitude of the
oscillation is growing up to a value which is inherent to the system
and the applied cooling function. Note that the velocity of the rear
\bfi[htp]
\centerline{
 \psfig{figure=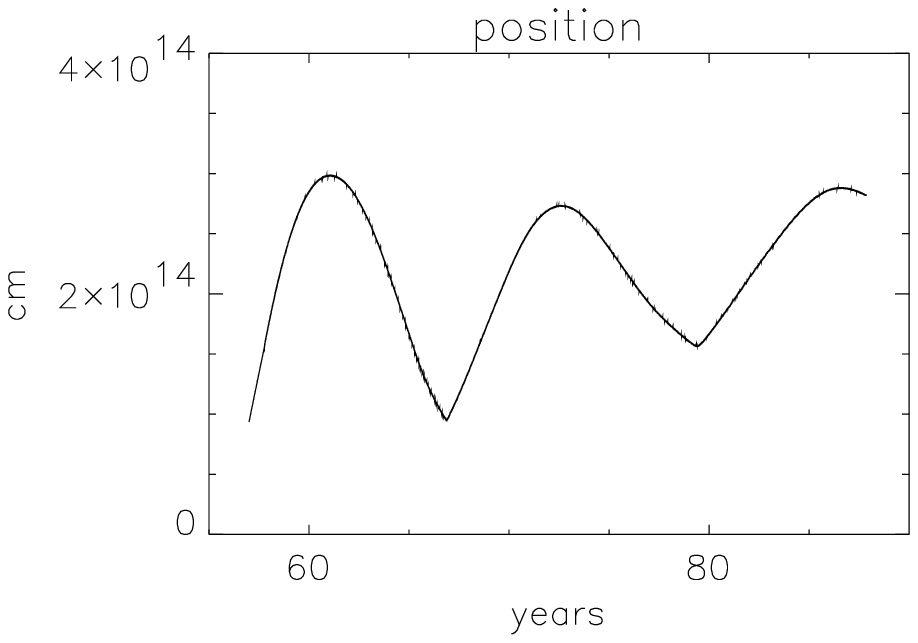,height=3.1cm,width=8.4cm} 
           }
\centerline{
 \psfig{figure=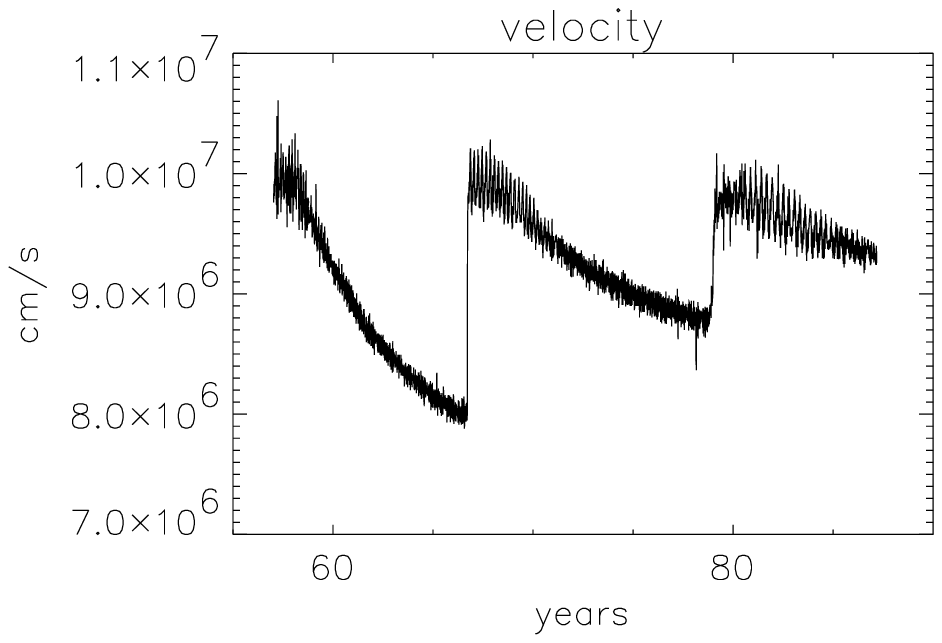,height=3.1cm,width=8.4cm} 
           }
\centerline{
 \psfig{figure=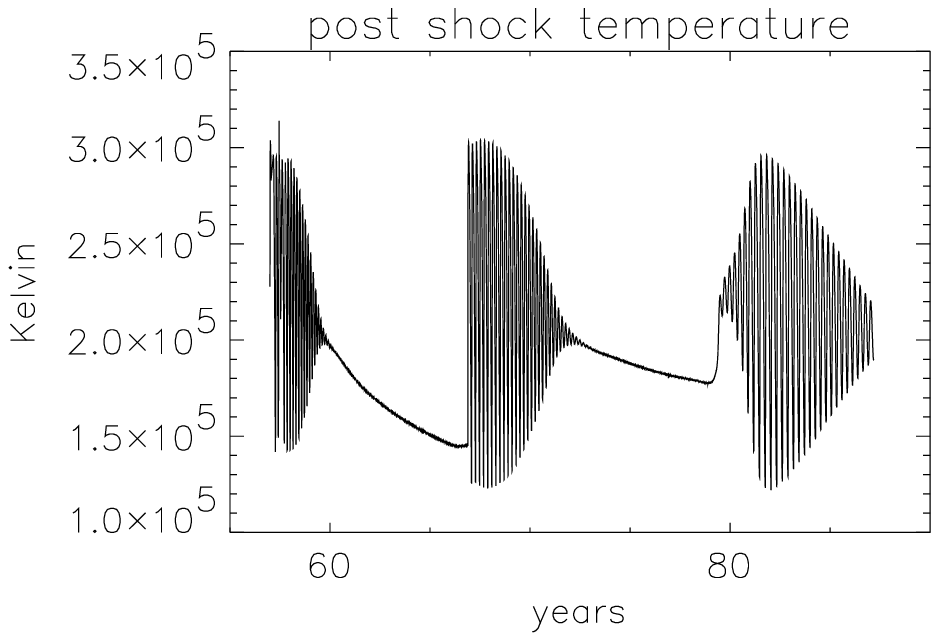,height=3.1cm,width=8.4cm} 
           }
\caption{The overstability switches on and off under the influence
         of an oscillating CDL. 
         {\bf Top:}   Size of the CDL.
         {\bf Middle:} Velocity of the interface between 
                        the CDL and the HLS.
         {\bf Bottom:}   Post-shock temperature of the leading shock.
         }
\label{fig:system-pn}
\efi
boundary of the cooling layer is approximately constant or is even
decreasing while the amplitude of the oscillation is growing. The
stationary amplitude of this oscillation is nearly the same as the one
in the second instable time period. The period of the limit cycle
increases with time due to the decreasing density in the slow wind.

%
%
%
%
\section{Discussion}
\label{sec:discussion}
\subsection{What determines the types and modes?}
The large sample of computations presented in this paper strongly
indicates: If parameterized loss functions are applied to smooth
colliding flows, a differently disturbed radiative shock wave always
relaxes to the same asymptotic solution, whether this solution is
stable or overstable. However, the time scale of the relaxation may
strongly depend on the size and the sign of the excitation. The
stationary post-shock temperature T$_{st}^{ps}$ completely fixes the
stability properties (stability limit, mode, type) of a radiative
shock for a {\it given} RLF.

The cooling overstability in a {\it C-type (I-type)} manifests itself
as runaway cooling. This requires that a significant amount of the
thermal energy of the cooling layer is radiated on a time-scale which
is short compared to the hydrodynamical time-scale. Thus, $\beta$ has
to be sufficiently negative at T$_{st}^{ps}$ and in a sufficiently
wide temperature range below.  For example, this condition may be
fulfilled at temperatures where line cooling of a strong coolant just
sets in and abruptly enhances the cooling efficiency. {\it For these
types, the stability properties are completely determined by the
strongly negative logarithmic slope of the RLF around
T$_{st}^{ps}$}. At least, we see hardly any difference between RLF1
and RLF2 even though the two loss functions are different for lower
temperatures, except for the absolute values of the oscillation period
which are due to the shift in $\Lambda$-direction by half a magnitude.

Runaway cooling and, therefore, negative $\beta$ are also necessary
for the occurrence of the {\it M-type}. Here, however, only the low
temperature part at the rear end of the shell collapses. To prevent
the shell from complete collapse (as in the C-type), the cooling time
in the hotter front part of the shell has to be sufficiently longer
than the time needed to collapse and reconstruct the rear part. This
requires a sufficiently positive $\beta$ at temperatures $T\approx
T_{st}^{ps}$. These conditions may be fulfilled at temperatures
between regions with efficient line cooling for different ions. {\it
For this type, not only the shape of the RLF at T$_{st}^{ps}$ is
important for the stability properties, but the entire shape of the
valley region}.

For the {\it S-types} it is necessary that the cooling time-scale at
every point of the cooling layer remains of the same order as the
hydrodynamical time-scale. {\it With the exception of the rear
boundary layer, the cooling layer must have temperatures associated
with moderately positive $\beta$.} For the temperatures we encounter
in the rear boundary layer, the shape of the RLF may be arbitrary. In
particular, it can be strongly negative (e.g. for smooth types with
post-shock temperatures in a valley region). This does not necessarily
mean that the shape of the RLF for temperatures in the rear boundary
layer is unimportant for the properties of the instability. On the
contrary, we notice {\it the importance of the conditions in the rear
boundary layer for the stability limit and the mode} in many examples.

For the modes the situation seems even more puzzling. We notice the
absence of the overtone modes in the strong form of the F-mode
oscillation. We ascribe this to the total destruction of information
associated with each collapse of the shell. Overtone modes just have
no time to grow. We emphasize the existence of a strong form of the
1O-mode which is able to radiate a substantial amount of
energy. Therefore, the 1O-mode is potentially as important as the
F-mode even though the temperature range of its appearance is
considerably smaller. Finally, we find it remarkable that, despite the
highly nonlinear character of radiative cooling, we find weak and
strong forms of the overstability which, while appearing
phenomenologically so differently, are governed by the same modes.

From the insight into the working mechanism of the different types one
may find constraints for the transitions between them. However, there
is certainly no simple way to exactly determine the stability
properties at every T$_{st}^{ps}$.
\subsection{Temperature regions with different stability properties}
Based on the above discussion we propose five different stability
regions for RLFs. In Fig.~\ref{fig:stab_char} we sketch a RLF which
shows the typical behavior of most high-temperature RLFs (see e.g. 
\cite{coolfunc:Schmutzler} or GEC). We note again that the boundaries 
between the different regions can be given only approximately, since
they cannot be derived by a simple rule.

$\bullet$ Region I: There is always a {\it low-temperature stable
region}, which is also present for fully time-dependent cooling. IGF
and GEC found that shocks with speeds below about 130~km/s are stable.

$\bullet$ Region II: Here we encounter {\it smooth types of the
overstability}. In this region, we expect the shape of the entire
cooling function to be important for the stability properties. In
particular, the rear boundary layer attached to the cold, compressed
sheet can play a significant role. The upper boundary of this region
lies at a temperature which is slightly above the point where the
logarithmic slope of the RLF becomes strongly negative.

$\bullet$ Region III: We find strong {\it runaway instabilities} (C-
or I-type, always F-mode) when T$_{st}^{ps}$ lies in a region of the
RLF with negative logarithmic slope {\bf and} T$_{st}^{ps}$ is
sufficiently above the low-temperature edge of the negative slope
region. Under these conditions, the overstability is nearly
independent of the shape of the RLF at lower temperatures.

$\bullet$ Region IV: This region is associated with a {\it valley in
the RLF} which determines the stability properties. Note, that the
valley is larger than stability region IV (e.g. between $1.6\cdot10^5$
and $1\cdot10^6$~K in RLF2). The shape of the RLF below the valley
seems of minor importance. The strong form of the 1O-mode (M-type) can
only occur in such a region. In addition, we find in this region the
strong form of the F-mode as well as the smooth form of the 1O-mode
and even stable behavior. For this region it seems hardly possible to
predict whether or not, and in which mode and type, the overstability
is present. Any real shock which is slightly disturbed can abruptly
change into any of these types. In this sense, the shock reacts truly
chaotic in this region.
\bfi[tp]
\centerline{\psfig{figure=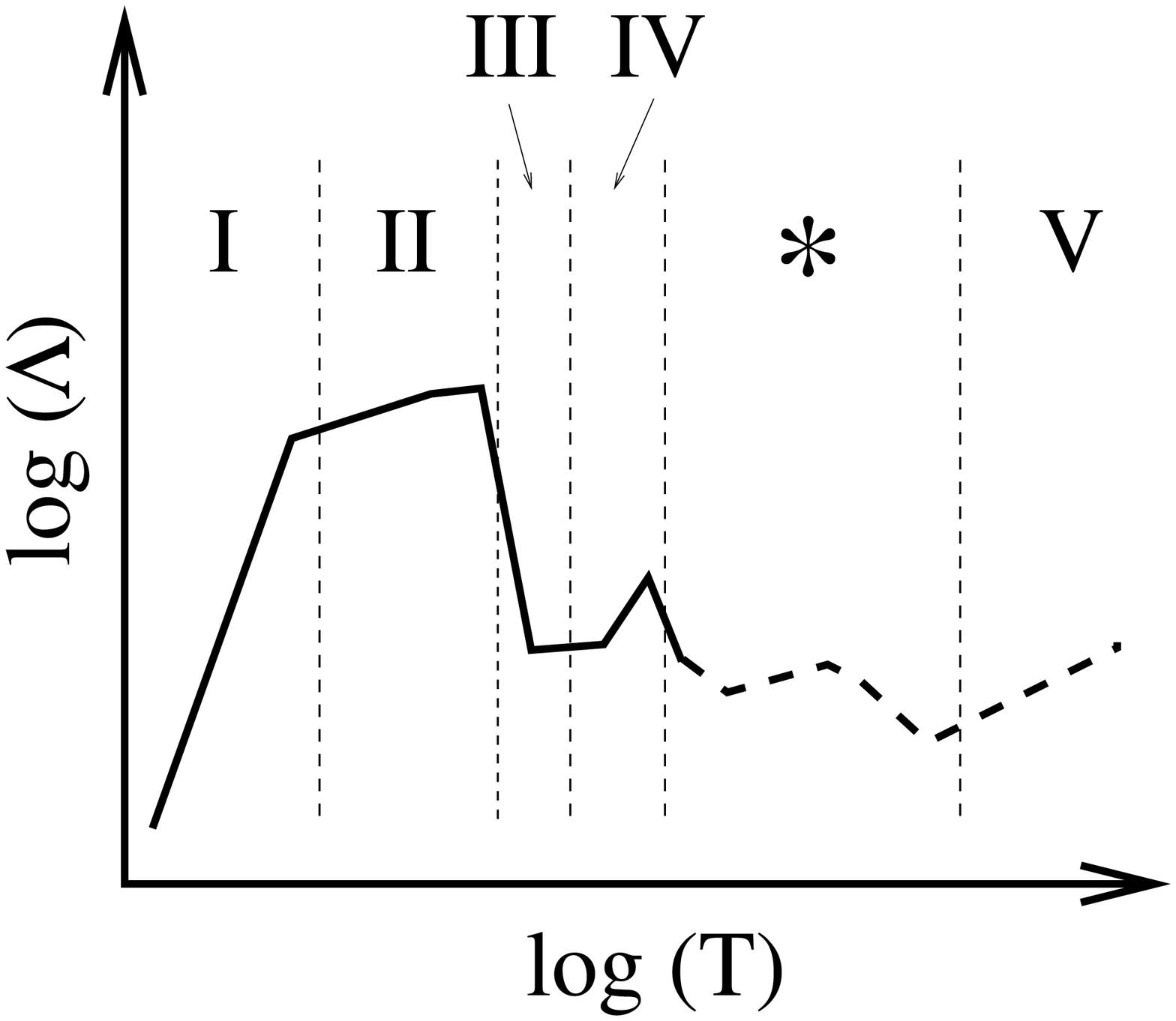,width=6.cm}}
\caption{Sketch of a typical radiative loss function $\Lambda(T)$ on
         a log--log scale. The solid line denotes the region we 
         have extensively investigated, the dashed line denotes the
         high-temperature region we have not investigated. Shown are 
         the five qualitatively different stability regions.
         region I: stable; region II: smooth types; 
         region III: runaway cooling; region IV: valley region;
         region *: behavior not known;
         region V: Bremsstrahlung region.}
\label{fig:stab_char}
\efi

$\bullet$ Region V: {\it High-temperature Bremsstrahlung region} with
$\beta=0.5$. Although we have not done any experiment in this region
we predict a smooth S2-type oscillation if we are sufficiently far
away from the low-temperature boundary where the logarithmic slope
flattens to $\beta \lapprox 0.5$. Near the low-temperature boundary
the existence of a F-mode may be possible. Also in this region the
rear boundary layer of the hot shell may be important for the
stability property.
\subsection{Parameterized radiative loss function versus more 
            elaborated models}
IGF and GEC emphasize the need for fully time-dependent radiation
hydrodynamical calculations which explicitly include ionization,
recombination and non-equilibrium line-cooling, and we completely
agree with them. However, such computations are very costly. At
present and even in the near future, systematic longtime evolution and
multi-dimensional computations cannot be done using such an
approach. One has, therefore, to deal with the use of parameterized
loss functions. Some of the major short comings of parameterized RLFs
compared to fully time dependent computations have already been
mentioned in Sect~\ref{subsec:model-cool}, and were discussed in some
more detail by GEC and IGF. A key point is that the ionization
structure of the plasma does not necessarily correspond to the gas
temperature. The cooling history is often crucial for the
determination of the ionization structure. Consequently, the correct
cooling rate cannot be described by a function of the temperature
alone as in the case of a RLF.
\begin{table*}[t]
\begin{center}
\btab{|c|c||c|c|c|c|c|c|} 
\hline 
  \multicolumn{8}{|c|}{Radiative cooling overstability in colliding flows} \\
\hline 
  \multicolumn{2}{|c||}{types}                                &
  \multicolumn{6}{c|}{characteristics}                       \\
\hline
  F-mode                                                     &
  1O-mode                                                    &
  strength                                                   &
  oscillation amplitude                                      &
   $\tau_{\mbox{cool}} $ /                                   &
  collapsing                                                 &
  cools temp. to                                             &
  secondary                                                  \\

                                                             & 
                                                             & 
                                                             & 
  F-mode                                                     & 
  $\tau_{\mbox{hydro}} $                                     & 
  shell                                                      & 
  nebular T                                                  & 
  shocks                                                     \\ 
\hline  
%
  S1-type                                                    & 
  S2-type                                                    &
  weak                                                       &
  order one                                                  &
  $ \approx $ 1 always                                       & 
  no                                                         &
  no                                                         &
  none                                                       \\
  I-type                                                     &
                                                             &
  moderate                                                   &
  one magnitude                                              &
  temp. $ \lapprox $ 1                                       & 
  slow collapse                                              &
  no                                                         &
  weak                                                       \\
  C-type                                                     & 
  M-type                                                     &
  strong                                                     &
  up to two magnitudes                                       &
  temp. $ <<     $ 1                                         &  
  yes                                                        &
  yes                                                        &
  strong                                                     \\
\hline
\etab
\end{center}
\caption{Schematic summary of the classification and the characteristics for
         the spherically symmetric cooling overstability. T~stands for
         temperature and temp. for temporarily. Note, the oscillation amplitude
         of the 1O-mode is always of order one, also for the strong form.}
\label{tab:summary}
\end{table*}

In the runaway cooling range (u$_{in} \approx 200$~km/s), GEC directly
compare one example of the fully time-dependent model with simulations
applying two differently parameterized models (their Fig.~3). The
global dynamics resulting from the two different approaches 
qualitatively coincide. The oscillation period and amplitude differ by
some percent. {\it For the runaway cooling range} this result is not
surprising. However, in cooling regions~II and IV we assume that the
correct shock dynamics is less well mirrored when using a RLF. Here,
the stability properties cannot be pinned down to the shape of the
loss function around T$_{st}^{ps}$. A time dependent computation in
this regime may yield different overall dynamics.

Generally, newly shocked matter is likely to be underionized, whereas
gas on its cooling track tends to be overionized. {\it In regions~II
and III underionization tends to stabilize the shock} since the
effective cooling function flattens compared to the parameterized
function at this temperature (see Fig.~2 of GEC). The influence of a
possible overionization at lower temperature will not change this
picture, because lower temperatures are of minor importance for
runaway cooling. On the other hand, {\it in the valley stability
region~IV, underionization tends to decrease the logarithmic slope of
the effective loss function at T$_{st}^{ps}$, and has probably a
destabilizing effect}. Overionization along the cooling track may
slightly reduce the size of the collapsing part at the rear end of the
cooling layer.

Partly based on the global similarity between fully time dependent
computations and RLF computations GEC suggested that the overall
stability properties of a radiative shock can be determined by the
local $\beta$ at the high temperature end of the recombination zone,
at least if only small perturbations of the steady state shock are
considered. As has been demonstrated in this paper, this local
condition applies only in runaway cooling regions. For other stability
types we assume that the importance of the cooling behavior at lower
temperatures persists also in a time-dependent computation.

Also GEC have shown that {\it some important details are missed or are
different in the parameterized model,} even in cases where the overall
dynamics are relatively well captured by parameterized cooling. In
addition, IGF have found {\it considerable differences between
calculated spectra for equilibrium and non-equilibrium cooling
shocks}. In general, we feel that the question of how well the use of
parameterized radiative loss functions can replace fully
time-dependent calculations badly needs further investigation. Not
only as far as the global dynamics are concerned but also with regard
to diagnostics. A future perspective may be the use of improved
parameterized models, e.g. the use of several different radiative loss
functions, to better account for the different cooling histories
%
%
%
%
\section{Conclusion}
\label{sec:conclusion}
Before drawing conclusions from this systematic study of the long-term
evolution of the radiative cooling instability in colliding flows we
emphasize that these results hold for parameterized cooling of the form
$N^2 \cdot \Lambda(T)$. Possible relevances for fully time-dependent
cooling are discussed in Sect.~\ref{sec:discussion}. Even though most
of the results are obtained by computing spherically symmetric flows,
the conclusions are only reliable for planar flows, since the size of
the radiative shock is much smaller than the radius of the structure.

We emphasize the many {\it different manifestations of the cooling
overstability}. For both practically important modes, the fundamental
and first overtone mode, we find a strong and a weak form of the
overstability. The key point for the distinction between these two forms
is the ratio of the two governing time scales,
$\eta=\tau_{\mbox{cool}}/\tau_{\mbox{hydro}}$, which in the strong
forms may be temporarily very small but which is always of order one
in the weak forms. We, therefore, propose the introduction of
different phenomenological types as summarized in
Table~\ref{tab:summary}.

The system always relaxes to the {\it same asymptotic oscillatory or
stable solution} which only depends on the stationary post-shock
temperature. This asymptotic solution is independent of the sign and
the size of the disturbance. It is possible to distinguish five
qualitatively different stability regions with different stability
properties. Even though the importance of the different shapes of the
RLF in different temperature regions can be roughly estimated, we find
no easy criterion which determines the stability properties, neither
the stability limit, nor the mode, nor the type. In particular, we
find many examples in which the logarithmic slope of the RLF at the
stationary post-shock temperature is not the only relevant parameter
for the stability behavior.

Probably {\it not all radiative shocks are unstable above a certain
threshold velocity} of the stationary shock. Due to the existence of
'valleys' in the radiative loss function we find stable islands or
smooth types of the instability even for high velocities. A necessary
but not sufficient condition for their occurrence is a logarithmic
slope of the radiative loss function which according to linear theory
is stable at the stationary post-shock temperature. We also find a
strong form of the first overtone mode, which exists in such valley
regions.

We find different {\it evolutionary scenarios when radiative shocks
run into a single density disturbance}. The system reacts aperiodic
only if the length and the mass of the disturbance are appropriately
scaled. If they are too small we find only a moderate, short reaction
but the system remains in the original limit cycle. If the length is
too big the system has time to adjust itself and the oscillations
are modulated according to the new pre-shock conditions. If the mass
is too large, the shell collapses immediately and its history is
destroyed.

The {\it cold compressed gas} downstream of an interstellar radiative
shock can be important for the stability of the shock. When smooth
flows collide, this layer is to a minor degree (ten percent effects)
dynamically important in the strong forms of the
instability. However, a density disturbance in the flow on the order
of a few percent of the mass of the layer leads to a significant
dynamical influence of the layer. As long as we encounter strong forms
of the overstability, the cold dense layer shows a rich internal
structure.

The {\it numerical method} and the {\it coarseness of the numerical
grid} can significantly influence the computed solutions. The entire
instability or any particular mode can vanish only due to a too coarse
mesh combined with an inappropriate numerical method. This effect
is most probably present in all multidimensional calculations.
%
%
%
\vspace{1.cm}
\appendix

\section{Appendix}
\vspace{-0.2cm}
\btabl
\bc
\btab{|c|c||c|c|c|} 
\hline 
number  & range   &  $\beta$ & $ 3 k_B / 2\Lambda_0 $ &$\tau\cdot N$ \\ \hline
 1      & 0.70 -- 1.35 (4) &   4.64   &   4.6 (25) &   4.2 (10) \\ 
 2      & 1.35 -- 2.00 (4) &   2.43   &   3.4 (16) &   2.5 (10) \\  
 3      & 2.00 -- 2.69 (4) &  -0.30   &   6.3 ( 4) &   3.7 (10) \\ 
 4      & 2.69 -- 7.94 (4) &   1.48   &   4.8 (12) &   2.2 (10) \\ 
 5      & 0.79 -- 2.63 (5) &   0.15   &   1.5 ( 6) &   6.0 (10) \\ 
 6      & 2.63 -- 4.17 (5) &  -2.23   &   1.9 (-7) &   2.7 (11) \\ 
 7      & 4.17 -- 5.62 (5) &   0.16   &   5.4 ( 6) &   3.5 (11) \\ 
 8      & 0.56 -- 2.00 (6) &  -0.56   &   4.0 ( 2) &   9.7 (11) \\ 
 9      & 2.00 -- 3.31 (6) &  -2.27   &   1.9 (-8) &   3.8 (13) \\ 
 10     & 0.33 -- 10.0 (7) &   0.00   &   1.2 ( 7) &   1.2 (15) \\
 11     & 0.10 -- 5.00 (9) &   0.50   &   1.2 (11) &   8.2 (15) \\ \hline
\etab
\ec
%
%
%
%
\caption{\bf RLF1 \rm based on the loss function by Cook et al. (1989).}
%
\label{tab:scal1}
\etabl
\vspace{-0.8cm}
\btabl
\bc
\btab{|c|c||c|c|c|} 
\hline 
number   & range       &  $\beta$ & $3 k_B/2\Lambda_0$ &$\tau\cdot N$ \\ \hline
 1       & 0.79 -- 1.82 (4) &   4.06   &    7.2 (23) &     6.9 (10) \\ 
 2       & 1.82 -- 2.51 (4) &   1.71   &    7.6 (13) &     5.5 (10) \\  
 3       & 2.51 -- 3.80 (4) &   0.56   &    6.1 ( 8) &     6.5 (10) \\ 
 4       & 0.38 -- 1.58 (5) &   0.79   &    7.1 ( 9) &     8.8 (10) \\ 
 5       & 1.58 -- 2.19 (5) &  -0.50   &    1.4 ( 3) &     1.4 (11) \\ 
 6       & 2.19 -- 3.98 (5) &  -1.92   &    3.5 (-5) &     8.2 (11) \\ 
 7       & 3.98 -- 5.37 (5) &   0.00   &    2.1 ( 6) &     1.1 (12) \\ 
 8       & 5.37 -- 9.55 (5) &   1.12   &    5.4 (12) &     1.0 (12) \\ 
 9       & 0.95 -- 1.41 (6) &  -0.82   &    1.3 ( 1) &     2.1 (12) \\ 
 10      & 1.41 -- 2.40 (6) &  -2.35   &    5.4 (-9) &     1.2 (13) \\
 11      & 2.40 -- 3.98 (6) &  -0.77   &    6.1 ( 1) &     3.1 (13) \\
 12      & 3.98 -- 8.91 (6) &  -0.09   &    2.1 ( 6) &     7.3 (13) \\
 13      & 0.89 -- 1.74 (7) &  -0.45   &    2.0 (10) &     2.0 (14) \\
 14      & 1.74 -- 3.02 (7) &   0.13   &    8.9 ( 7) &     3.1 (14) \\
 15      & 0.30 -- 1.00 (8) &   0.39   &    7.9 ( 9) &     6.5 (14) \\
 16      & 0.10 -- 5.00 (9) &   0.50   &    6.5 (10) &     4.6 (15) \\ \hline
\etab
\ec
%
%
%
%
%
\caption{ \bf{RLF2} \rm based on the time-dependent loss function by 
          Schmutzler and Tscharnuter (1993).}
%
\label{tab:scal2}
\etabl
\vspace{-0.2cm}
\begin{table*}[[htp]
\bc
\btab{|c||c|c|c||c|c||c|c|c||c|c|c|} 
\hline 
Model  &   \multicolumn{11}{c|}{Wind bubble (WB)}  \\ \hline
       & \multicolumn{3}{c||}{geometry} & \multicolumn{2}{c||}{characteristics}
                   & \multicolumn{3}{c||}{wind of central star}
                   & \multicolumn{3}{c|}{interstellar medium}  \\ \hline
       & r$_{\mbox{min}}$ & r$_{\mbox{max}}$ & $\Delta$ r  
       &   RLF    & T$_{\mbox{\small C}}$/T$_{\mbox{\small H}}$
                   & massloss & vel. & temp. 
                   & N   & vel. & temp.  \\ \hline
       &  cm      &   cm         &   cm
       &          & $\degr$kK
                  & M$_{\odot}/y $         & km/s & $\degr$kK
                  & cm$^{\mbox{\small 3}}$ & km/s & $\degr$kK
\\ \hline
WB1     & 1.6 (17)        & 1 (20)           &  2.4 (13) 
        &    1            & 15/10           
                   & 1 (-4)    & 1.6 (3)  & 15
                   & 1.5       & 0        & 10  \\ \hline
WB1e    & 1.6 (17)        & 1 (20)           &  2.4 (13) 
        &    1            & 10/10           
                   & 1 (-4)    & 1.6 (3)  & 15
                   & 1.5       & 0        & 10  \\ \hline
WB2     & 1.6 (17)        & 1 (20)           &  2.4 (13) 
        &    2            & 15/10           
                   & 1 (-4)    & 1.6 (3)  & 15
                   & 1.5       & 0        & 10  \\ \hline
WB1f    & 1.6 (17)        & 1 (19)           &  3.0 (11) 
        &    1            & 15/10           
                   & 1 (-4)    & 3.0 (3)  & 15
                   & 75        & 0        & 10  \\ \hline
WB2f    & 1.6 (17)        & 1 (20)           &  2.4 (13) 
        &    2            & 15/10
                   & 1 (-4)    & 2.8 (3)  & 15
                   & 15        & 0        & 10  \\ \hline \hline
       & \multicolumn{11}{c|}{Supernova remnant (SNR)}  \\ \hline
       & \multicolumn{3}{c||}{geometry} & \multicolumn{2}{c||}{characteristics}
                   & \multicolumn{3}{c||}{stellar blast}
                   & \multicolumn{3}{c|}{interstellar medium}  \\ \hline
        & r$_{\mbox{min}}$ & r$_{\mbox{max}}$ & $\Delta$ r  
        & RLF      & T$_{\mbox{\small C}}$/T$_{\mbox{\small H}}$
                   & mass & energy & temp. 
                   & N   & vel. & temp.  \\ \hline
        &  cm      &   cm         &   cm
        &          & $\degr$kK
                   & M$_{\odot}$              & ergs & $\degr$kK
                   & cm$^{\mbox{\small 3}}$   & km/s & $\degr$kK
 \\ \hline
SNR1     & 1 (16)        & 1 (21)           &  4.9 (14) 
         &    1             & 15/10           
                   & 50     & 1 (51)      & 15
                   & 1      & 0           & 10  \\ \hline
SNR2     & 1 (16)        & 1 (21)           &  6.1 (13) 
         &    2             & 15/10           
                   & 50     & 1 (51)      & 15
                   & 1      & 0           & 10  \\ \hline \hline
       &   \multicolumn{11}{c|}{Colliding winds (CW)}  \\ \hline
       & \multicolumn{3}{c||}{geometry} & \multicolumn{2}{c||}{characteristics}
                   & \multicolumn{3}{c||}{present wind of CS}
                   & \multicolumn{3}{c|}{earlier wind of CS}  \\ \hline
        & r$_{\mbox{min}}$ & r$_{\mbox{max}}$ & $\Delta$ r  
        & RLF      &  T$_{\mbox{\small C}}$/T$_{\mbox{\small H}}$
                   & massloss & vel. & temp. 
                   & massloss & vel. & temp. \\ \hline 
        &  cm      &   cm         &   cm
        &          & $\degr$kK
                   & M$_{\odot}/y$ & km/s & $\degr$kK
                   & M$_{\odot}/y$ & km/s & $\degr$kK 
\\ \hline
CW      & 5 (15)        & 1 (17)           &  2.3 (10) 
        &    2          & 15/15           
                   & 1.1 (-7)  & 2.5 (3)   & 15
                   & 2 (-6)    & 20        & 15  \\ \hline \hline
        &   \multicolumn{11}{c|}{Colliding planar flows (CPF)}  \\ \hline
        & Instability  & \multicolumn{2}{c||}{geometry} 
                       & \multicolumn{2}{c||}{characteristics}
                       & \multicolumn{3}{c||}{driving wind}
                       & \multicolumn{3}{c|}{interstellar medium}  \\ \hline
        & type     & domain           & $\Delta$ r  
        & RLF      &  T$_{\mbox{\small C}}$/T$_{\mbox{\small H}}$
                   & density   & vel. & temp. 
                   & density   & vel. & temp. \\ \hline 
        &          &   cm         &   cm
        &          &  $\degr$kK
                   & cm$^{\mbox{\small -3}}$ & km/s &  $\degr$kK
                   & cm$^{\mbox{\small -3}}$ & km/s &  $\degr$kK
\\ \hline
CPF\_25 &    C      & 1 (20)           &  2.4(13) 
        &    2      & 15/10           
                    & 0.025  & 2800    & 15
                    & 15     &   0     & 10  \\ \hline
CPF\_35 &    M      & 1 (20)           &  2.4(13) 
        &    2      & 15/10           
                    & 0.035  & 2800    & 15
                    & 15     &   0     & 10  \\ \hline
CPF\_45 &    S2     & 1 (20)           &  2.4(13) 
        &    2      & 15/10           
                    & 0.045  & 2800    & 15
                    & 15     &   0     & 10  \\ \hline
CPF\_55 & stable    & 1 (20)           &  2.4(13) 
        &    2      & 15/10           
                    & 0.055  & 2800    & 15
                    & 15     &   0     & 10  \\ \hline
\etab
\ec
\caption{Sample of computed colliding smooth flows.
         As a representative for the {\it wind bubble} flow we take 
         NGC~6880 (WR~136). We assume model parameters as derived
         by Schmutz et al. (1989). In addition, we compute models
         with faster winds (models .f). For the {\it supernova 
         remnant} we assume the same parameters
         as Bertschinger (1986). 80 percent of the initial
         energy is thermal in our models. For the 
         {\it colliding wind} we take the same
         parameters as Icke et al. (1992) took for their (2D) 
         model of an (aspherical) planetary nebulae for the 
         {\it colliding wind} model. r$_{\mbox{min}}$ and 
         r$_{\mbox{max}}$ denote the minimum and maximum extent
         of the computational domain. $\Delta$r denotes 
         the finest spatial discretization of the computational
         mesh. T$_{\mbox C}$ denotes the cooling limit, T$_{\mbox H}$
         the heating limit. CS stands for central star.}
\label{tab:sample}
\end{table*}
\nocite{hotstar:schmutz-hamann-wesselowski}
\nocite{pn:icke-balick-franck}
\begin{table*}[htp]
\bc
\btab{|c|c|c|c|c|c|c|c|c|c|c|} 
\hline 
 number                          &
 model                           & 
 $\xi$                           & 
 $\lambda_0$                     & 
 $\lambda_0 / \ell_{c} $         & 
 M$_{\mbox{d}}$                  & 
 M$_{\mbox{d}}$/M$_{\mbox{hls}}$ &
 M$_{\mbox{d}}$/M$_{\mbox{cdl}}$ &
 $\phi_0$                        &
 response                       &
 $\tau_{rel} / \tau_{0} $         \\
    &  WB1    & & $10^{17}$ cm &     & $10^{33}$ gr  &  
    &    &    & &  \\ 
\hline
0  &  \_d0  & 64   & 30   & 8.8 & 7000 & 1170 & 10    & 0.17 & co & \\ 
\hline
 1 & \_d2.1 & 0.75 & 0.67 & 0.2 & 1.83 & 0.305 & 0.0026 & 0.07 & ap & 2.4 \\  
 2 & \_d2.2 & 0.75 & 0.67 & 0.2 & 1.83 & 0.305 & 0.0026 & 0.14 & ap & 2.4 \\ 
 3 & \_d2.3 & 0.75 & 0.67 & 0.2 & 1.83 & 0.305 & 0.0026 & 0.52 & ap & 2.5 \\ 
\hline
 4 & \_d3   & 0.5  & 0.67 & 0.2 & 1.22 & 0.203 & 0.0017 & 0.14 & ap & 1.6 \\ 
 2 & \_d2.2 & 0.75 & 0.67 & 0.2 & 1.83 & 0.305 & 0.0026 & 0.14 & ap & 2.4 \\ 
 5 & \_d4.1 &  1   & 0.67 & 0.2 & 2.44 & 0.407 & 0.0034 & 0.14 & ap & 3.0 \\ 
\hline
 6 & \_d4.0.2 & 0.5 & 0.028 & 0.0083 & 0.10 & 0.017 & 0.00014 & 0.14 & nop & \\
 4 & \_d3   & 0.5   & 0.67  & 0.2 & 1.22 & 0.203 & 0.0017 & 0.14 & ap & 1.6 \\ 
14 &  \_d5 & 0.5  &  4.34 & 1.28  & 7.91 & 1.318 & 0.0111  & 0.14 & mp & \\ 
\hline
 7 & \_d4.0.1 & 1  & 0.0566 & 0.0166 & 0.20 & 0.033 & 0.00028 & 0.14 & nop & \\
 8 & \_d4.0.0 & 1 & 0.17 & 0.05 & 0.61 & 0.102  & 0.00086 & 0.14 & ap & 1.6 \\
 9 & \_d4.0 & 1  &  0.33 & 0.1  & 1.22 & 0.203  & 0.0017  & 0.14 & ap & 1.7 \\
 5 & \_d4.1 & 1  &  0.67 & 0.2  & 2.44 & 0.407  & 0.0034  & 0.14 & ap & 3.0 \\
10 &  \_d4  & 1  &  1.34  & 0.4  & 4.88 & 0.813 & 0.0069  & 0.14 & mp & \\ 
11 &  \_d6  & 1  &  4.02  & 1.18 & 14.6 & 2.433 & 0.0206  & 0.14 & mp & \\ 
12 &  \_d7  & 1  & 12.06  & 3.55 & 43.9 & 7.317 & 0.0618  & 0.14 & mp &  \\ 
\hline
13 &  \_d1 & 4.14 & 30.00 & 8.8  &  437 & 73    & 0.6155  & 0.17 & mp & \\ 
14 &  \_d5 & 0.5  &  4.34 & 1.28 & 7.91 & 1.318 & 0.0111  & 0.14 & mp & \\ 
 \hline
\etab
\ec
\caption{Sample of the disturbed flows discussed in Sect.~6. We disturb 
         model WB1 (see Table~4) according to Eq.~13 at a time of 
         approximately 69'000~years where we encounter typical I-type 
         overstability. The period of the limit cycle, $\tau_0$, 
         is approximately 1100 years. The mean mass  M$_{\mbox{ls}}$ 
         of the leading shell is approximately $6\cdot 10^{33}$~gram.
         The rear boundary of the leading shell has a mean velocity 
         of $9.95\cdot10^6$~cm/s. Multiplication with the period
         of the limit cycle yields a characteristic length 
         $\ell_{c}\approx 3.4\cdot 10^{17}$~cm. The mass M$_{\mbox{cdl}}$ 
         of the cold dense 
         layer is $7.1\cdot10^{35}$~gram. M$_{\mbox{d}}$ denotes
         the mass, $\lambda_0$ the length and $\xi$ the amplitude of the 
         disturbance. $\phi_0$ denotes the phase of the limit cycle 
         at which the leading shock hits the disturbance. $\tau_{rel}$
         denotes the relaxation time for the aperiodic cases ({\it ap}). 
         The other cases are: {\it nop:} noisy original periodic evolution, 
         {\it mp:} modulated periodic evolution, ({\it co:}) collapse
         of the leading shell. Note that some models may be
         listed several times to better illustrate dependences.}
\label{tab:disturb-i}
\end{table*}
%

%
%
%

%
\begin{acknowledgements}
 We thank the crew of ETH running the Cray supercomputer for 
 steady support, Harry Nussbaumer for his comments on the
 manuscript, and the referee, Roger Chevalier, for useful
 comments which improved the manuscript.
 R.W. was supported by the ETH Forschungskredit, D.F. by the 
 Schweizerischer Nationalfond.
\end{acknowledgements}
\bibliographystyle{aabib} 
%

\begin{thebibliography}{}

\bibitem[\protect\astroncite{Berger}{1985}]{berger:mesh1}
Berger M.~J., 1985, Lectures in applied Mathematics 22, 31

\bibitem[\protect\astroncite{Berger \& LeVeque}{1989}]{amr:be-rjl-amr-1}
Berger M.~J., LeVeque R.~J., 1989,
\newblock in AIAA 9th Computional Fluid Dynamics Conference, Buffulo, NY, No.
  89--1930--CP in AIAA--Papers

\bibitem[\protect\astroncite{Bertschinger}{1986}]{radshock:bertsch-stab2}
Bertschinger E., 1986, ApJ 304, 154

\bibitem[\protect\astroncite{Chevalier \&
  Imamura}{1982}]{radshock:cheva-ima-stab1}
Chevalier R.~A., Imamura J.~N., 1982, ApJ 261, 543,
\newblock (CI)

\bibitem[\protect\astroncite{Chevalier \& Imamura}{1983}]{sim:ci1}
Chevalier R.~A., Imamura J.~N., 1983, ApJ 270, 554

\bibitem[\protect\astroncite{Colella \& Glaz}{1984}]{meth:col-gl-real}
Colella P., Glaz H.~M., 1984, Journal of Computional Physics 59, 264

\bibitem[\protect\astroncite{Colella
  et~al.}{1986}]{sources:colella-majda-Roytburd}
Colella P., Majda A., Roytburd V., 1986, SIAM J.Sci. Stat. Comput. 4, 1059

\bibitem[\protect\astroncite{Cook et~al.}{1989}]{coolfunc:Cook}
Cook J.~W., Cheng C.-C., Jacobs V.~L., Antiochos S.~K., 1989, ApJ 338, 1176

\bibitem[\protect\astroncite{Dgani et~al.}{1995}]{hstab:dgani-soker-cadavid}
Dgani R., Soker N., Cadavid M.~L., 1995, AJ 110(4), 1894

\bibitem[\protect\astroncite{Falle}{1975}]{radshock:falle-1}
Falle S. A. E.~G., 1975, MNRAS 172, 55

\bibitem[\protect\astroncite{Falle}{1981}]{radshock:falle-2}
Falle S. A. E.~G., 1981, MNRAS 195, 1011

\bibitem[\protect\astroncite{Gaetz
  et~al.}{1988}]{radshock:gaetz-edgar-chevalier}
Gaetz T.~J., Edgar J., Chevalier R.~A., 1988, ApJ 329, 927,
\newblock (GEC)

\bibitem[\protect\astroncite{Houck \&
  Chevalier}{1992}]{radshock:houck-chevalier}
Houck J.~C., Chevalier R.~A., 1992, ApJ 395, 592

\bibitem[\protect\astroncite{Icke et~al.}{1992}]{pn:icke-balick-franck}
Icke V., Balick B., Frank A., 1992, A\&A 253, 224

\bibitem[\protect\astroncite{Imamura \&
  Wolff}{1990}]{radshock:imamura-wolff-stab1}
Imamura J.~N., Wolff M.~T., 1990, ApJ 355, 216

\bibitem[\protect\astroncite{Imamura
  et~al.}{1984}]{radshock:imamura-wolff-durisen}
Imamura J.~N., Wolff M.~T., Durisen R.~H., 1984, ApJ 276, 667

\bibitem[\protect\astroncite{Innes}{1988}]{radshock:innes}
Innes D.~E., 1988,
\newblock in W. Kundt (ed.), Supernova Shells and Their Birth Events, Vol. 316
  of {\em Lecture Notes in Physics\/}, p.~74

\bibitem[\protect\astroncite{Innes
  et~al.}{1987}]{radshock:innes-giddings-falle-2}
Innes D.~E., Giddings J.~R., Falle S. A. E.~G., 1987, MNRAS 226, 67,
\newblock (IGF)

\bibitem[\protect\astroncite{Klingenstein}{1994}]{source:klingenstein}
Klingenstein P., 1994,
\newblock Hyperbolic Conservation Laws: Errors of the shock location,
\newblock Technical Report 94--07, Seminar f\"{u}r {A}ngewandte {M}athemtik,
  Eidgen\"{o}ssische {T}echnische {H}ochschule, {Z}\"{u}rich, Switzerland

\bibitem[\protect\astroncite{Langer
  et~al.}{1981}]{radshock:langer-chanmugam-shaviv-stab1}
Langer S.~H., Chanmugam G., Shaviv G., 1981, ApJ 245, L23

\bibitem[\protect\astroncite{Langer
  et~al.}{1982}]{radshock:langer-chanmugam-shaviv-stab2}
Langer S.~H., Chanmugam G., Shaviv G., 1982, ApJ 258, 289

\bibitem[\protect\astroncite{Le{V}eque \&
  Yee}{1990}]{meth:rjl-yee-stiffsources}
Le{V}eque R., Yee H., 1990, Journal of Computional Physics 86, 187

\bibitem[\protect\astroncite{Plewa}{1995}]{radshock:plewa-1995}
Plewa T., 1995, MNRAS 275, 143

\bibitem[\protect\astroncite{Ryu \& Vishniac}{1991}]{sim:ryvi1}
Ryu D., Vishniac E.~T., 1991, ApJ 368, 411

\bibitem[\protect\astroncite{Schmutz
  et~al.}{1989}]{hotstar:schmutz-hamann-wesselowski}
Schmutz W., Hamann W.-R., Wessolowski U., 1989, A\&A 210, 236

\bibitem[\protect\astroncite{Schmutzler \&
  Tscharnuter}{1993}]{coolfunc:Schmutzler}
Schmutzler T., Tscharnuter W.~M., 1993, A\&A 273, 318

\bibitem[\protect\astroncite{Strickland \&
  Blondin}{1995}]{radshock:strickland-blondin-1995}
Strickland R., Blondin J.~M., 1995, ApJ 449, 727

\bibitem[\protect\astroncite{Walder}{1993}]{walder-thesis}
Walder R., 1993,
\newblock {\em Ph.D. thesis\/}, ETH {Z}\"{u}rich

\bibitem[\protect\astroncite{Wolff
  et~al.}{1989}]{radshock:wolff-gardner-wood-stabshock1}
Wolff M.~T., Gardner J.~H., Wood K.~S., 1989, ApJ 346, 833

\end{thebibliography}
%

%
%
%
%
\end{document}